# Improved Volterra Kernel Methods with Applications to the Visual System

Richard T. Miller

Vladimir Y. Vildavski

Anthony M. Norcia

The Smith-Kettlewell Eye Research Institute

Short Title:  Improved kernel methods

Acknowledgements: Supported by EY06579, EY015790, and the Smith-Kettlewell Eye Research Foundation and the Louis and Gladyce Foster Family Foundation.  We thank Mark W. Pettet for critical comments on the manuscript.

# ABSTRACT

**Background**

Volterra analysis and its variants have long been prominent among methods for modeling multi-input non-linear systems. The product of Volterra analysis, the Volterra kernels, are particularly suited to quantifying intra- and inter-input interactions. They are also readily interpretable, which means that they can be related directly to physical behaviors, and more distantly, to the underlying processing mechanisms of the system being tested. However, accurate estimation of a sufficient set of classical kernels is often not possible for complex systems because the number of kernels that need to be determined, and hence experiment time, increases radically with system memory, response frequency bandwidth, and non-linear interaction order. Practical approaches to kernel estimation often involve forced reductions of the generality of the analysis that in turn compromise interpretability.

**Principal Findings**

Here we illustrate the effects on kernel interpretability of two common reductions, slow-stimulation and the use of binary inputs, using both numerical simulations and data from a Visual Evoked Potential experiment. We show how a non-standard version of binary analysis, involving a different coding of the inputs and the use of particular groupings of kernel slices, improves kernel interpretability.

**Significance**

We bring together, in a comprehensive fashion, all of the mathematical considerations needed to apply and interpret the results of Volterra kernel analyses. The input-coding method we describe allows one to correctly quantify multi-input interactive effects that occur both within the separate input channels and across them.

# 1 Introduction

Sensory systems display a number of non-linear behaviors such as adaptation and spatio-temporal interaction. Because of this, non-linear systems analysis approaches have been used to study their behavior. One purpose of this paper is to present a method for non-linear analysis based on the Volterra series that has certain interpretative advantages over other commonly used approaches.

As a prelude to introducing this method, we describe the Volterra series formalism in detail, expanding upon a number of points that are important for correct interpretation of experimental results. We do this in Sections 2 and 3. Some of this material not been made explicit previously, and much of it will be unfamiliar to readers who are not experts in Volterra analysis. The introductory material has been formulated with an eye to its utility in Sections 5 and 8, which illustrate potential pitfalls of commonly used simplifications of the general Volterra model. In Section 9 we provide an alternative framework that alleviates some interpretive problems. Our theoretical discussions are illustrated throughout with examples from simulated systems. In Section 10, we apply our analysis methods, for the first time, to real experimental data. Taken as a whole, our goal is to present a tutorial in Volterra and associated methods that takes the reader through all the steps necessary to correctly setup experiments and to interpret the results.

*1.1 Modeling and interpreting non-linear interactions*

Interaction is the essential ingredient of complexity in system behavior. We are interested in techniques by which it may be quantified, and, even better, understood. Take, for example, a 2-input system capable of responding to pulse trains presented to each of its inputs. A measure of interaction is obtained by taking the difference between the response to simultaneous pulses on the two inputs and the sum of the responses to pulses presented separately. Because the input lines are physically separated, this interaction may be considered "spatial". Interaction may also be "temporal". For example, a single-input system might show interaction between pulses presented at various delays. Such interaction is sometimes called "adaptation", since it has to do with response changes dependent on whether or not the system experienced previous pulses.

In this paper, we examine interaction by means of signal processing techniques that are a part of the general class known as "non-linear" analysis. That the analysis must be non-linear is forced – the more traditional and simpler methods of "linear" analysis do not support the notion of interaction as we use it.

Analysis in the sense of the previous paragraph means modeling -- the development of a *theoretical* structure that captures some or all of the key features of the actual system being studied. These structures are typically parameterized frameworks that are made specific by fixing the parameters. Models come in many flavors. Many are "predictive", providing accurate computation of system responses to specific system inputs, without regard to whether the details of the computation are in any way analogous to the workings of the actual system. Predictive models are sometimes referred to as "black-box" models – inputs are inserted, and outputs are generated, but the works are hidden. Black box models, also called "functional" models, are common in psychophysics (Wilson, 1999), single-unit physiology (Carandini, Heeger & Movshon, 1997, Cavanaugh, Bair & Movshon, 2002, Simoncelli & Heeger, 1998), evoked potentials (Candy, Skoczenski & Norcia, 2001, Hou, Pettet, Sampath, Candy & Norcia, 2003, Victor & Conte, 2000) and functional imaging (Zenger-Landolt & Heeger, 2003). The most extreme example of black-box modeling comes from undifferentiated neural nets, where the testing and connection updating procedure can produce parameters that allow essentially perfect reconstruction, but because the parameters are distributed throughout the network, their interpretation is obscure.

Other models are abstractly "quantitative". In these, analysis components capture interesting system properties numerically, but again, not necessarily analogously. The "kernel slices" that are our main subject, are real-valued functions, and thus, quantitative objects. They are related, but not typically equal to, actual system responses. Still other models are "structural", describing the system organization in terms of idealized components that are closely analogous to the real system components. Finally, we have the "metaphorical" models. These depict a *fictional* structure that, in useful ways, works like the real system.

Often models combine the flavors. Although it might seem that a structural model would be the most desirable, that is not necessarily the case. The ultimate structural model is the system itself, and the complexity can be overwhelming. One is often better served by reducing the full complication of the real system to the simplest metaphor that captures behaviors of interest.

*1.2 The Volterra model*

In this work, we seek to achieve a characterization of spatial and temporal interactions, so multi-frequency Fourier methods {Victor, 1977 #2288}{Regan, 1988 #1084} are precluded. Among time-domain methods, the classic technique is Volterra analysis. Somewhat more recent is the statistics-based Wiener variant. Both methods have great theoretical power, each being able to characterise very general groups of stationary (that is, non-adapting, in a sense we make precise below, and different from the usage cited above) systems having continuous, multiple, real-valued inputs and outputs. Volterra kernels also quantify important aspects of these systems and can be used to predict system responses to a wide range of inputs. In this paper, we will develop a reduced variant of Volterra analysis that has less theoretical reach than the classic Volterra or Wiener methods, but that has the benefit of being actually applicable in real-world circumstances. The reduction we describe preserves interpretational features of classic Volterra analysis that are better than those directly available in Wiener analysis, while eliminating the main drawbacks of the classical Volterra approach. In particular, our reduction, like classic Volterra analysis and unlike Weiner analysis, provides an easy to grasp metaphor for the input-output behavior of the system.

All variants of Volterra analysis are based on the "Volterra expansion" shown in Equation (1). This fundamental tool of systems analysis is designed to provide a general description of real-input, real-output, time-independent systems (Doyle, Pearson & Ogunnaike, 2002, Marmarelis, 2004, Nelles, 2001, Westwick & Kearny, 2003). These systems are active in time, receiving inputs and returning outputs, both of which are time progressions of real numbers. "Time independence" means that the relationship between the input and output does not depend on absolute time. This assumption implies that the system properties themselves, as opposed to the particular outputs, do not change with time – long-time adaptation is precluded. Initially, we restrict to single-input, single-output (SISO) systems. That keeps the notation, which has a tendency to the cumbersome, as simple as possible. The extension to multiple-input systems will be easy since it constitutes, as will be shown, a reduction of the SISO case.

In the equation, $t$ represents time, and $r(t)$ the system response at time $t$ driven by input $s(t)$.

$$r(t) = L_0 + \int_0^\infty L_1(d_0)s(t-d_0)dd_0 + \int_0^\infty \int_{d_0}^\infty L_2(d_0,d_1)s(t-d_0)s(t-d_1)dd_1 dd_0 +$$
$$\int_0^\infty \int_{d_0}^\infty \int_{d_1}^\infty L_3(d_0,d_1,d_2)s(t-d_0)s(t-d_1)s(t-d_2)dd_2 dd_1 dd_0 + \ldots$$
(1)

The multi-variate functions $L_n$, traditionally called the "kernels" embody the parameters of the system, which are the individual function values $L_n(d_0, d_1, d_2, \ldots)$. We call these "kernel elements". Our assumption of system time independence is reflected in the fact that the kernels are not functions of time, but rather of dummy variables $d_j$, which, based on their appearance in the combination $t - d_j$ as arguments to the input function, may be thought of as "relative delays". Note that there is no requirement that the kernel functions be particularly regular. They need not even be continuous; merely being integrable would satisfy the functional form.

The expansion as shown embodies another assumption about the system, namely that it is "causal". This means that the output at any given time is uniquely determined by the inputs that have arrived at or before that time. The lower bound of each outer integral being zero implements this assumption. It follows that the delays are all zero or positive.

While the generality of Equation (1) is theoretically useful, no real experiment can fix the infinitely many free parameters represented by the kernel elements. We need to make the determination task finite, which is usually done by reducing to the discrete-time approximate expansion shown in Equation (2), adapted from (Benardete & Victor, 1994).

$$r(t) = L_0 + \sum_{d_0=0}^{T-1} L_1(d_0)s(t-d_0) + \sum_{d_0=0}^{T-1} \sum_{d_1=d_0}^{T-1} L_2(d_0,d_1)s(t-d_0)s(t-d_1) +$$
$$\sum_{d_0=0}^{T-1} \sum_{d_1=d_0}^{T-1} \sum_{d_2=d_1}^{T-1} L_3(d_0,d_1,d_2)s(t-d_0)s(t-d_1)s(t-d_2) + \ldots$$
(2)

Here $t$ is discrete time, and again, $s$ is the input, and $r$ the output. Both $s$ and $r$ are taken to be functions of $t$. $L_0$ is the zeroth-order kernel, $L_1$ the first-order kernel, etc. The kernels are functions of discrete relative delays $d_0$, $d_1$, $d_2$, etc. As shown, the expansion embodies another commonly made assumption about the system, namely that it has "finite memory", which mean that the outputs depend only on sufficiently recent inputs. The summation bound $T$ limits the magnitude of the discrete delays that can affect the value of the response, and thus represents the extent of "memory" in the expansion. Discrete time, the discrete delays, and the summation bound are integers. In applying Equation (2) to a real system, these numbers are multiplied by a real timestep $\Delta_t$ to yield real time, delays, and system memory.

Taken together, our previous assumptions imply that each term of (2) depends on a finite number of parameters. To make the full parameter set finite, we must assume that the expansion comprises only a finite number of terms. This has important ramifications since the $n^{th}$ term in the expansion quantifies the interaction between $n$, possibly repeated, input samples. This can be seen in the fact that $n$ delays comprise the argument to the $n^{th}$ kernel and that the multiplier of each $L_n(d_0, d_1, \ldots)$ is a product of $n$ input samples. We shall examine interaction in more detail later. Our point here, however, is that in truncating the expansion we are tacitly assuming that the system has no interactions above a certain order. This is justified by the expectation that effects dependent on *multiply-interacting* input samples eventually become insignificant as the number of participating samples increases. Typically this expectation is tested after the fact. The system is analyzed assuming a bound on the order, and if the resulting predictions correspond to observed behavior, the assumption is taken to be valid.

The discrete kernel of order $n$ may be represented as the graph of a function on a compact region of an n-dimensional grid. The upper panel of Figure 1 shows an example kernel where $n = 2$. In this depiction, the delay pairs specifying the kernel elements are treated even-handedly except for the restriction coming from the lower summation bounds in (2) that imply $d_0 \leq d_1$. Note that the constraint on the delays means that the domain of the kernel occupies the part of a square grid on and above the "main diagonal", the points defined by the condition $d_0 = d_1$. The lower panel of the figure shows the same kernel elements graphed by "kernel slice". The elements in a slice have delay pairs of the form $(0, d^*) + d$, where $d^*$ is held constant while $d$ varies. These sets of pairs define the so-called "super-diagonals" of the grid square, the 1-dimensional grid arrays equal or parallel to the main diagonal. The pair $(0, d^*)$ is called the "slice signature", and the function defined by the restriction of the kernel to the corresponding super-diagonal is called the "slice function". As we shall show below (Section 2.1), these objects are interesting because the signatures quantify the *type* of an interaction among system inputs, while the functions quantify its *temporal course*.

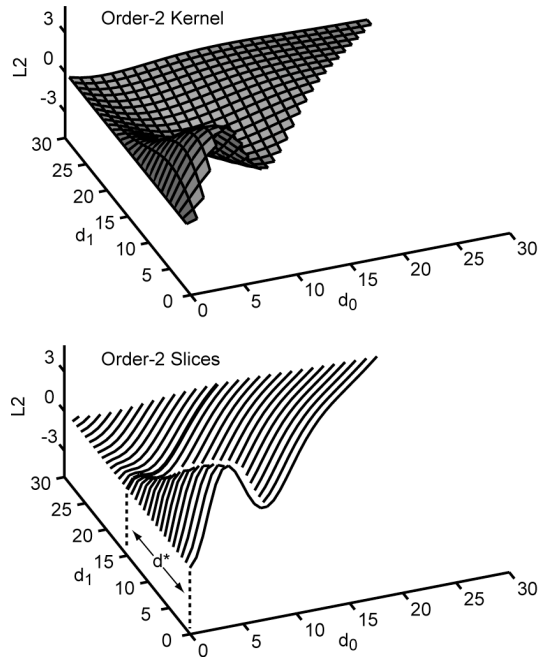

*Figure 1.* Example of an order-2 kernel and associated slices. The top panel shows a surface representing the graph of the order-2 kernel in Equation 2. The kernel is a function of the pair of delays, $d_0$ and $d_1$. The effect of the double summation in Equation 2 is the (two-dimensional) convolution of the kernel with copies of the inputs. The bottom panel shows the same surface but depicted as a set of traces, corresponding to "slices" of the surface taken over sets of delay pairs running parallel to the main diagonal. The heavy trace is the slice corresponding to the signature $(0, d^*)$. The double summation can be rewritten as first, a summation along these parallel diagonals, effectively convolving each slice with an offset product of the input s with itself, and then as a summation over slices. This "slice representation" is helpful for kernel interpretation.

*1.3 Sources of error*

Extraction of kernels is essentially an exercise in high-dimensional curve fitting in that there is a parameterized class of approximating systems from which one is to choose the particular member that in some sense best fits the system being analyzed. A measure of the difference between the best fitting member and the underlying system is the fitting error. Traditionally, one distinguishes two sources of fitting error: First, the underlying system typically is not contained in the approximating class, and so there is a non-zero measure between the underlying system and its best approximation. Choices in the experimental design, for example, the distribution of input values, can affect the difference measure, and thus the best approximation. Changing the approximating class will also be expected to affect the approximation. Theoretical results about kernel analysis commonly take the form of providing a sequence of classes (usually having more and more members) and showing that underlying systems having particular properties may be approximated by a sequence of members of the classes for which the fitting error goes to zero.

A second source of fitting error occurs when there is intrinsic system noise or the system behavior can only be known through noisy measurements. In these cases, the best approximation will vary from experiment to experiment. The expectation is that as the experiment is made longer, fitting errors of this sort will be reduced. If the specific measurement noise distribution is known, fitting procedures can sometimes be adjusted with beneficial results (Wu, David & Gallant, 2006). When fitting error as described above can be reduced to zero, one says the approximations "converge" to the underlying system. Convergence is, by its nature, a theoretical property.

The focus of this paper is on matters of interpretation that are independent of questions of convergence. The interpretive difficulties that we highlight exist for systems that are *defined* by finite Volterra expansions, which are themselves members of the approximating class. Thus, our considerations take precedence over matters of convergence. The problems we describe can exacerbate convergence errors, but cannot be ameliorated by improved convergence procedures.

In the theory of Volterra and Wiener analysis, questions of convergence of the integrals and the infinite sequence of Equation (1) loom large and are afforded great consideration (Marmarelis, 2004). The hypothetical experiments by which the kernels are to be fixed are critical. For example, the Volterra expansion is classically taken to be analogous to the Taylor expansion for functions of a single variable, with coefficients (kernels in Volterra) fixed from knowledge of infinite sets of derivatives, found by presenting particular time-dependent input functions. In contrast, Wiener analysis, which has different convergence properties, depends on the presentation of quite another sort of input function (see Section 4).

In biological practice, and especially in studies of large-scale brain operation, details of convergence are completely out of reach. Faced with a finite number of kernels we can determine, we are happy to find system models that elucidate the most important features of the system. And to do that, we must select the kernels that are to be retained very carefully. In particular, we often feel constrained to trade interaction order for discrete memory. Kernel choice means giving up the ability to examine certain details in order to see others more clearly, or even, adequately.

Once we have agreed to reduce to a particular finite version of Equation (2), we can set about choosing inputs that will let us solve for the kernels we have left. How we do this is discussed below.  Here, however, we point out that our methods are like Wiener stimulation in that the input values are distributed over the most important operating range of the system being studied.  In particular, even though we retain the name "Volterra" for our analysis method, we specifically *do not* present inputs clustered about some base amplitude as we might if we were trying to determine our kernels via derivative considerations.

Before leaving this section, we note that the impossibility of using of convergence methods in practical experiments leaves us in quite a difficult situation since it removes the only fully systematic means of validating extracted kernels.  What validation we can do is typically *ad hoc*, and often not very satisfying.  Considerable art is required in order to obtain reliable kernels.

*1.4 Specific utility*

Volterra analysis of a system consists of performing experiments in which various inputs are presented to the system and its responses noted.  Based on elaborate comparison of the inputs and the responses, the kernel slices are computed.  If the experiment is sufficiently exhaustive, all the slices may be determined, with contamination of them by various sources of error reduced to tolerable levels.  One then has in hand a collection of possibly highly featured but reproducible curves, which fix the model.  But, so what?  What has one gained?

First, we have a test of the parameter reductions embodied in Equation (2).  If they are justified, the slices provide enough information to allow the reconstruction of the system responses to any possible system inputs. We examine this question further in Section 3.

Second, the Volterra slices encapsulate specific systemic behavior in a *relatively* understandable and accessible way.  In particular, they are closely related to system responses generated by compact groups of non-zero inputs that we call "impulse packets".

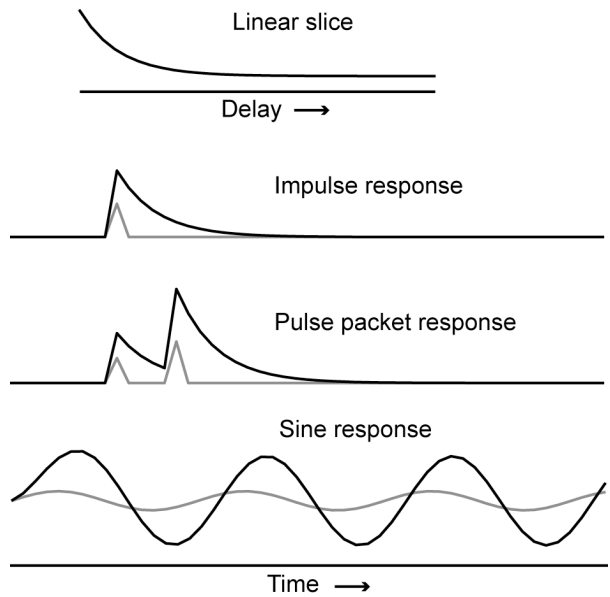

*Figure 2. Example of an order-1 kernel and associated response contributions. The top plot shows the $L_1$ kernel as a function of delay. The second plot shows the system response to a pulse input. The input is shown in gray. The congruence of the black traces in the top two plots justifies calling $L_1$ the "impulse response" of the system. The third plot shows the system response to two pulses of different amplitudes. The last plot shows the system response to a sinusoidal input.*

The simplest illustration of this comes from linear systems, where Volterra analysis always yields a single non-zero slice, $L_1(d_0)$, which may be identified with the classical "impulse response" of the system. That function completely characterizes the system. But the impulse response also has eponymous meaning: It is the actual system response to a unit-valued pulse input. One can easily visualize a pulse coming along and causing the system to "ring", as a bell does when hit by the clapper. Because the system is linear, this is all that can really happen. Successions of pulses, merging perhaps into continuous, real-valued inputs, each affect the system independently. The total response is just the sum of the responses to the individual pulses. There is no *interaction* between the pulses; there is nothing beyond the independent effects. This total picture, illustrated graphically in Figure 2, is so powerful that we feel that we really *understand* the system.

If the system is non-linear, impulse packets elicit analogous, but more complex, behavior that is captured by collections of slices. We describe the result for the simplest packets. Figure 3 provides an illustrative example.

Begin with the empty packet -- that is, input constant with value zero for all time -- the input is entirely "quiescent". Under these conditions, a system liable to Volterra analysis must produce constant, although, possibly non-zero, output. This value is fully captured in a constant-valued "bias" slice. This slice always has at most one non-zero value, equal to the constant system output, which is located at delay 0. In reconstructing the system response to the empty packet, the bias slice is added at every timestep, yielding a constant contribution. The bias slice is shown in the top left trace of Fig. 3, and its contribution to the reconstruction is shown in the top right trace.

The next simplest input comprises a single "pulse" having some amplitude $a$ – quiescence followed by one step's worth of input value $a$, the shortest possible active interval, followed again by quiescence. There is an extra response, beyond the bias contribution, of the system to this input, and is it expressed as the sum, with coefficients $a^k$, integral $k$, of a family of slices, the "diagonals". This sum, in effect a Taylor series with coefficients in the diagonal slices, represents non-linear behavior of the system as a function of the single pulse amplitude $a$. In the illustration, there are two such slices, and they are shown in the middle two left-hand traces. Among the diagonals is the "linear" slice, which fully captures the linear part of the system's behavior. The linear slice, as the name would imply, contributes to the response an amount that depends linearly on $a$. The linear slice is shown in the second trace from the top, left side. The second trace from the top on the right shows the sum of the contributions from the diagonal slices to the reconstruction of the response to a single pulse of amplitude $a$ arriving at step 2. In the example, $a = 3/4$. The full response to this pulse is the sum of this contribution and the bias contribution. The third trace from the top, right side, shows the diagonal contribution to the single pulse $b = 4/3$, arriving at step 4. Note that the $a$ and $b$ pulse contributions are not simply scaled versions of one another. This can be

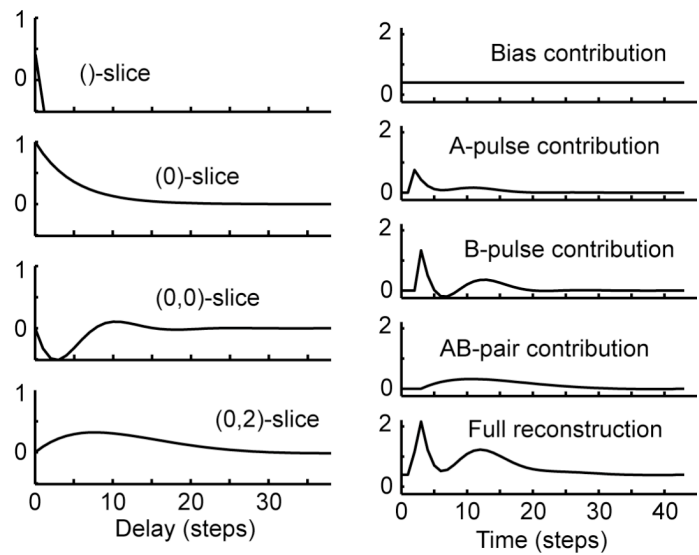
12

*Figure 3.* Slices and packet contributions associated with the full reconstruction of a 2-pulse input comprised of pulses having different amplitudes ($a$ and $b$). Contributions from each subpacket, appropriately offset in time, sum to make up the full system response.

seen, for example, in the fact that trough following the peak in the $a$ contribution remains positive, while that in the $b$ contribution goes clearly negative. The failure of scaling is an interaction effect, mediated by the fact that the two diagonal slices are multiplied by different factors ($a$ and $a^2$, verses $b$ and $b^2$) before being added to produce the diagonal contributions shown on the right side of the figure.

An input comprising a pair of pulses having a particular temporal separation and amplitudes $a$ and $b$ causes a response that is the sum of the bias response, the contributions of the diagonals to the responses to the two pulses treated separately, and a Taylor series in $a$ and $b$ with coefficients in another family of slices. In this sense, the new family represents the part of system response due to the *interaction* of the two pulses. In the illustrative example, there is only one such slice, and it reacts to a packet having inter-pulse delay of 2 steps. It is shown in bottom-left trace of Figure 3. That trace was multiplied by $ab = 1$ and offset to step 4 to make the $ab$ contribution shown in the 2nd from bottom right trace. The offset is 4 because that is when the system is first aware that it is seeing the *pair* of pulses whose individual contributions are shown in the previous traces.

Similar statements hold connecting the system responses to arbitrarily complicated impulse packets and multi-variate Taylor series with coefficients in associated slice families. The details are given in Sections 2.1 and 2.2. The authors appreciate that the idea of manipulating Taylor series with coefficients in slices can appear daunting, but suggest that with exposure, familiarity grows, and one can become quite comfortable. In addition, we never deal with fully general Volterra expansions, and typically the Taylor series we actually must face truncate to polynomials having just a few terms.

*1.5 Further simplifications and restrictions*

Examination of the multiple summations in each term of Equation (2) yields an estimate of the number of kernel elements of each order given in Equation (3). The numerator on the right side is simply the number of indices that would be summed over if there were no constraint on the relative sizes of the $d_j$, and the denominator is an estimate of the effect of enforcing the constraint.

$$\text{count}(n) \sim T^n / n! \tag{3}$$

Thus, the total number of kernel elements grows exponentially with the maximum interaction order. Clearly if $T$ is not very small, the count of kernels of order less than $n$ is dominated by the count of order equal to $n$. The dependence on $T$, although less critical than that on $n$, is often limiting.

One would like to have a small parameter count, but analysis is ultimately constrained by the system at hand: "You must analyze the system you have, not the one you want." The number $T$, the "discrete memory", is dictated by two real system quantities, the system memory and the "system resolution", both measured in real time. The memory tells how long after an input is seen that the system response can be affected by it, and the resolution measures how closely spaced inputs may be and still be distinguished by features appearing the system response. In applying Volterra-based methods, one cannot, without cost, just give up resolution, memory, or interaction order by simply pretending that these numbers are other than what they actually are. For example, if one "cheats" on the resolution by probing the system with inputs that do not excite the system over its full temporal response bandwidth, then one is actually studying a different system. This procedure and problems associated with it will be discussed later in this section (slow-stimulation) and in Section 5.

Assuming that the system memory is shorter than it really is, while sometimes possible, is a very delicate business (Sutter, 1992). Except in quite special circumstances, effects from past stimulation that one has assumed to be out of memory and simply ignored contaminate even the early portions of slices that remain in the analysis. There is no *general* technique that allows one make a shortened experiment that finds, say, just the early parts of slices.

Arbitrary restriction of interaction order invariably results in loss of accuracy in reconstruction. Sometimes this occurs relatively gracefully, as, for example, in Wiener analysis, where order restriction is central and systematic (Schetzen, 1989). In practice, one is frequently forced to severe reduction in order, often to order 2, to make experimental determination of the slices feasible.

Ideally, the generality of the analysis should be adjusted to match the complexity of the system being analyzed. One should use an analysis that is as simple as possible, but one that is still capable of capturing all the important behavior of the system. Unfortunately, it may happen that the system being investigated has discrete memory and interaction order so large that there is no feasible experiment. In estimating feasibility, the relevant facts are that one noise-free data point is required to solve for each parameter, and noise reduction goes with the inverse of the square root of the number of replications.

When further reduction is needed, experimenters have resorted to the use of "slow-stimulation" (Klein, 1992) and (Sutter, 1992) . In slow-stimulation, one samples the inputs and the system responses to them at one rate, usually rather fast, but changes the input values at another, slower, rate. Typically inputs are changed only at regularly spaced, non-adjacent samples. The time between adjacent samples is the "sampling interval", and the time between input changes is the "stimulus interval", also called the "stimulus step". Slow stimulation is a very effective means of parameter reduction. However, as we point out below, it can lead to confusion and must be used with considerable caution.

Finally, there is another important way one can achieve parameter reduction: The inputs may each be restricted to having values among a limited collection of distinct real numbers. When we make this restriction, we shall say we are doing "multilevel analysis". As we shall show in Section 2.3, multi-level inputs force a reduction in the number of parameters that can independently influence the behavior of the system. It will generally turn out that there is a degree of arbitrariness in how the reduced parameters may be chosen, and that, in turn, can raise interpretive questions. The reduction of parameters is most radical in "binary analysis", where the number of levels in the inputs is precisely two. In that special case the redundant parameters coalesce into disjoint collections that cannot be distinguished by any experiment involving only the restricted inputs. The practical consequence of this is that all parameters except one in each collection may be removed from the analysis without a loss of descriptive power. And, there is a particular "canonical" choice for the retained parameter. This removes the arbitrariness present when the analysis involves more than two levels. We present this reduction in Sections 8.1 and 9.

Multi-level analysis is sensible in two cases. First, when the system is such that the input values are *inherently* limited, as for example, in a digital circuit, whose input values are low and high voltage states. Second, when a system that allows and processes real-valued inputs is *presented* only an artificially reduced set of input values. In this case, the combination of the real system with the artificially restricted set of inputs implements a reduced system. This procedure is useful if the reduced system is still interesting in its own right. For example, it might comprise a well-defined part of the general system behavior. We will illustrate this below with a Visual Evoked Potential (VEP) study where the inputs were two disjoint regions of the visual scene, each of which displayed, alternately, two spatially complicated pattern stimuli. The reduced system comprising the whole brain and the restricted inputs formed a 2-input, binary system, and we sought evidence of interaction between the inputs in the details of the analysis parameters.

*1.6 Problems arising from restricted forms of Volterra kernel analysis*

The goal of reducing parameters as described above its to be able to perform experiments adequate to determine a full set of sufficiently noise-free kernels of the reduced system. To be useful, the retained kernels should provide enough information to allow the reconstruction of the system responses to any possible system inputs. Slice collections having this property are said to be "complete". If our collection of slices is incomplete, we cannot claim to have characterized the system by the analysis. And this presents the first problem: Although there are theoretical proofs of completeness for the unreduced Volterra (including Wiener) and binary systems, the reduced systems will fail to be complete if slices that really are needed have been excluded in the reduction. In practice, one tests completeness by reconstructing the system response to a long and varied input sequence, typically the one used in the experiment that determined the kernels, and compares the reconstruction with the measured response. If the two agree, one tends to believe the analysis is complete.

However, we want to stress that while completeness of the analysis is *necessary*, it is not always *interesting*. In the case where there is *a priori* reason to know that a reduced analysis is complete, we get to reconstruct system responses to inputs that have not actually been tested, and this is interesting and useful information. The most mundane example of this behavior occurs when the system is known to be linear: We may then reduce to the linear slice, traditionally known as the impulse response, and from it, be able to reconstruct the system response to all possible inputs. In the more typical case, where we do not have *a priori* knowledge, we test for completeness by sampling responses to at least what we believe are a representative collection of inputs. But, for the tested inputs, it is pointless to do the analysis just to reconstruct what we have measured directly, and for the untested ones, we cannot be sure the reconstructions are accurate.

The slow-stimulation approach to parameter reduction has proven very attractive to experimenters because it offers marked reduction of discrete memory, which can be parlayed into somewhat higher interaction order.  The cost to be paid is not exacted in reconstruction, but rather in interpretation:  The slowly-stimulated Volterra analysis is complete on the class of all inputs that change only on the step boundaries, which is to say that the system response to any such input may be accurately reconstructed.  Unfortunately, when one attempts to go further, to attribute meaning to individual slow-stimulation slices, one strikes upon serious difficulties.  To the extent that the underlying system really is *fast* Volterra with behavior that responds to stimuli changing at the sample rate (or at least at a rate faster than the step rate), and also requires the full discrete memory, effects due to a single (non-linear) Volterra slice will usually make contributions to a number of slow-stimulation slices, with the precise distribution of the contributions depending on the details of the Volterra signature and on the chosen step duration.  We call this phenomenon "boundary crossing" (named for the actual mechanism by which this happens, the boundaries being those of the steps; we address it in detail in Section 5.1, and again in Section 8.4).  The contributions can occur in a way that produces strong intra-step features that are not apparent in the original Volterra slice function.  Correct interpretation of these features can call for considerable subtlety.  If the fast system comprises multiple slices having unknown functional forms, teasing those forms out of the slow slices is typically not possible.

Figure 10 shows a telling and typical example of boundary crossing in the context of binary analysis that we will discuss in detail in Section 8.4. The two lower traces depict slow slices, and the cusps that appear in mid-step in each are the result of the interplay between the fast Volterra slice function and its particular signature, shown in the top trace, and the stimulus/sampling rate ratio.  In the experimental context, where we presume that the underlying system is invisible to us, and we can only see the slow, binary reduction, the cusps must seem mysterious.

In our exposition of the VEP study (Section 10), we show evidence for boundary crossing effects in real experimental data from a biological system restricted to two slow, binary inputs and expand on the interpretational difficulties, in particular, of inferring interaction effects from the particular slices.  We then show how, using a variant of the traditional binary analysis, boundary crossing problems can be partially overcome in a way that allows us to clearly infer and quantify interaction between the inputs.

*1.7 B1m vs. B01 analysis*

The traditional binary reduction (Sutter, 1992) proceeds by assuming an underlying Volterra model that is driven by two-level inputs taking the specific values 1 and -1. We shall use the designation "B1m1" (binary, 1, minus 1) for this analysis. If the system has two-level inputs that are numerical, but are not 1 and -1, application of B1m1 analysis entails "normalizing" the input levels to 1 and -1. If the system has inputs comprising two configurations that are not expressly numerical, application of B1m1 analysis requires assigning one configuration to the value 1 and the other to -1. As we shall show below, the slices that result from B1m1 analysis, even when complete, and so able to reconstruct responses to arbitrary binary stimuli, are, in themselves, difficult to interpret. One reason is that *every* non-zero slice enters into the reconstruction of the response to any given stimulus, a phenomenon which we refer to as "all the kernels, all the time", or AKAT. How a slice enters is encapsulated by a sign derived from the interplay of its signature and the current (binary) memory state. Unfortunately, the particular logic of the sign determination tends to lead to complication rather than clarity in the interpretation of interaction among samples. This problem is revisited in Section 8.3.

Another binary analysis, which we designate "B01", results from driving an underlying Volterra model with inputs restricted to the values 0 and 1 (Yokota & Usui, 1999). This apparently trivial change in coding turns out to produce slices that have clear meaning that is expressed in terms of response reconstruction from increasingly complex impulse packets. It is easy to see how system behavior is captured by individual B01 slices. These beneficial properties result from the specific logic of signature and memory-state interplay, which is quite different from that in B1m1 analysis. In the slow-stimulation condition, the interpretation of boundary crossing effects is more tractable using B01 than B1m1. Another benefit of B01 analysis comes from the way slices from an underlying Volterra system coalesce into B01 slices. As we shall show in Section 9, it allows "grouping" of slices in a way that lets us quantify, among other things, single-input effects and inter-input interactions, and to do so in a way that further mitigates the interpretive complications of boundary crossing. Similar grouping of B1m1 slices is not helpful, either for input-effect quantification or for boundary crossing.

## 2. Volterra details

Although the usual scientific task is to start with a physical system and derive its Volterra expansion, it is useful for purposes of understanding to reverse the procedure -- to suppose that our system is *defined* by a Volterra expansion, and then develop intuition for the roles played by various summands. We can think of the principal summands of Equation (2) as defining parallel model subsystems whose individual outputs are simply added to produce the full system response. We call a system defined by a Volterra expansion a "Volterra system".

Begin with the $L_0$ term. Its contribution to the response is always present and fixed, independent of input. It represents a constant subsystem, and sets what is commonly known as the system "bias".

The term containing $L_1$ is more interesting. If the input $s$ is replaced by $a \cdot s$ where $a$ is a real constant, by commutativity of multiplication and by distribution over summation, the factor $a$ can be moved to the left of the summation sign, meaning that the contribution of the $L_1$ term is multiplied by $a$. Also, if $s$ is replaced by the sum $s_1 + s_2$ of two separate inputs, the contribution is replaced by the sum of the separate contributions. A system where a constant times an input yields that constant times the output, and the sum of two inputs yields the sum of the separate outputs, is said to be "linear". Thus, the $L_1$ term represents a linear subsystem.

It is instructive to examine how the $L_1$ term acts on some simple specific inputs. Begin with the "zero" input, $s = 0$ for all $t$. Substitution of $s(t - d_0) = 0$, $d_0 = 0, \ldots, T-1$ into the $L_1$ term makes it zero; the values of the function $L_1$ do not come into play.

Look next at a "unit-pulse" input. Fix a particular time $t^*$. Following common practice, suppose $\delta_{t^*}$ denotes the "parameterized delta function" $\delta_{t^*}(t) = \begin{cases} 1, & t = t^* \\ 0, & t \neq t^* \end{cases}$. Let $s = \delta_{t^*}$. What is the contribution of the $L_1$ term at the time $t$? If $t < t^*$ or $t \geq t^* + T$, just as above, all the values $s(t - d_0) = \delta_{t^*}(t - d_0)$ are zero, so the contribution is zero. However, for $t \geq t^*$ and $t < t^* + T$, we may get an effect. In particular, the single summand $L_1(t - t^*)s(t - (t - t^*)) = L_1(t - t^*)\delta_{t^*}(t^*) = L_1(t - t^*)$ is non-zero and represents the entire contribution from the $L_1$ term. Said another way, the contribution from the pulse at time $t^*$ is the value $L_1(0)$, the contribution from that same pulse a little later, at $t^*+1$ is $L_1(1)$, and so forth. Thus, $L_1$, viewed as a function of delay, specifies the way the contribution from a single pulse input plays out in time. The contribution begins no earlier than the arrival of the pulse (reflecting the physical property of causality), and endures no longer than the memory $T$. In light of this discussion, we identify $L_1$ as the "impulse response" of the linear part of the system. Figure 2 shows an example that illustrates the situation. In the figure, inputs are shown in gray, responses in black, and time is shown, as usual, increasing to the right, the direction of the future; delay, the argument of the kernel $L_1$, is shown also increasing to the right. The top plot of Figure 2 show the graph of the function representing the $L_1$ kernel. The next plot shows the same functional form as it represents the response to a unit-pulse input. In that case, time rather than delay is the independent variable.

By the linearity of the $L_1$ term, we know that a single pulse of amplitude $a$ contributes the impulse response amplified by $a$ and that a pair of pulses contribute the time-wise sum of their individual contributions. If the pulses are separated by more than the memory, the total contribution is just the contribution for the first followed by that for the second. If the pulses are less separated, the contributions overlap and are summed where they overlap. The third plot of Figure 2 depicts this situation.

Since any input can be represented as the sum of distinct pulses, one at each timestep, each having its own amplitude, the total $L_1$ contribution is simply the time-wise sum of the amplified, delayed, impulse responses. The $L_1$ contribution to the system response to a sine-wave input is shown in the bottom plot.

The specific mathematical form of the $L_1$ term in Equation (2) is known as a "convolution". A convolution converts a function into another function by means of a helper function, traditionally called the "kernel". Here, the input $s$ is converted to the linear contribution to the response $r$ with the help of the kernel $L_1$. The use of the word kernel in systems analysis is an extension of its use in convolution. The argument given above that the $L_1$ subsystem is linear just uses properties of the convolution form. It therefore shows that any system defined by a convolution is linear. In signal processing, the effect of convolving a signal with a function is known as "linear filtering". We adopt that terminology here.

We now turn to the $L_2$ term, which merits particular scrutiny since its behavior, more complex than that of the lower terms, is definitive -- it captures all the *essential* complication of the higher terms. We examine the contribution of the $L_2$ term to the response to impulse inputs. Figure 1 is relevant to our discussion. The top panel depicts the graph of $L_2$ as a function of the delays $d_0$ and $d_1$. As with the $L_1$ term, there is clearly no contribution from the zero input. What is the contribution to the response at time $t$ from a single unit-amplitude pulse at time $t^*$? As before, $s = \delta_{t^*}$. The input enters the $L_2$ term via the product $s(t-d_0)s(t-d_1)$ where $d_0 = 0, \ldots, T-1$, and $d_1 = d_0, \ldots, T-1$. The product is non-zero only when both its factors are non-zero. But since $s(t-d_0) = \delta_{t^*}(t-d_0)$ and $s(t-d_1) = \delta_{t^*}(t-d_1)$, that happens precisely when $t - d_0 = t - d_1 = t^*$, that is, when $d_0 = d_1 = t - t^*$. Thus, the total contribution of the $L_2$ term is $L_2(t-t^*, t-t^*)\delta_{t^*}(t^*)\delta_{t^*}(t^*)$. Note that the product $\delta_{t^*}(t^*)\delta_{t^*}(t^*) = \delta_{t^*}(t^*)$. If we define the 1-dimensional function $\mathcal{L}_{(0,0)}(d_0) \equiv L_2(d_0, d_0)$, (this notation will be explained later) we return to a convolution form reminiscent of the interaction of a unit pulse with the $L_1$ term; here the restriction $\mathcal{L}_{(0,0)}$ to the "diagonal" of the domain of the 2-dimensional function $L_2$ plays the role of the impulse response (the trace closest to the viewer in the bottom panel of Figure 1). The analogy, however, is very delicate, and breaks with the slightest perturbation. In particular, suppose $a$ is a real number not equal to 1. Then the contribution to the response at time $t$ of the *non*-unit impulse $a \cdot s$ from the $L_2$ term is $L_2(t-t^*, t-t^*)a\delta_{t^*}(t^*)a\delta_{t^*}(t^*) = a^2 \mathcal{L}_{(0,0)}(t-t^*)$, which exposes the essentially *quadratic* nature of the $L_2$ contribution.

Pressing on, what about the contribution from $L_2$ when the input is a *pair* of pulses? In the first instance, suppose both pulses have unit value, and occur at times $t^* - d^*$ and $t^*$, $d^* > 0$. (Think of $t^*$ as "now". Then $t^* - d^*$ is in the past, and when the pulse at t* occurs, the pulse at $t^* - d^*$ will already have occurred, and so the system will be aware that it is in a two-pulse situation.) In delta function notation we have $s(t) = \delta_{t^*}(t) + \delta_{t^* - d^*}(t)$. Generally, $d_1 \geq d_0$. We shall see below that we have already sufficiently addressed the case $d_1 = d_0$. Suppose then that $d_1 > d_0$. As above, there is a contribution only when the product $s(t-d_0)s(t-d_1)$ is non-zero. By the definition of $s$ and the distributive law, we have

$$s(t-d_0)s(t-d_1) = \left(\delta_{t^*}(t-d_0) + \delta_{t^*-d^*}(t-d_0)\right)\left(\delta_{t^*}(t-d_1) + \delta_{t^*-d^*}(t-d_1)\right)$$
$$= \delta_{t^*}(t-d_0)\delta_{t^*}(t-d_1) + \delta_{t^*}(t-d_0)\delta_{t^*-d^*}(t-d_1) \quad (4)$$
$$+ \delta_{t^*-d^*}(t-d_0)\delta_{t^*}(t-d_1) + \delta_{t^*-d^*}(t-d_0)\delta_{t^*-d^*}(t-d_1)$$

But since $t^*-d^* < t^*$ and $d_0 < d_1$, only the $\delta_{t^*}(t-d_0)\delta_{t^*-d^*}(t-d_1)$ term can ever be non-zero. This happens only when $t-d_0 = t^*$ and $t-d_1 = t^*-d^*$, that is when $d_0 = t-t^*$ and $d_1 = t-t^*+d^* = d_0 + d^*$. Thus, the contribution is $L_2(t-t^*, t-t^*+d^*)\delta_{t^*}(t^*)\delta_{t^*-d^*}(t^*-d^*)$, which equals $L_2(t-t^*, t-t^*+d^*)$. The requirement that $d_1 < T-1$ implies $d^* < T-1$. Note that only the one-dimensional restriction $L_2(d_0, d_0+d^*)$ of the two-dimensional function $L_2(d_0, d_1)$ plays a role. The bottom panel of Figure 1 shows a portrayal of $L_2$ as the collection of super-diagonal restrictions. As we said earlier, the restrictions have come to be called slices. Each allowed $d^*$ determines the unique slice $\mathcal{L}_{(0,d^*)}(d_0) \equiv L_2(d_0, d_0+d^*)$. The diagonal restriction is the slice $\mathcal{L}_{(0,0)}$. The cumbersome subscript $(0, d^*)$ previews a notation whose usefulness will become apparent later.

The important result is that the slice $\mathcal{L}_{(0,d^*)}$ is the contribution of the $L_2$ term of the Volterra expansion *explicitly* evoked by the pulse *pair* with separation $d^*$. The contribution begins when the second pulse appears, and continues for the length of the slice. As is apparent in Figure 1, slice length gets shorter as the pulse separation $d^*$ increases, reflecting the fact that memory is bounded; to respond to a pulse pair, the system needs to be able to remember both pulses in the pair. When the first pulse goes out of memory, the system can no longer respond to the pair. Of course, it can continue to respond to the second pulse, but only through the $L_0$, $L_1$, and, with important consequences, the diagonal restriction of $L_2$.

In the case of non-unit pulse pairs, each pulse has its own non-zero value, $a_0$ and $a_1$, and the contribution, again reflecting the quadratic nature of $L_2$, is $a_0 \cdot a_1$ times the response of the unit pulse pair with the same separation.

If $L_0$, $L_1$, and $L_2$ happen to be the only non-zero terms in our system, all the response contributions to a pulse pair are now at hand, and they need only be summed to yield the full system response. Note that beyond the constant term $L_0$, which is always present, each pulse of the pair contributes separately to the response via $L_1$ and via the diagonal of $L_2$, and they contribute jointly via the single super-diagonal of $L_2$ determined by the delay $d^*$. The important thing is that no other terms enter into the calculation of the response to the pulse pair with delay $d^*$. In particular, no $\mathcal{L}_{(0,d)}$, $d \neq 0$, $d \neq d^*$, play a part.

We end this discussion by noting that the $L_2$ term in Equation (2) has the form of a 2-dimensional convolution, specifically, the 2-dimensional function $L_2$ is convolved with two input factors. However, the explicit contribution from the pair is just the 1-dimensional convolution of a single pulse with the slice $\mathcal{L}_{(0,d^*)}$. We shall expand on this below.

*2.1 Impulse packets and the slice representation*

The situation summarized in the previous paragraph reflects a general phenomenon, which we call "subpacket-correction", wherein the response of a Volterra system to an impulse packet is the sum of contributions or "corrections" for each of the subpackets of the packet. A "subpacket" is an inpulse packet that precisely fits onto a part of the original packet, perfectly matching delays and amplitudes. Subpackets by convention include the empty packet and the packet itself. Subpacket-correction will turn out to be very helpful in the matter of interpreting the structure of Volterra systems, and we shall develop our understanding of it more fully. It formalizes the notion of the interaction between stimulus pulses that is familiar to systems neuroscientists and psychophysicists, who often perform experiments with single and double pulses (Breitmeyer & Ogmen, 2000, Macknik & Livingstone, 1998, Rashbass, 1970, Schiller, 1968). Volterra analysis is a nice way to approach these experiments in a complete and systematic fashion.

Subpacket-correction is intimately related to the subdivision into slices of the Volterra kernels. This was illustrated for $L_2$ in Figure 1, but can clearly be done to any kernel. The $L_2$ term in Equation (2) is shown as a 2-dimensional sum over first one delay $d_0$ and then another delay $d_1$, (with the constraint that $d_1 \geq d_0$). Rearrangement of the summands permits summation first along the super-diagonals, and subsequent summation of these sums over the various inter-pulse delays, $d^*$. Making the analogous rearrangement in each of the principal summands of (2) yields the "slice representation" of the Volterra expansion:

$$r(t) = \mathcal{L}_{(\varnothing)} + \sum_{d=0}^{T_{(0)}-1} \mathcal{L}_{(0)}(d) S_{(0)}(t-d) + \sum_{d_1=0}^{T-1} \left( \sum_{d=0}^{T_{(0,d_1)}-1} \mathcal{L}_{(0,d_1)}(d) S_{(0,d_1)}(t-d) \right) +$$
$$\sum_{d_1=0}^{T-1} \sum_{d_2=d_1}^{T-1} \left( \sum_{d=0}^{T_{(0,d_1,d_2)}-1} \mathcal{L}_{(0,d_1,d_2)}(d) S_{(0,d_1,d_2)}(t-d) \right) + \ldots \quad (5)$$

where

$$\begin{aligned}
\mathcal{L}_{(\varnothing)} &\equiv L_0, \\
\mathcal{L}_{(0)}(d) &\equiv L_1(d), \\
\mathcal{L}_{(0,d_1)}(d) &\equiv L_2(d, d+d_1), \\
\mathcal{L}_{(0,d_1,d_2)}(d) &\equiv L_3(d, d+d_1, d+d_2), \quad etc.
\end{aligned} \quad (6)$$

and

$$\begin{aligned}
S_{(0)}(t) &\equiv s(t), \\
S_{(0,d_1)}(t) &\equiv s(t)s(t-d_1), \\
S_{(0,d_1,d_2)}(t) &\equiv s(t)s(t-d_1)s(t-d_2), \quad etc.
\end{aligned} \quad (7)$$

The slices need to be uniquely labeled, and we use a "primary signature" scheme to accomplish this. A signature is simply an n-tuple of non-negative integer delays $(d_0, d_1, d_2, \ldots, d_{n-1})$ with $d_0 \leq d_1 \leq d_2 \leq \ldots \leq d_{n-1}$. The arguments of the kernel function $L_n$ are signatures comprising $n$ delays. A "primary signature" is one with $d_0 = 0$. The primary signature for a slice of $L_n$ is defined as the unique primary signature that lies on the super diagonal that forms the domain of the slice. Given a primary signature, the domain of the slice has the form $(0+d, d_1+d, d_2+d, \ldots, d_{n-1}+d)$, $d = 0, 1, \ldots, T_{(0,d_1,d_2,\ldots,d_{n-1})} - 1$, where $T_{(0,d_1,d_2,\ldots,d_{n-1})}$ is the slice length. For consistency, we take the "empty signature", $()$ or $(\varnothing)$, to be primary as well. The primary signature for the linear term is $(0)$. In the slice representation, the term of order $n$ is the summation first over primary delay along slices, and then over primary signatures across slices. We use the symbol $\mathcal{L}_{\text{signature}}$ to denote the slice having the given primary signature. Thus the bias term is $\mathcal{L}_{(\varnothing)}$ and the linear slice is $\mathcal{L}_{(0)}$. Note that the order of the kernel from which a slice is extracted is implicit in the signature.

The slice representation given in Equation (5) implies a decomposition of the Volterra system into a set of parallel subsystems that may be more enlightening than that implied by the $L_n$ terms in Equation (2). Each subsystem in the decomposition acts on the common input $s(t)$ and consists of a non-linear element followed by a linear one. There is one subsystem for each primary signature, defined as follows: The function $S_{\text{signature}}(t)$ defined in Equation (7) represents a "delayed static non-linearity" or DSNL, being the product of a number of copies of the input, each delayed according to the signature. This generalizes the commonly used term "static non-linearity", which is simply a DSNL with all delays zero. The (1-dimensional) convolution of $S_{\text{signature}}$ with $\mathcal{L}_{\text{signature}}$ that comprises an innermost summand in Equation (5) represents the effect of passing the output of the DSNL through the linear filter whose characteristic is the slice. The outer summations in Equation (5) collect the subsystem outputs to produce the full system output. Figure 4 shows a block diagram that represents the decomposition. Each subsystem is depicted in the figure as a pair of blocks connected by a horizontal line. These are fed by the common input and feed the sum block that delivers the total response.

The interaction of impulse packets with delayed static non-linearities is important. Consider a packet defined by delays $(0, d_1, d_2, \ldots, d_{n-1})$, $0 < d_1 < d_2 < \ldots < d_{n-1}$, based at time $t^*$ and having pulse amplitudes $(a_0, a_1, \ldots, a_{n-1})$; by definition $s = a_0 \delta_{t^*} + a_1 \delta_{t^*-d_1} + \ldots + a_{n-1} \delta_{t^*-d_{n-1}}$. We shall call the delays the "packet signature" and the amplitudes, the "packet amplitudes". Inserting $s$ in the definition

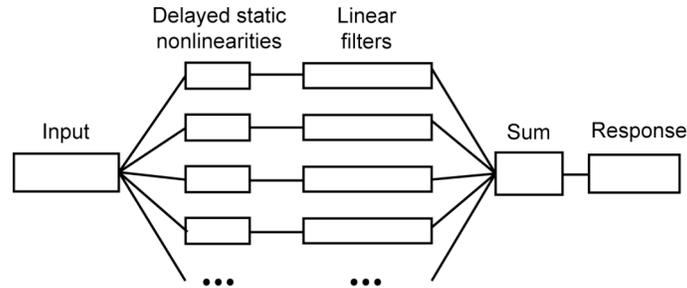

*Figure 4. Block diagram of the system decomposition implied by the slice representation. Each of the parallel channels corresponds to a primary signature. It comprises a cascade consisting of a DSNL followed by a linear filter. The full system response is the sum of all of the parallel channel responses.*

$S_{(0,d_1,d_2,\ldots,d_{n-1})}(t) = s(t)s(t-d_1)s(t-d_2)\ldots s(t-d_{n-1})$ and repeating, in expanded form, the reasoning consequent to Equation (4), we find that $S_{(0,d_1,d_2,\ldots,d_{n-1})}(t^*) = a_0 \ldots a_{n-1}$ and $S_{(0,d_1,d_2,\ldots,d_{n-1})}(t) = 0$, $t \neq t^*$. That is, the output of the DSNL is just a single pulse whose amplitude is the product of the sub-pulse amplitudes. Figure 5 shows a graphical version of the argument. The figure illustrates the case of the static non-linearity defined by the primary signature $(0, d_1, d_2)$ illustrated in the second line from the top. In the example illustrated, the packet delay pattern on the input line agrees exactly with the delays comprising the primary signature. Time steps are marked along the bottom of the plot. Pulse widths shown have no significance since data is defined only at the timesteps. The top plot represents the input packet, and the shorter plots below represent the memory window moving toward the future, carrying the primary signature delay pattern with it. At each timestep, the input is sampled at just the signature locations for that time, and the values obtained are multiplied to yield the output of the non-linearity, which is shown in the bottom plot. In this illustration, only one window has a signature pattern that that exactly matches the input packet, and thus contributes a non-zero to the output; it is shown in black fill. The corresponding single pulse output has amplitude equal to the product of the amplitudes of the individual input pulses. More generally, the output is identically zero unless the input pulse pattern contains the signature pattern, since otherwise, whatever time translation is applied, there is at least one sampled input value that is zero, and it makes the entire product zero. Signatures with duplicate delays use duplicate copies of the sampled input in forming the product.

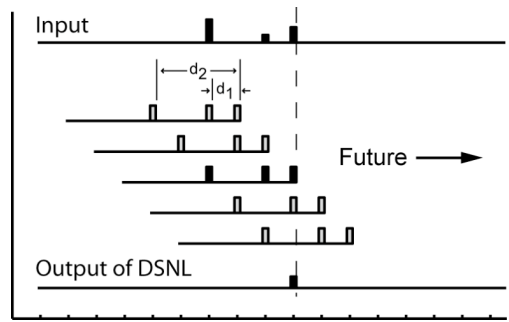

*Figure 5. The effect of passing an impulse packet through a DSNL. The figure illustrates the case the static non-linearity defined by the primary signature $(0, d_1, d_2)$ illustrated in the second line from the top. In the example illustrated, the packet delay pattern on the input line agrees exactly with the delays comprising the primary signature. Time steps are marked along the bottom of the plot. The top plot represents the input packet, and the shorter plots below represent the memory window moving toward the future, carrying the primary signature delay pattern with it. At each timestep, the input is sampled at just the signature locations for that time, and the values obtained are multiplied to yield the output of the non-linearity, which is shown in the bottom plot. See text for details.*

A slight elaboration of this argument shows that the only primary signatures $(0, d'_1, d'_2, \ldots, d'_{m-1})$ for which our packet can pass through $S_{(0, d'_1, d'_2, \ldots, d'_{m-1})}$ with non-zero result are those such that all the delays $d'_k$ are contained among the delays $(0, d_1, d_2, \ldots, d_{n-1})$. It is important to note that this is not the same as saying that the slice signature is contained in the packet signature. Packet signatures must necessarily be defined by a set of *distinct* delays since for each timestep there can be just one amplitude in the input stream. Slice signatures suffer no such restriction – they may include *duplicate* delays. The condition for the non-zero passing of an impulse packet through a delayed static nonlinearity can be restated compactly in terms of the following formalism, which is useful in its own right. The process of replacing a pair of duplicate delays in a signature by a single copy of the delay defines an equivalence relation, which we call "2-for-1" equivalence", on the set of signatures. Since duplicate replacement may be repeated as long as duplicate delays are present, we see that each 2-for-1 equivalence class contains exactly one *duplicate-free* signature. In light of this observation, one might prefer to think of the relation as "n-for-1" equivalence. Recall that a primary signature is one having $d_0 = 0$. These are the signatures that label slices. Since 2-for-1 reduction of a signature is primary if and only if the signature itself is, 2-for-1 equivalence defines by restriction an equivalence relation on the subset of primary signatures. One more definition will lead us to our immediate goal. We say that a signature is "primitive" if it is primary and also duplicate-free. It is clear that each primary equivalence class contains a unique primitive. In this language, our result is: An impulse packet drives all slices whose associated equivalent primitive signatures are contained in the packet signature.

Equivalence classes are often referred to as "partitions" of the underlying set. The partitions of the set of primary signatures imply partitions of the set of slices. We shall typically be lax about the distinction, letting the context determine which we mean. A couple of examples: Slices with primary signatures $(0, \ldots, 0)$, comprise a partition whose primitive is $(0)$. The two slices with signatures $(0, d_1, d_1, d_2, d_2)$ and $(0, d_1, d_1, d_2, d_2, d_2)$ belong to the same partition, and its primitive is $(0, d_1, d_2)$.

Signatures with duplicate delays have been loosely called "diagonal" by extension from the situation for 2$^{nd}$ order kernels, where the elements with two equal delays are plotted over the main diagonal of the matrix of delay pairs. (See Figure 1). Slices corresponding to these signatures tend to complicate situations that one would prefer simple. But these slices are there and must be dealt with. Actually, as we show below, slice terms with duplicate delays in their signatures gather via 2-for-1 equivalence into partitions that encapsulate the non-linear response of the system to impulse packets. The picture that arises is ultimately rather satisfying.

*2.2 The primitive representation*

Equation (5), the slice representation of the Volterra system, may be rearranged into its "primitive form"

$$r(t) = \mathcal{P}_{(\emptyset)}(t) + \mathcal{P}_{(0)}(t) + \sum_{d_1>0}^{T-1} \mathcal{P}_{(0,d_1)}(t) + \sum_{d_1>0}^{T-1}\sum_{d_2>d_1}^{T-1} \mathcal{P}_{(0,d_1,d2)}(t) + \ldots \qquad (8)$$

where

$$\mathcal{P}_{(\emptyset)}(t) \equiv \mathcal{L}_{(\emptyset)},$$

$$\mathcal{P}_{(0)}(t) \equiv \sum_{d=0}^{T_{(0)}-1} \mathcal{L}_{(0)}(d)S_{(0)}(t-d) + \sum_{d=0}^{T_{(0)}-1} \mathcal{L}_{(0,0)}(d)S_{(0,0)}(t-d) +$$

$$\sum_{d=0}^{T_{(0)}-1} \mathcal{L}_{(0,0,0)}(d_0)S_{(0,0,0)}(t-d) + \ldots$$

$$\mathcal{P}_{(0,d_1)}(t) \equiv \sum_{d=0}^{T_{(0,d_1)}-1} \mathcal{L}_{(0,d_1)}(d)S_{(0,d_1)}(t-d) + \sum_{d=0}^{T_{(0,d_1)}-1} \mathcal{L}_{(0,0,d_1)}(d)S_{(0,0,d_1)}(t-d) + \qquad (9)$$

$$\sum_{d=0}^{T_{(0,d_1)}-1} \mathcal{L}_{(0,d_1,d_1)}(d)S_{(0,d_1,d_1)}(t-d) +$$

$$\sum_{d=0}^{T_{(0,d_1)}-1} \mathcal{L}_{(0,0,d_1,d_1)}(d)S_{(0,0,d_1,d_1)}(t-d) + \ldots$$

etc.

Note that the signatures labeling the terms on the right side of each equation in (8) correspond to the elements of the partition whose primitive labels the left side.

We apply the expansions of Equation (9) to impulse packet inputs and compute response corrections $\mathcal{P}_{\text{primitive}}$. First recall that we showed in the previous section that if an impulse packet with signature $(0,d_1,d_2,\ldots,d_{n-1})$ enters the DSNL $S_{(0,d_0,d_1,\ldots,d_{n-1})}$, it emerges as a single pulse whose amplitude is the product of the packet amplitudes. A similar result holds for some more elaborate non-linearities. If the DSNL signature $(0,d'_1,d'_2,\ldots,d'_{m-1})$ has 2-for-1 primitive equal to the packet signature $(0,d_1,d_2,\ldots,d_{n-1})$, then essentially the same argument shows that the output of $S_{(0,d'_1,d'_2,\ldots,d'_{m-1})}$ is a single pulse of amplitude $a_0^{m_0} \cdot a_1^{m_1} \cdot a_2^{m_2} \cdots a_{n-1}^{m_{n-1}}$, where $m_j$ is the multiplicity of the $d_j$ among the $d'_k$. This pulse amplitude passes through the convolution with $\mathcal{L}_{(0,d'_1,d'_2,\ldots,d'_{m-1})}$. $\mathcal{P}_{(\varnothing)}$ contains just the bias slice, which comprises the single value $L_0$. Thus, if the input is the empty packet and $t^*$ is any time, we have

$$\mathcal{P}_{(\varnothing)}(t^*) = \mathcal{L}_{(\varnothing)} = L_0. \tag{10}$$

This is the only contribution to the total response. $\mathcal{P}_{(0)}$ contains the linear slice. Its other slices are the higer-order diagonals. If $s(t)$ is a single pulse, arriving at time $t^*$ and having amplitude $a$, we see that

$$\mathcal{P}_{(0)}(t^*+d) = a \cdot \left( \mathcal{L}_{(0)}(d) + a \cdot \mathcal{L}_{(0,0)}(d) + a^2 \cdot \mathcal{L}_{(0,0,0)}(d) + \ldots \right) \tag{11}$$

The sum of this contribution and the bias term is the total response. Note that the non-linearities represented in (11) are the classical static, as opposed to delayed-static, type. $\mathcal{P}_{(0,d_1)}$ contains the slices reducing to $\mathcal{L}_{(0,d_1)}$. Its contribution to the response to a two-pulse packet having separation $d_1$ and amplitudes $a$ and $b$ is

$$\mathcal{P}_{(0,d_1)}(t^*+d) = ab \cdot \begin{pmatrix} \mathcal{L}_{(0,d_1)}(d) + a \cdot \mathcal{L}_{(0,0,d_1)}(d) + \\ b \cdot \mathcal{L}_{(0,d_1,d_1)}(d) + ab \cdot \mathcal{L}_{(0,0,d_1,d_1)}(d) + \ldots \end{pmatrix} \tag{12}$$

The total system response to this pulse packet is the bias, plus the single pulse corrections for the pulses separately, plus this correction.

The corrections shown in both (11) and (12) have the form of a product of *distinct* pulse amplitudes, times a power series with coefficients in the slices, expanded in powers of the individual pulse amplitudes. This form neatly separates time and amplitude effects in the impulse packet corrections.

*2.3 Kernel calculation by subpacket-correction*

The primitive representation described in the previous section is the formulaic embodiment of the subpacket-correction property of the Volterra slices. As we shall show here, the formulas of that section, especially (10), (11) and (12), and their higher order analogs (which we denote (10), etc), give a recipe for constructing an experiment from which the slice functions $\mathcal{L}_{\text{signature}}$ can be calculated. The experiment is not generally an *efficient* means of finding the slices, but it has considerable *theoretical* utility. In particular, it shows how to choose a collection of packets that fully determine any given Volterra reduction (Schetzen, 1989).

Begin by presenting the empty packet. Practically, this means one sets the input level to 0 and waits one full memory duration for the system to settle down to the constant value required by the Volterra assumptions. Let $t^*$ be a time when the settling is complete. From the discussion of Equation (10) we have that $\mathcal{P}_{(\varnothing)}$ is the only contribution to the response. Thus, we have

$$r(t^*) = \mathcal{L}_{(\varnothing)} \tag{13}$$

which determines $r(t^*) = \mathcal{L}_{(\varnothing)}$.

Next, present a single pulse having some non-zero amplitude $a$. Again set the input level to 0 and wait a full memory duration before passing the pulse. Let $t^*$ be the time the pulse first enters memory. From the discussion of Equation (11), only $\mathcal{P}_{(\varnothing)}$ and $\mathcal{P}_{(0)}$ contribute to the response of the system to our pulse. Since we already know the value of $\mathcal{P}_{(\varnothing)}$, we can subtract it out, getting

$$r_a(t^* + d) - \mathcal{P}_{(\varnothing)} = a \cdot \left( \mathcal{L}_{(0)}(d) + a \cdot \mathcal{L}_{(0,0)}(d) + a^2 \cdot \mathcal{L}_{(0,0,0)}(d) + \ldots \right) \tag{14}$$

where the left-hand side is known for each $d < T$. The response $r$ is subscripted to remind us that the response depends on the pulse amplitude. Since $a$ is fixed, for each $d$, (14) may be construed as a linear equation in the unknowns $\mathcal{L}_{(0)}(d)$, $\mathcal{L}_{(0,0)}(d)$, $\mathcal{L}_{(0,0,0)}(d)$, etc. If there are no terms involving duplicate delays, then we can solve for $\mathcal{L}_{(0)}(d)$, and there is nothing more to do. If there are duplicate-delay terms in (14), we don't yet have enough information to pin them down. We proceed by taking the only path available, which is to repeat the presentation of the pulse, but using a different amplitude. In general, if there are $k$ slices on the right side of (14), we need to present $k$ distinct non-zero amplitudes. These presentations lead to a set of simultaneous equations in the unknowns whose coefficient matrix is

$$\begin{pmatrix} a_0 & a_0^2 & a_0^3 & \cdots \\ a_1 & a_1^2 & a_1^3 & \cdots \\ a_2 & a_2^2 & a_2^3 & \cdots \\ \cdots & \cdots & \cdots & \cdots \end{pmatrix} \tag{15}$$

which, as we show in Appendix I, is invertible for distinct non-zero values $a_k$, allowing solution of the equations for the unknowns. In practice, one chooses amplitudes so that matrix (15) becomes well conditioned, since that makes the numerical solution for the kernels more robust. Such choices can be found by trial and error. Repeating this procedure for each $d$, we determine the slices $\mathcal{L}_{(0)}$, $\mathcal{L}_{(0,0)}$, $\mathcal{L}_{(0,0,0)}$, etc.

In a similar way, present pulse pairs of various separations and amplitudes, and use Equation (12) to determine the slices $\mathcal{L}_{(0,d_1)}$, $\mathcal{L}_{(0,0,d_1)}$, etc., using the known values of $\mathcal{L}_{(\varnothing)}$, $\mathcal{L}_{(0)}$, $\mathcal{L}_{(0,0)}$, $\mathcal{L}_{(0,0,0)}$, etc. to form the left-hand sides of the equations. Again note that if there no duplicate-delay terms, we need only present pulse pairs involving the single amplitude $a$ to be able to solve for the unique duplicate-free slice $\mathcal{L}_{(0,d_1)}$. If there are terms with duplicate delays, we must present pulses having multiple non-zero amplitudes. We obtain a coefficient matrix of the form

$$\begin{pmatrix} a_0 b_0 & a_0^2 b_0 & a_0 b_0^2 & a_0^2 b_0^2 & \cdots \\ a_1 b_1 & a_1^2 b_1 & a_1 b_1^2 & a_1^2 b_1^2 & \cdots \\ a_2 b_2 & a_2^2 b_2 & a_2 b_2^2 & a_2^2 b_2^2 & \cdots \\ a_3 b_3 & a_3^2 b_3 & a_3 b_3^2 & a_3^2 b_3^2 & \cdots \\ \cdots & \cdots & \cdots & \cdots & \cdots \end{pmatrix}. \qquad (16)$$

The main result of Appendix I shows that if the maximum monomial degree appearing among the elements of (16) is $n$, then appropriately chosen states formed from zero and $n$ distinct non-zero amplitudes substituted in for the $a_i$ and $b_j$ will guarantee the invertibility of the matrix.

Continuing in this manner, one can generate matrices corresponding to the higher analogs of (10), etc. The general situation is analogous to the special cases just described. If there are no duplicate delay terms, pulse packets having a single non-zero amplitude suffice for solution. Otherwise, we must present at least as many non-zero amplitudes as the maximum monomial degree.

We can turn the preceding arguments around and see that the forms of the matrices (15) and (16) limit the number of slices that can be solved for in a multi-level experiment. This is because the $a_k$ and $b_j$ in terms of which the matrices are defined can each take on only the same, limited, number of values, and so eventually we run out of distinct rows that can be produced. But there can be no more slices found than the number of rows available. In effect, we need to remove columns from the analysis so that matrices (15), (16), and their higher order analogs are *invertible* and square. Beyond that, which columns we chose retain is arbitrary, although we would have a natural prejudice toward retaining, as far as possible, those associated with the lowest order signatures.

*2.4 Adequacy of assuming finite order*

As we have said above and will make very clear below, the feasibility of experiments to determine the Volterra kernels depends critically on there not being too many of them that are non-zero. The primitive form of the Volterra expansion, Equations (8) and (9), and especially the application to pulse packets, Equations (10), (11), and (12), make the kinship with Taylor expansion explicit, and we look to them for hints about the effectiveness of modeling with only a limited number of terms. The simplest case is shown in Equation (11), where the dependence on pulse amplitude at each delay is just that of a Taylor series. This leads us to investigate approximation by truncated Taylor series, that is, by polynomials. As an example, we look at polynomial approximations of a function that has a sigmoid dependence on its independent variable. Sigmoid dependence on input amplitude is very common in biological systems. How does the quality of the approximation depend on the number of terms in the polynomial?

Figure 6 illustrates the general answer. A sigmoid here defined by the formula $y = x^{15} / (50^{15} + x^{15})$ is shown in gray. The dashed black curve shows that an approximation of the sigmoid can be gotten from a degree-3 polynomial, and the solid black curve shows that going to degree-9 makes the approximation quite accurate. However, there is a major caveat: The accuracy holds only for a bounded range of independent values; outside that range the match deteriorates extremely rapidly.

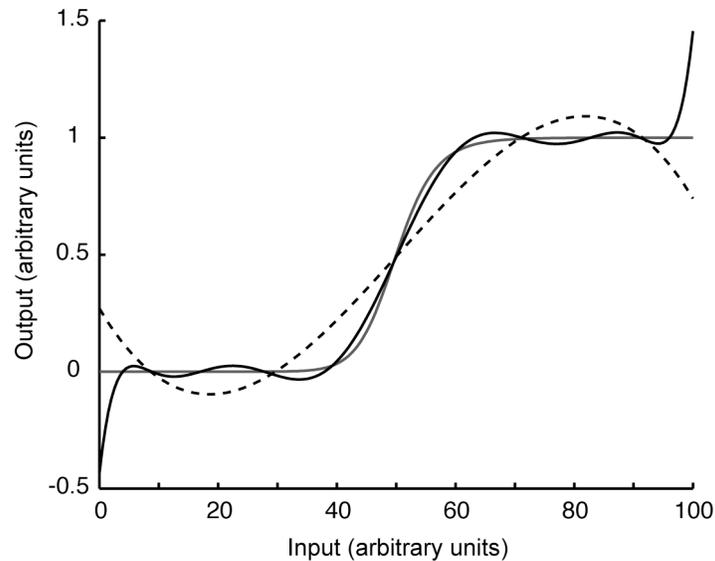

*Figure 6*. An illustrative sigmoidal non-linearity (gray) is least-squares fit with polynomials of degree-3 (dashed black) and 9 (black).  The degree-9 curve matches the sigmoid quite well over the central portion of the domain, but breaks away near the edges.  This reflects the fact that any polynomial (which by definition has finite degree) becomes monotonically unbounded as the absolute value of the independent variable goes to infinity.  The break is abrupt since it goes with the degree of the polynomial.  The match by the degree-3 curve is, of course, worse, but even it captures the flavor of the sigmoid.

From a practical point of view, limited range is not usually a problem since the physical systems we deal with typically function normally only with restricted inputs – they break down or turn off under too-violent stimulation.  Unless we are specifically exploring the breakdown, we feed our systems inputs they can handle; that is, we investigate a system as it normally functions.  In determining the Volterra expansion for this normal system we would naturally expect to provide inputs representing the acceptable range. For some systems, the input range is limited by *definition*, so the range issue is moot.  For example, input amplitude might represent percent contrast, a number between 0 and 100.  Or, a system might accept only two input values, *on* and *off*.

In the above discussion we have finessed an important point.  The approximating curves in Figure 6 were fit by least-squares over the functional domain.  It is more usual, when fitting a function with a power series, to match derivatives at some particular point in the domain.  That procedure requires knowing that the function actually has the requisite derivatives, and also that the radius of convergence of the approximation includes the region of interest.  The standard theorems on the applicability of Volterra expansion assume this latter fitting method.  However, in practice, other fits are typically more useful.  As Klein points out (Klein, 1992), derivative fitting may be appropriate where the response is likely to be very small, as in threshold determination, but will often not do a good job over a full range of system responses. The Wiener expansion, sometimes used in place of the Volterra expansion (Victor, 1992) has, built-in, the notion of global fitting. The kernel calculation techniques described below, even as applied to Volterra expansions, also fit globally.

Generally, the kernels one gets depend on the details of the fitting procedure. In particular, the input amplitudes used can affect the resulting kernels. In the example of Figure 6, the particular polynomial approximations obtained depended on the range of inputs and fitting was done by integrating the squared difference between the sigmoid and the polynomial over the whole range. In an experiment involving only a finite number of different amplitudes, one would naturally use the squared difference at the amplitudes used. If the question were one of approximating a polynomial rather than a sigmoid, the best approximation would of course equal the underlying polynomial as long as the degree of the approximating polynomials were as large as that of the target polynomial and the number of amplitudes used were also at least as large as the target degree. Thus, in that case, which particular amplitudes are used does not change the result. In a similar way, if we are dealing with an underlying system that is itself finite Volterra, it turns out that any finite set of amplitudes that allows for kernel determination will find the same thing, the actual underlying system. This result is implicit in the theorem of Appendix I.

**3. The S-matrix**

We saw in Section 2.3 how an experiment based on the systematic presentation of impulse packets allows the determination of the kernel slices for any reduced Volterra system that comprises finitely many kernels. Such experiments are not particularly efficient, primarily since a lot of time is spent clearing memory in order to present clean packets. It has long been known that experiments involving sustained sequential presentation of random or pseudo-random valued inputs can also be used to extract the kernels. These experiments are typically more efficient than those based on impulse packets. In this section we develop a matrix-based framework in which they may be understood and the calculations carried out.

We begin by again recasting Equation (2). Our previous reformulations have used specific groupings of kernels that enhance interpretation. Here we use a formulation that is better suited to the matrix manipulations that underlie our present methods. We begin with the inputs, which, in the context of Equation (2) are a sequence of values indexed by time. Since the summations in Equation (2) run over delays $0$ to $T-1$, it is clear that at any time $t$ only the inputs $s(t), s(t-1), \ldots, s(t-T+1)$ can affect the response. These input values, the one at time $t$ and those immediately preceding up to the limit of memory, define the "input state". For each signature $(d_0, d_1, d_2, \ldots, d_{n-1})$ and each time $t$, we define the "signature-state product" (SSP) by the equation $S(d_0, d_1, \ldots, d_{n-1}; t) \equiv s(t-d_0)s(t-d_1)\ldots s(t-d_{n-1})$. That is, the SSP at each time is the product of input values that arrived at delays comprising the signature, relative to that time. By convention, the SSP for the empty signature, which appears implicitly in the bias term of Equation (2), has the value 1 for every state. Note that except for the fact that here the signatures here are not necessarily primary, so $d_0$ may be non-zero, considered as a function of $t$, the SSP is the same as the DSNL defined in section 2.1.

Besides the SSPs, the other players in Equation (2) are the kernels $L_n(d_0, d_1, d_2, \ldots, d_{n-1})$. The subscript attached to $L$ is redundant since it is implied by the number of arguments. If we define $L(d_0, d_1, d_2, \ldots, d_{n-1}) \equiv L_n(d_0, d_1, d_2, \ldots, d_{n-1})$, we can rewrite Equation (2) as

$$r(t) = \sum_{\text{signatures}} L(\text{signature})S(\text{signature}; t) \qquad (17)$$

When we run an experiment, we know the inputs and gather the responses. Thus, for each time $t$, $S(\text{signature}; t)$, for all signatures, and $r(t)$ are known real numbers. On substitution of these numbers, Equation (17) becomes a simple linear equation in the unknowns $L(\text{signature})$. This is, of course, analogous to the situation in Section 2.3. Here, as there, the linearity is the key to the computability of the kernels. By performing the substitution at various times, we get a system of linear equations in the kernels $L(\text{signature})$ that *constrain* their possible values. The larger the system of equations, the tighter the constraint. The plan is that with sufficient enlargement, the system of equations will actually *determine* the kernels.

Here is the matrix formalism that corresponds to above scheme. Choose a particular ordering of signatures. It doesn't matter what the ordering is. Form a column vector by taking the kernel elements $L(\text{signature})$ in signature order. For a particular time, form a row vector whose elements are the SSPs $S(d_0, d_2, \ldots, d_{n-1}; t)$ also written in signature order. Then, the formula for the matrix product of the row vector on the left by the column vector on the right is precisely Equation (17). That is, the response at time $t$ is the matrix product of the SSPs vector at time $t$ with the kernels vector. If we have responses for a multiplicity of times, we can stack the corresponding SSP rows to form what we call a "signature-state-product-" or "S"-matrix. If we denote the S-matrix by $\mathbf{S}$, denote the column vector of kernel elements by $\mathbf{L}$, and if we stack the responses in the same order as the individual SSP rows to form the response column vector $\mathbf{R}$, the system of constraining equations becomes

$$\mathbf{S} \cdot \mathbf{L} = \mathbf{R}. \tag{18}$$

We know the responses $\mathbf{R}$, and we wish to know the kernel elements $\mathbf{L}$. The applicable theorem says we can solve for the elements of $\mathbf{L}$ if the matrix $\mathbf{S}$ is square and non-singular; that is, if its determinant is non-zero, or equivalently if its rows and columns are linearly independent. Since we control the inputs, we control the states, and via the SSPs, the S-matrix rows. It turns out that it is not hard to find a collection of states that make $\mathbf{S}$ non-singular (cf., Section 2.3). We can use the freedom to choose states to advance our purpose in a number of ways.

Note that we have presented Equation (18) as the prescription for finding the kernels from the states and the corresponding system responses. However, once the kernels are determined, Equation (18) becomes a formula for response *reconstruction*. To reconstruct the system response to a single state or a sequence of states, build the corresponding S-matrix row or S-matrix and multiply it by the known kernels vector.

Although in the previous discussion we have used time $t$ as the parameter in the SSPs and the responses, our fundamental assumption that the system response depends only on the input state, and not the absolute time, means that we might also use abstract input states to parameterize these quantities. Thus, in particular, we may write a SSP as $S(\text{signature}, \text{state})$ and a response as $r(\text{state})$. Indeed, we shall prefer this formulation, and unless otherwise indicated, our SSPs will be of this form.

*3.1 A special example*

We illustrate the calculation procedure with an example that, in addition to being very simple, provides our first encounter with the kernels of a binary, in fact B01, system. Binary systems will be presented in detail beginning in Section 8. Suppose that the slices in our Volterra system are non-zero only if they have primitive signatures, that is, ones comprising duplicate-free delays. Then, revisiting the argument in Section 2.1, we see that the strongest form of the subpacket-correction property holds – a slice is driven by an impulse packet only if its signature itself is contained in the packet signature. This implies, in particular, that a slice is driven only if its order is less than or equal to the number of delays in the packet, the "packet order". The assumption that we have only primitive signatures implies that the order of the system, that is, the highest $n$ for which $L_n \neq 0$, is no greater than $T$.

We proceed to carry out the impulse packet experiment described in Section 2.3, noting that because there are no repeated delays, we need only use pulses all of which have a single amplitude, which we may take to have the value 1. However, instead of solving for the kernels piecemeal as in Section 2.3, we now set up an S-matrix. Order the *unique* memory states as they appeared in the experiment: first the all-zero state; then the memory states generated in time by a single unit-amplitude pulse; then those generated by a unit-amplitude pulse pair having adjacent pulses; then the pulse separated by a quiescent step; etc, followed by unit-amplitude pulse triples, played out in time, in any reasonable order; etc. In general, we always add all the time translations of each packet, in time order. Since all the impulses in our packets have unit amplitude, the states described above are characterized by packet signatures, which by our convention are primary, that is, have smallest delay equal to 0, together with a single delay indicating where the packet is on its passage through the memory window. That is, each state is uniquely characterized by a duplicate-free, not generally primary, signature. Order the kernels so that their signatures match the order of the state delay patterns.

To make our example completely explicit, we suppose that both system memory $T$ and interaction order equal 3. Then, the S-matrix obtained by computing the individual SSPs is

$$\begin{pmatrix}
 & (\emptyset) & (0) & (1) & (2) & (0,1) & (1,2) & (0,2) & (0,1,2) \\
(\emptyset) & 1 & 0 & 0 & 0 & 0 & 0 & 0 & 0 \\
(0) & 1 & 1 & 0 & 0 & 0 & 0 & 0 & 0 \\
(1) & 1 & 0 & 1 & 0 & 0 & 0 & 0 & 0 \\
(2) & 1 & 0 & 0 & 1 & 0 & 0 & 0 & 0 \\
(0,1) & 1 & 1 & 1 & 0 & 1 & 0 & 0 & 0 \\
(1,2) & 1 & 0 & 1 & 1 & 0 & 1 & 0 & 0 \\
(0,2) & 1 & 1 & 0 & 1 & 0 & 0 & 1 & 0 \\
(0,1,2) & 1 & 1 & 1 & 1 & 1 & 1 & 1 & 1
\end{pmatrix}. \quad (19)$$

In (19), the S-matrix proper is to the right and below the solid lines. The column and row, to the left and above the solid lines indicate the state and kernel signatures. The matrix decomposes into blocks, as indicated by the dashed lines, associated with packets and slices. The first matrix column represents the bias slice, comprising a single kernel element. The SSP value corresponding to each state is 1, reflecting our convention for the empty signature $(\varnothing)$. The second group of 3 columns represent the kernel elements comprising the linear slice, signatures $(0)$, $(1)$, and $(2)$. The third group of 2 columns represent the kernel elements comprising a $2^{nd}$ order slice, signatures $(0,1)$, and $(1,2)$. The next column represents the single kernel element that comprises the $2^{nd}$ order slice with signature $(0,2)$, and the last column the single kernel element that comprises the $3^{rd}$ order slice with signature $(0,1,2)$. The first row represents the quiescent state, all entries zero. The next three rows show the states comprising the time motion of a single pulse through memory. The next two rows, the states comprising the time motion of an adjacent pulse pair through memory. The next-to-last row represents the state having two pulses separated by a quiescent step. This state cannot move at all without going out of memory. Similarly, the last row represents the state comprising 3 adjacent pulses, which again cannot move without going out of memory.

The lower-triangular form with 1's on the diagonal clearly implies that the determinant is non-zero, which means the matrix is uniquely invertible, which reconfirms that an experiment that shows the states corresponding to the rows will be complete in the sense that, from knowledge of the system responses, Equation (18) can be solved to determine all the kernels.

We note in passing that the particular ordering of the rows of the S-matrix is unimportant, as it does not affect invertibility. If we do reorder the matrix rows, we must, of course, reorder the rows of the responses column vector in the same way, before solving for the kernels. The principle is that the state that produces an S-matrix row must be the one that produces the corresponding response.

Equations (20) below correspond to Equation (18) with all substitutions made explicit.

$$\begin{aligned}
L_0 &= r(\emptyset) \\
L_0 + L_1(0) &= r(0) \\
L_0 + L_1(1) &= r(1) \\
L_0 + L_1(2) &= r(2) \\
L_0 + L_1(0) + L_1(1) + L_2(0,1) &= r(0,1) \\
L_0 + L_1(1) + L_1(2) + L_2(1,2) &= r(1,2) \\
L_0 + L_1(0) + L_1(2) + L_2(0,2) &= r(0,2) \\
L_0 + L_1(0) + L_1(1) + L_1(2) + L_2(0,1) + L_2(1,2) + L_2(0,2) + L_3(0,1,2) &= r(0,1,2)
\end{aligned} \quad (20)$$

These equations graphically reveal the strong subpacket-correction property of our system, that a state drives a kernel only if the kernel signature is contained in the packet signature, and each driven kernel contributes a correction to the total response equal to its value. Moreover, these corrections work slice by slice. The 2$^{nd}$ through 4$^{th}$ equations can be read to say that the time response to a single pulse is the bias plus the linear slice; the 5$^{th}$ and 6$^{th}$ equations, that the response to an adjacent-pulse-pair, w*hile the entire pair* is in memory, is the bias plus two adjacent copies of the linear slice plus a copy of the second order slice with signature $(0,1)$.

*3.2 More generality*

The discussion of the preceding section may be expanded to handle the Volterra system of Section 2.3. The states that index the S-matrix and response vector rows would have entries comprising multiple input levels, and the S-matrix entries, being products of these levels, would have values other than just 0 and 1. Nonetheless, the S-matrix formed in analogy to that in Section 3.1 would have blocks corresponding packets and slices. It would be a square matrix, which the arguments of Section 3.1 implicitly show to be invertible. Of course, for any particular bound on interaction order, the Volterra matrix will be much bigger than the B01 matrix having that same order. Still, the Volterra matrix and the response vector may be solved for the kernels, and the kernels may be used to reconstruct responses to arbitrary inputs.

*3.2 Noise control*

Until now, our entire discussion has been based on the assumption that our system *exactly* satisfies the conditions laid out at the beginning of Section 2, and so the Volterra expansion, Equation (2) holds *precisely*. Real systems, of course, are always to some degree uncertain. They can't be given exactly replicated inputs, and if they could, the responses would always show some amount of variation. How uncertainty, which we shall simply call "noise", arises is dependent on the system at hand. As mentioned previously, considerable research has gone into controlling its effects. We shall content ourselves here with dealing with only the simplest noise model, in which we suppose an underlying, purely deterministic, noise-free system operates in parallel with a noise generator that acts independent of the system inputs, and the outputs of these two channels is simply added before being recorded. At each sample, the noise generator chooses a value from some particular probability distribution. We take the distribution to be Gaussian with zero mean and some given variance.

Our goal is to determine the kernels of the underlying system. The usual strategy is to simply replicate the experiment some number of times, and, for each input state, form the average response to that state. Since, under our assumptions, the underlying part of the response to the state doesn't change with repetition, our procedure is the same as adding the average noise part of the responses to the true response. But, the average over the noise samples converges to the distribution mean as the number of repetitions increases. Since the mean of the Gaussian is zero, the average response converges to the underlying response. The convergence is known to go inversely with the square root of the number of repetitions. We compute kernels based on the average responses, which by the continuity of the matrix inversion, converge to those of the underlying system. In fact, because of the linear connection between the responses and the kernels (mediated by the S-matrix inverse), Gaussian noise added to the response causes Gaussian noise in the kernels, and there are effective means of computing the variance of the latter using the covariance matrix constructed from the S-matrix (Press, Flannery, Teukolsky & Vetterling, 1988). We return to these considerations in Section 11.

*3.3 Over-determination and singular-value decomposition*

We can take a slightly different computational tack in dealing with our noisy system. Instead of averaging the responses, state by state, we can use the individual responses to the repeated inputs to build an S-matrix that is no longer square, but rectangular, having more rows than columns. In that case, many rows of the S-matrix would be identical to one another (the ones built from repetitions of the same state), but the corresponding responses would be slightly different (because of the noise). Matrices that are not square cannot be invertible in the usual sense. Even so, there is a general matrix procedure that, in the present case allows us to find exactly the same kernels that we find by the response averaging of the previous section, and that will prove to be very helpful when we take up the statistically-based kernel determination that we promised at the beginning of Section 3.

A matrix $\mathbf{S}$ is invertible when its rows and columns form independent sets of vectors. If there are more rows than columns, it is not possible that the rows be independent. However, the columns may still be. In that case, the system of equations associated with the matrix is "over-determined", and while it cannot generally be solved exactly, there is a sense in which a unique "best approximate solution" exists, and this is frequently quite useful. There is a mathematical method called "Singular Value decomposition" (SVD) that can be used to find the solution (Press et al., 1988). Given the overdetermined matrix $\mathbf{S}$, SVD can be elaborated to return a matrix $\mathbf{T}$ having the property that.

$$\mathbf{T} \cdot \mathbf{R} = \mathbf{L} \tag{21}$$

That is, $\mathbf{T}$ functions as a left inverse to $\mathbf{S}$, delivering the approximate kernels $\mathbf{L}$ from the noisy responses $\mathbf{R}$ by simple matrix multiplication.

There is an enlightening description of the approximate solution of an overdetermined system of equations: Suppose that $\mathbf{S}$ is an $m \cdot n$ matrix whose columns define $n$ vectors in $\mathbf{R}^m$, the m-dimensional Euclidean space. The columns being independent means that the subspace $\mathcal{S}$ of $\mathbf{R}^m$ determined by the column vectors is n-dimensional, and the column vectors form a basis for it. That is, the vectors lying in $\mathcal{S}$ are precisely those that may be written exactly as a sum (linear combination) of the columns. The experimentally determined response column vector $\mathbf{R}$ defines a vector $\mathbf{R}$ in $\mathbf{R}^m$, but, because of system noise, it generally doesn't lie in $\mathcal{S}$. However $\mathbf{R}$ can be projected into $\mathcal{S}$, and the projection $\mathrm{proj}(\mathbf{R})$, which is unique, is the vector in $\mathcal{S}$ that is closest to the vector $\mathbf{R}$ in the sense of Euclidean distance, the square root of the sum of component-difference squares. Minimal Euclidean distance implies minimal "least-squares difference", a common measure in statistics. The vector $\mathrm{proj}(\mathbf{R})$, since it lies in $\mathcal{S}$, can be written exactly as a sum of the columns. The coefficients in the sum form the approximate kernels $\mathbf{L}$.

In the case of overdetermination due to exactly repeated inputs, it is not hard to see that the approximate kernels formed by response averaging have precisely the same values as those delivered by the projection procedure. We show this in Appendix II.

We close this section with a pair of theoretical observations that hold when our response measurements are noise free: If we have a system that is actually finite Volterra in the sense of satisfying a finite truncation of Equation (2) *exactly* for *all* possible inputs, then, if we do an experiment for which the S-matrix columns corresponding to the reduced set of kernels are independent as discussed above, SVD inversion will find the kernels *exactly*. This is because the response vector $\mathbf{R}$ will actually lie in the space $\mathcal{S}$, so the approximation engendered by the projection into $\mathcal{S}$ is removed. On the other hand, if our system is not of the above sort, the response vector will not lie in $\mathcal{S}$, and in that case the projection will cause a fitting error; indeed, it will be the sole cause of that error. If another experiment is done using different inputs, typically the subspace $\mathcal{S}'$ generated by the columns of the S-matrix for the second experiment will be different from the subspace $\mathcal{S}$. However, as the response vector will also almost certainly be different in the two experiments, the kernels extracted may or may not be the same. What happens depends on the details. This issue is addressed a bit more in Appendix III.6.

*3.4 Experimental and computational time estimation*

As is apparent from Equation (3), the number of kernel elements required to complete even a reduced fast-stimulation analysis can be astronomical. For kernel extraction to be feasible, the reduction must be quite radical. For example, it is not uncommon to find a biological system that requires a step rate of 10 ms, has a memory of 1 sec, and has kernels of interaction order 3. For such a system, Equation (3) yields an estimate of $100^3/6 \simeq 170,000$ kernels. In the best circumstance, where noise is not a factor and where the states are input with perfect efficiency, the experiment would take 1,700 secs, or about one-half hour. If the interaction order were 4, the minimal experiment would 12 hours. If, as is always the case in biology, noise needs to be controlled by repetition, these time estimates would need to be multiplied by the square root of the number of repetitions.

Moreover, in assessing the feasibility of an experiment, more than the total experiment time needs to be considered. In particular, the question of system stability must be addressed. Even a half-hour experiment exceeds the time many biological preparations deliver stable behavior. But, stability time is not always critical. If multiple *identical* preparations may be made, it can be worked around, since each separate preparation may be exposed to a new subset of states, and the responses pooled.

Assuming an experiment can be done, are we able to actually calculate the kernels? Press et al., (1988) provides particularly clear discussion of the methods available for the solution of matrix equations and the applicable results and costs are summarized here. General techniques for solving matrix equations require, assuming the matrix is square of size $p \times p$, on the order of $p^3$ floating-point operations (FLOPS). Equations involving matrices having specialized form can often be solved more quickly. In particular, orthogonal matrices can be solved by projection methods. Each projection requires an inner product evaluation, which uses on the order of $p$ operations. Since there are $p$ of these to be done, the cost is order $p^2$ FLOPS, a very considerable saving over the general case. SVD is a fully general method, so its computation cost grows with the cube of the number of kernel elements. If there are 100,000 of these, this means some small integer multiple of $10^{15}$ FLOPS are required to do the calculation. At $10^9$ FLOPS/sec, extraction would take hundreds of hours.

These numbers clearly push the envelope. Fortunately, in addition to the impulse packet method of presenting inputs, there is another way to proceed that is helpful in reducing experiment time, but whose main utility is to produce orthogonal, or, at least, mostly orthogonal, S-matrices, and so allow kernel extraction at a cost that rises with only the square of the number of kernels to be determined. If we apply this alternative method to our system, we find that the balance changes, and the experiment time, rather than the computation time becomes limiting. We shall present the alternative method the next section.

**4. Random and pseudo-random inputs**

The explicit technique for kernel extraction we described in Section 2 involved system stimulation by impulse packets. In Section 3, matrix extraction methods were introduced, and by way of example, applied to the packet-stimulation experiments of the previous section. This approach elucidated the tight connection between particular inputs and slices that we call subpacket-correction. Historically, quite different extraction methods were developed and applied. Most important were the "white-noise" techniques, where the system input comprises continuous sequences whose values are taken at random from specific probability distributions (see Appendix B; (Schetzen, 1989)). White-noise techniques are theoretically complete, in that they allow, in the infinite-time limit, exact solution for all kernels of the continuous Volterra expansion. In addition, they are instructive, since there exists a clear pictorial analog of the extraction calculation.

We apply white-noise techniques by constructing discrete, finite-length approximations to the continuous white-noise inputs. This is analogous to our replacement in Section 1.2 of the infinite, continuous, theoretical form of the Volterra expansion (1) by the finite, discrete, practical one (2). The discrete inputs are gotten by simply extracting a finite number of random samples from the probability distribution used in the continuous case. We present these inputs to the system and measure responses as usual. Since we know the input sequence, we know the sequence of states, and so, as in Section 3, can form the S-matrix corresponding to the inputs and the assumed reduced set of kernels. If the number of distinct states presented is greater than the number of kernels to be found, there is at least the possibility that the columns of the S-matrix are independent. If they are, we can solve for the kernels, for example, by applying the SVD method of Section 3.3. Since our input values are random, if the number of distinct input values is sufficiently large and appropriately chosen, the columns are likely to be independent. Moreover, as the length of the input sequence increases, the likelihood of independence tends toward certainty. We present the arguments that support this and other technical assertions of this section in Appendix III.

White-noise analysis as introduced by Wiener used inputs whose values were drawn from a Gaussian distribution. That much variation in input values is not necessary when seeking the kernels of finite expansions. As we saw in Section 2.3, expansions having higher order terms involving duplicate delays require variable amplitude input pulses for kernel extraction, but limited interaction means that a limited number of different amplitudes is sufficient. Similar considerations apply here. As is made clear in Appendix III, distributions appropriate for white-noise analysis must have zero mean. In practice, the use of a finite symmetric distribution whose values span the typical input range of the system is common. (Marmarelis, 2004) champions this procedure and explains it in considerable detail. In fact, the distribution need not be symmetric.

Beyond column independence, which is necessary for kernel extraction, there is a stronger limiting property of the columns from which the real power of white-noise technique devolves: As the input sequence grows, the S-matrix columns, seen as vectors as in Section 3.3, tend toward having fixed, non-zero angles relative to one another. Indeed, most vector pairs tend toward being orthogonal. Which vector pairs these are can be read from the kernel signatures – orthogonal pairs are those whose signatures do not differ by paired, duplicate delays. This condition is an equivalence relation on signatures; different from the 2-for-1 equivalence we looked at earlier. As it is obtained by simply canceling duplicate delays in pairs, we call it "mod-2 equivalence". Mod-2 shares with 2-for-1 equivalence the property that each partition contains a unique duplicate-free representative, and that it restricts to an equivalence on the set of primary signatures. We call the duplicate-free mod-2 representative of a primary signature its "mod-2 primitive". The number of vectors in each mod-2 class is small compared to the total number of column vectors. We note that from these signature considerations, we may conclude that columns corresponding to distinct kernels in any given slice are in different mod-2 classes, so tend toward orthogonality.

If the vectors in different mod-2 class were actually, as opposed to approximately, orthogonal, we could solve for the kernels by independently projecting, as in Section 3.3, the response vector into the subspaces of $\mathbf{R}^m$ spanned by the vectors belonging to the various classes, and then separately solving from the projections, say by SVD, for the individual kernels in each class. Since projection into a low dimensional vector space is an order $p$ FLOPS operation, where $p$ is the length of the vector, as is each SVD, the entire solution would require order $p^2$ FLOPS.

Typically, the failure of orthogonality goes with $1/\sqrt{p}$. Frequently, the procedure of the previous paragraph is simply carried out as if the class subspaces were orthogonal, and this leads to errors in the estimated kernels. Rather remarkably, it turns out that we can eliminate this problem by replacing finite-length, truly random input sequences by certain very special "pseudo-random" sequences (described below). These have the property that the subspaces that we would wish to be orthogonal are indeed, orthogonal. In that case, the projection process delivers the kernels exactly.

Pseudo-random sequences are *deterministic*, algorithmically generated, thus repeatable, sequences of numbers that, according to various measures, *look* random. Since these sequences are not random, there will necessarily be other measures in which the failure of randomness is apparent. For example, the so-called random Gaussian sequence generated by computer programs is always really a very long periodic, and therefore, non-random, sequence. If one extracts enough sequence elements, the repetition will be seen. Of course, the particular algorithms used are chosen such that the computer, and probably the universe, will break down before this happens. The pseudo-random sequences of interest here have relatively short periods, ideally a modest multiple of the number of kernels to be determined. Multi-level, symmetric, pseudo-random sequences having useful lengths are available (see Appendix IV).

In working with finite input sequences, we need be careful that the system memory state is fully loaded with known inputs before response collection begins. A convenient way to do this, which avoids computational "edge effects", is to treat our finite input sequence as though it were a single period extracted from an infinitely-repeating sequence. In practice, this amounts to using the last $T-1$ inputs (discrete memory equal to $T$) as a "run-up" to the first input, and beginning response recording when the first input is presented.

We conclude this summary of white-noise stimulation with the observation that no method of providing inputs affects the combinatorial explosion of the number of possible non-zero Volterra kernels with increasing interaction order and discrete memory, and so cannot ameliorate the impossibly long experiments required to extract them. Thus, as we indicated in Section 3.4, the main utility of random or pseudo-random techniques is to reduce computation time enough so that experiment time becomes the factor limiting analysis.

**5. Slow stimulation**

A common method of further reducing the number of kernels is to remove the restriction, hitherto implicitly enforced, that stimulation and response sampling share the same clock ticks. The "slow stimulation" technique uses fast sampling of the output of a system that has been driven by relatively slowly changing inputs. This procedure generates a set of highly-sampled slices. With a sufficiently slow stimulation rate, a complete presentation of stimulus states becomes feasible, while the fast sampling avoids aliasing and reveals *some* aspects of the fast-response detail of the system. The slow stimulation technique has been much applied. It was introduced by (Klein, 1992) and (Sutter, 1992) who referred to it as "sparse stimulation". It was also discussed by (Chen, Aine, Best, Ranken, Harrison, Flynn & Wood, 1996) who referred to "inserted" sequences.

Typically, in slow stimulation, the sampling rate is chosen to avoid aliasing, and the stimulation rate chosen as an integer divisor of the sampling rate. A sampling clock is set up. Samples are taken on its ticks. Time intervals delineated by the sample ticks are "sample intervals". The stimulus clock runs slower, but in synchrony with the sample clock, so that its ticks occur only on sample ticks. Stimulus values change only on stimulus ticks. The time intervals delineated by these ticks are the "stimulus intervals" or "steps". With this setup, we may label the samples in each stimulus interval by their delays relative to the stimulus ticks. Analysis is done by collecting the samples coincident with stimulus ticks (that is, at relative delay = 0), processing them in the standard way, collecting the samples obtained at relative delay = 1, processing these, etc. The resulting "sample kernels" are interleaved in relative-delay order to obtain "slow kernels" whose temporal resolution is the sampling rate. This process is illustrated in Figure 7. Each stemmed symbol represents a sample, and the square-headed ones denote the initial sample in each step, that is, the one at relative delay 0. In the figure, the responses at relative delay 3 have been pulled out of line. It is these responses that are collected and processed to produce the pulled-out sample kernels at relative delay 3. Note that slow kernels are not single values, but rather, finite sequences of values, one for each relative delay. Just as previously, kernels are combined into slices, with slow kernels forming "slow slices".

For biological systems, slow stimulation looks extremely attractive because it can radically reduce the number of kernels to be determined, reducing in turn experiment and computation times. The signatures that label slow slices have delays measured in stimulus steps, which means that the discrete memory used in Formula (3), which counts the number of signatures, is the ratio of the system memory to the stimulus step. Because of the interleaving, each slow signature corresponds to a number of sample kernels equal to the "sampling factor", the ratio of sample ticks to stimulus ticks. Taken

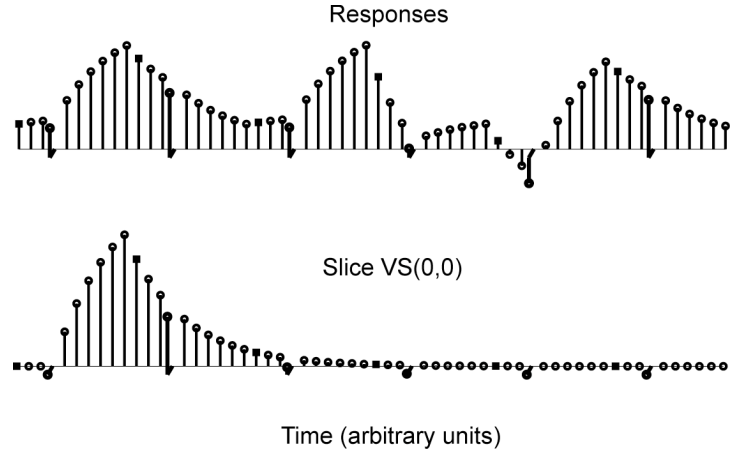

*Figure 7.* Slow slices are produced by separately processing responses at samples having the same relative delay in each time step. The top panel shows a portion of the response of a simulated Volterra system. The square-headed stems represent the initial sample in each step. The bottom panel shows a slice computed from the responses. Here, the square-headed stems represent the initial sample in each delay step. The pulled-out elements are taken at the fourth relative sample in each step.

together, the estimate of the number of kernels to be extracted in a slow-stimulation reduction is

$$\text{count}(n) \sim \sigma \cdot T_{\text{slow}}^{n} / n! \qquad (22)$$

where $\sigma$ denotes the sampling factor, and $T_{\text{slow}}$ is the "slow discrete memory". Since the sampling factor part of the fast discrete memory is taken to the 1st rather than nth power, even moderate sampling factors cause large count reductions.

However, there are important limitations to the usefulness of slow stimulation. One is to be expected: since slow slices contain information only about inputs that change at the stimulus rate of the experiment used to elicit them, they can be used to reconstruct responses only to inputs of this sort. That slow slices are *adequate* for this purpose may be seen in the following way: The analysis on the samples collected at relative sample delay $d$ is just the standard discrete Volterra analysis that we have been discussing up until this section, and for which reconstruction from the slices is a fundamental property. Reconstruction means simply that the system response to a particular memory state is the sum of kernels each multiplied by the SSP of the state and the kernel signature. The present discretization, where the interval is the stimulus step, and inputs and responses are sampled regularly at delay $d$ relative to the step boundaries is as good as any other as far as the mathematics goes. Because slow inputs change only at step boundaries, for a given input step, the state associated with the various relative delays $d$ are all the same. Similarly, all the relative delay kernels that are interleaved to make the slow kernels have the same signature. Thus all the sample kernels that comprise a slow kernel enter the reconstruction of a particular slow state with the same multiplier, and that is the same multiplier gotten by taking the SSP of the slow state and slow kernel signature, each understood in terms of steps rather than samples. Thus, slow reconstruction works formally as it should.

Our general interest in kernel interpretation suggests that we should examine more closely the meaning of input and response samples, as it is quite different in the context of slow stimulation. Previously we left the precise meaning unspecified, but implicitly took it to be something that changed in such a way that there was no *interesting detail* between sample values. In that case, whether the sample reflects the average input or system response over the sampling interval, or an effectively instantaneous A-to-D conversion value, makes no important difference. The relative delay $d$ samples under discussion here, being one *step* apart, are so far separated in time that interesting detail is *expected* to occur between them. It is precisely that detail that slow analysis is supposed to elucidate. For the relative delay $d$ samples, the input sample means the input value that was presented first to the system $d$ samples ago, and that was held constant since then, and the response sample means the system response at the current sample given the particular history of input changes, each made at a step boundary and held constant until the end of that step, or in the case of the present step, held constant until now. Thus, even though the underlying system is time independent, the slow stimulus responses encode a very specific time locking relative to the step transitions, and the kernels at large relative delay would be expected to show a tendency to settling reflecting the longish stretch of unchanging input since the previous step boundary.

Before leaving these considerations, it is useful to draw a further conclusion. Because subpacket-correction *characterizes* slices in terms of system responses to input packets associated with particular collections of signatures, and because the same collection of signatures is used for each relative delay $d$ and for the slow slice, the fact established above that reconstruction holds as expected for slow kernels implies that subpacket-correction does too.

A second limitation of slow stimulation is more important and less obvious. It reflects effects that occur because of the way the slice generating procedure distributes the finely sampled response data into the various coarsely labeled slices. Resulting features confuse the interpretation of slow-sampled slices. We described this effect in the Introduction and attached the name boundary crossing to it.  In the next section we shall work through a simple example that illustrates the general phenomenon.

*5.1 Boundary crossing*

Figure 8 shows how boundary crossing affects the slow slices of a simple Volterra system. The top panel shows a hypothetical slice function, which we take to be associated with the signature $VF(0,4)$. The $VF$ here designates "fast Volterra", the underlying system.  The delays 0 and 4 in the signature count samples, as do the delays in the fast slice function.  Remember that the underlying system is oblivious to our decision to slowly stimulate it.  The Volterra system in our example is taken to be the one entirely defined by the single slice.  From the signature, we see that the system has interaction order equal to 2, and on the basis of the slice function shown, we conclude the (fast) discrete memory is effectively less than 30 sample ticks long.  The two lower panels show the slow slices of this system associated with the sampling factor 10.  We shall show that the two "slow Volterra" slices, having signatures $VS(0,0)$ and $VS(0,1)$ are *all* the non-zero slow slices.  However, that *two* slow slices arise from an underlying system that is defined by only *one* slice is the essence of boundary crossing.  The delays in the slow signatures count steps.

Our means of investigating the slow slices is subpacket-correction, which we showed above relates the responses to slow packets to the slow slices.  Begin by presenting the empty slow packet to our underlying system.  Since the slow empty packet is also the underlying empty packet, and since there is no explicit $VF(\varnothing)$, meaning that the underlying bias slice is identically zero, the system response is zero, which makes $VS(\varnothing)$ identically zero as well.

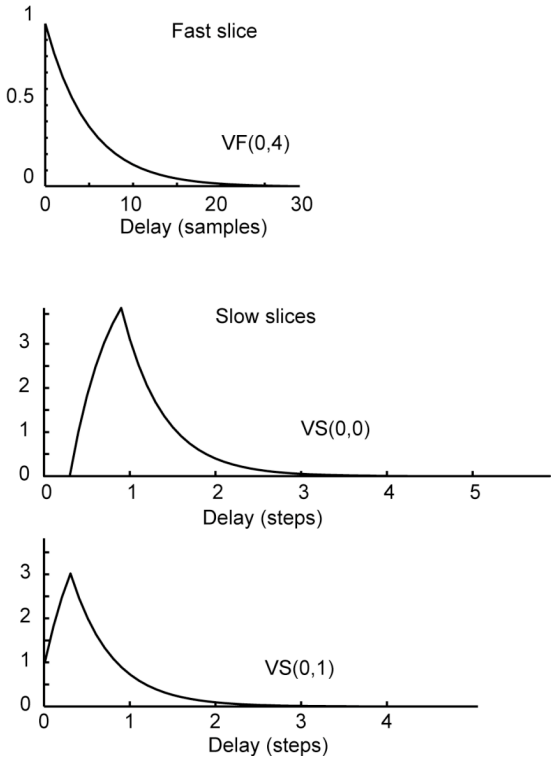

*Figure 8. The effects of boundary crossing on Volterra slices. The simple underlying (fast) system used in this illustration (top panel) is the same as in Figure 7. It is generated by a single slice having signature $VF(0,4)$ and function a decaying exponential. The lower two panels depict the two non-zero slow slices. The fact that both slice functions show features starting away from the step boundaries is a manifestation of the boundary crossing phenomenon.*

Next, present the slow single pulse having constant amplitude equal to 1. Recall from Section 2.2 that the DSNL produces only zero output until relative sample delay 4, when, for the first time, both sample delays, 0 and 4, are in the packet. When that happens, the DSNL output jumps to the value 1. It stays at value 1 for 5 *more* samples, while the signature remains contained in the 10 contiguous samples of the 1 step pulse. At the sample after that, one signature delay moves outside the packet, and the DSNL value drops to zero, where it remains. Thus, the DSNL converts the 10-sample unit pulse to a 6-sample unit pulse that arrives with the 4[th] sample following the rise of the step pulse. As shown in Section 2.1, the 6-sample pulse passes through the linear filter defined by the slice function associated with the signature, here shown in the top panel of Figure 8, to produce a contribution to the underlying system response. However, since we have supposed that $VF(0,4)$ designates the only underlying slice, our contribution is the whole response. This response is shown as the function depicted in the middle panel of Figure 8. Note how the quick rise of the fast slice in the top panel converts to a slow, saturating rise that continues while the 6 sample DSNL output pulse remains current in the filter, and which thereafter decays, disappearing completely by 30 samples after the peak. Note, too, how this response curve is located in time, beginning on the 4$^{th}$ relative sample, that is, when the original 10 sample long pulse first contained the $VF(0,4)$ delay pattern.

The discussion of the last paragraph delivered the function representing the underlying system response to the slow pulse. However, how do we conclude, as indicated by the labeling of the middle panel of Figure 8, that the response is the slice function for a single signature and that that signature is $VS(0,0)$? The answer is provided by the slow analog of Equations (10) and (11) of Section 2.2. The only slices driven by a single pulse correspond to the empty signature $VS(\emptyset)$, the linear signature $VS(0)$, and the diagonals $VS(0,0)$, $VS(0,0,0)$, etc. We distinguish among these, as we indicated previously, by passing pulses of varying amplitudes. However, since we are finding slow slices, a pulse having non-unit amplitude must have the same amplitude at each of the 10 samples. Suppose the chosen amplitude is $a$. Then the output of the DSNL is the pulse of length 6 having amplitude $a \cdot a = a^2$. On passing this through the linear filter, we find that the total system response is just what we had before, except multiplied uniformly by $a^2$. It is this that picks out the signature $VS(0,0)$, for that is the only one available that behaves exactly quadratically with input amplitude.

To pin down the other slow slice, we present a slow unit-amplitude pulse that is two steps, or equivalently, 20 samples, in duration. When this input passes through the DSNL, it comes out as a unit amplitude pulse that is 16 samples long. Fed into the linear filter, it comes out as the response shown in the top panel of Figure 9. Notice that this response has roughly the same form as the response to the single pulse, but the rising portion is longer, and so peaks closer to the asymptote before decaying. Now, according to subpacket-correction, since there is no bias slice, the difference between the system response and two appropriately placed copies of the response to the single slow pulse, which we have identified as the $VS(0,0)$ slice function, must be made up of order-2 slices that are not diagonal. The three known pieces are shown in the middle panel of the figure, and the difference is shown in the bottom panel. Note that the difference, by definition, cannot begin to rise before the second step arrives and must die out no later than a memory's worth of samples after the end of the second step.

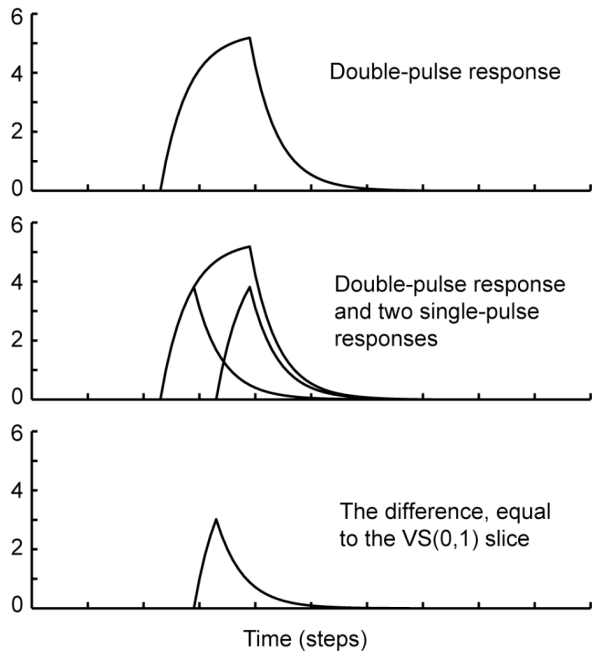

*Figure 9*. Determination of the $VS(0,1)$ slice function from the subpacket-correction property. When the bias slice is zero, the difference between the system response to the adjacent step-pair packet and the response to the steps considered as separate packets is a sum of step-amplitude multiplied slow slices having signatures 2-for-1 equivalent to $VS(0,1)$. In this example only the $VS(0,1)$ slice itself is non-zero.

Again there is the question of identifying the slow signature or signatures that are associated with this functional form. Since the pulse is 2 steps long, any such signature must have delays in two adjacent steps. The possibilities can be read from Equation (12). Since the steps are adjacent, $d_1$ must equal 1. As before, we vary the pulse amplitudes to further refine our search for the appropriate signature. Suppose the amplitude of the first step (10 samples) is $a$ and that of the second step (the next 10 samples) is $b$. From the hypothesized signature $VF(0,4)$ for the underlying system, we have that there is specific system response to samples in two different steps only when the 4-sample delay is in the (earlier arriving) amplitude $a$ step and the 0-sample delay is in the (later arriving) amplitude $b$ step. In that case, the contribution to the response is exactly $ab$ times the underlying slice function. This can happen only if the $VS(0,1)$-slice acts, and acts alone.

As shown in Fig. 8, $VS(0,0)$ is a cusp starting late in the step and $VS(0,1)$ is a cusp starting early, defined by the fact that it precisely fills in the gap needed to give the full 2-step reconstruction. Do we like this? Probably not much, since it is hard to imagine that there is any physical mechanism that directly produces the $VS(0,1)$-slice. It makes sense as a reconstruction difference, just a formal manipulation. We emphasize this example because it brings home the essentially kinematic (forced by the mathematics of the problem), as opposed to dynamic (coming from properties inherent in the physical system), nature of the Volterra slices. In the language of the Introduction, the Volterra model is quantitative and metaphorical rather than structural.

Before leaving this section, we point out the general and useful fact, exemplified here, that signature interaction order does not change under the boundary crossing distribution from fast to slow Volterra signatures.

*5.2 Flash stimulation*

There is a variant of the slow-stimulus paradigm in which the inputs are not held constant during the whole stimulus interval. Instead, the zero stimulus is displayed at all times, except possibly during the initial sample in each stimulus interval. There, either nothing changes, or a stimulus having some non-zero value is briefly flashed, and then, before the next sample is taken, replaced by the zero stimulus (see (Sutter, 2001), Figure 1). We use the term "flash stimulation" to distinguish this type of stimulation from our usual slow stimulation where inputs are held constant over each entire stimulus interval. Responses taken during a flash experiment can be processed in the slow-stimulus manner, yielding "flash slices". These slices have the property that they suffer no boundary crossing effects. This may be a considerable benefit, but, of course, flash slices may be used only to reconstruct flash stimuli, which may or may not be what you want. In any case, there is a warning here: you need to know what your stimulus is really doing in time (*cf* Keating, Parks, Malloch & Evans, (2001)). CRT's, with raster drawing and screen blanking may give one set of slices, and LCD's or LED's, which stay constant for the whole stimulus interval, quite different ones. Indeed, the implications for Volterra analysis of the complicated subframe behavior of CRT's, which shows many different momentary displays may be very difficult to understand, and the problems will probably be compounded when there is slow stimulation.

## 6. Multiple inputs

So far, we have restricted our attention to systems having a single input. Much of the interest and usefulness of non-linear kernel analysis lies, however, in its ability to handle multiple input systems, and to speak to the question of the interaction between the inputs. The vocabulary and techniques developed above, because they deal with non-linearities in general, carry over easily into the multi-input setting. Indeed, as will become clear below, the treatment of multiple inputs interacting with each other is even slightly less complicated than that of a single input interacting with itself.

For the moment, let us consider multiple-inputs in the context of full Volterra systems. The parallel subsystem description of Figure 4 expands to allow each of the inputs to feed each of the DSNLs, which become products across both delay and input. The result of the multiplication is passed to the linear filter as in the single-input case. Each DSNL is defined by a *collection* of single-input signatures, one for each input. This is the "multiple-input signature". We shall use curly bracket notation $\{\text{sig}_{input1}, \text{sig}_{input2}, \ldots\}$. Since multiplication is commutative, the ordering of the inputs is actually irrelevant, but of course, must be set once and for all in the analysis. The terms in the multi-input Volterra expansion can be gathered together in slices, and once again, for each primary signature, the characteristic function of the linear filter following the DSNL is the slice function.

In the multiple-input case, the formula for the approximate number of kernel elements is

$$\text{count}(n) \sim \frac{(c \cdot T)^n}{n!} \tag{23}$$

where $c$ is the number of inputs and $T$, as usual, is the discrete memory. The way to see this is to note that although the delays in the single-input signatures comprising the multi-input signature range from $0$ through $T-1$, the same delay value on different inputs has different meaning, and so must be counted separately.

## 7. Slice grouping and interactions among the inputs

Generally, completeness holds that the system response to any input can be *fully* reconstructed from the set of all slices. More particularly, each individual slice provides a *contribution* to reconstruction, and as we pointed out previously, and the meaning of the contribution is quite clear. In this section we develop the notion of partial reconstruction based on groups of slices, and indicate how such reconstructions can be helpful in quantifying inter-input interaction.

In case the system has two separate inputs, impulse packets are characterized by patterns of delays that can involve both inputs, and the appropriate *packet* signatures comprise pairs of single-input packet signatures, one for each input. As we saw in the previous section, 2-input slice signatures are similarly characterized. On the basis of how their signatures distribute over the two single-input parts, we may divide two-input slices into four "signature groups". The first group comprises the bias slice alone. Its signature is empty (no delays), and so its two parts are both empty. The second group comprises the slices with non-empty signatures, whose delays all lie in the input 1 part (input 2 delays empty). This is the "input 1 group". The third group, formed analogously, is the "input 2 group". Finally, there is the "mutual group", comprising slices both parts of whose signatures are non-empty.

It is easy to check that the subpacket-correction property implies that the following statements hold: The empty impulse packet requires only contribution from the bias slice to achieve full reconstruction. Impulse packets whose activity is limited to input 1 require only contributions from the bias slice and the input 1 group. Packets with activity limited to input 2 require contributions only from the bias slice and the input 2 group. Packets with activity on both inputs require contributions from all four groups for complete reconstruction.

However, the logic of subpacket-correction yields another conclusion, which starts rather trivially but gets better as it goes along: Given any impulse packet, the contribution from the bias slice to the full reconstruction is just the contribution of the empty subpacket, that is, the bias slice itself. The contribution from the input 1 group is just the same as its contribution due to the subpacket comprising its input 1 part. The input 2 group contribution is analogous. Finally, the contribution from the mutual group is precisely what's needed to complete the reconstruction after the separate effects of the empty, input 1, and input 2 subpackets are accounted for. That is, the reconstruction from the mutual group captures precisely the *interaction* between the inputs.

In Section 9.2 we illustrate slice grouping in the analysis of a simulated binary two-input linear-nonlinear (LN) system having a sigmoidal non-linearity. In Section 10, we investigate grouping in the analysis of real VEP data.

## 8. Binary analysis

The greatest parameter reduction is available for systems that either admit only binary (that is, two-level) inputs, or for which two-level inputs are adequate to drive the system to behavior interesting in its own right. For example, a visual stimulus might comprise either a patterned or a blank screen, shown in some sequence of time-dependent alternation. Restricting the entire visual system to such inputs makes it effectively a much simpler system, which, it may be hoped, isolates a significant structural or behavioral part of the whole.

### 8.1 B1m1 – the classical method

One proceeds by coding the two input possibilities in some numerical way, and feeding the numerical values into the Volterra expansion in one of the forms given above. Most often the inputs are coded as a sequence of plus and minus 1's. We described this coding in the introduction and denoted it by the term B1m1. Evaluation on B1m1 states results in the various SSPs in the Volterra expansion having values of 1 or -1. But, there is something more: SSPs with signatures differing by exactly paired delays have identical values for each input state, reflecting the fact that $1 \cdot 1 = -1 \cdot -1 = 1$. Thus, the Volterra terms containing these products can be combined by summing the kernels and multiplying the sum by a single copy of the product. We recognize this condition on the signatures as mod-2 equivalence defined in Section 4. The sum over mod-2 equivalent Volterra kernels will be called the "B1m1 kernel" of the signature class. The class signature is most compactly represented by its duplicate-free member, and its SSP is appropriate for use as the multiplier of the B1m1 kernel. The sum of products of B1m1 kernels and associated SSPs forms the "B1m1 expansion" of the binary system.

The identity of mod-2 equivalent SSPs implies identity of the slice forms of the products shown in Equation (7) that have mod-2 equivalent signatures. Explicitly,

$$\begin{aligned}
1 &= S_{(\varnothing)}(t) = S_{(0,0)}(t) = S_{(0,0,d^*_1,d^*_1)}(t) = \ldots \\
S_{(0)}(t) &= S_{(0,d^*_1,d^*_1)}(t) = S_{(0,d^*_1,d^*_1,d^*_2,d^*_2)}(t) = \ldots \\
S_{(0,d_1)}(t) &= S^*_{(0,d_1,d^*_2,d^*_2)}(t) = S^*_{(0,d_1,d^*_2,d^*_2,d^*_3,d^*_3)}(t) = \ldots \\
S_{(0,d_1,d_2)}(t) &= S^*_{(0,d_1,d_2,d^*_3,d^*_3)}(t) = S^*_{(0,d_1,d_2,d^*_3,d^*_3,d^*_4,d^*_4)}(t) = \ldots \\
&\quad etc.
\end{aligned} \qquad (24)$$

The symbol $S^*$ in (24) denotes the same function as $S$ but with the condition that the signature delays be written in order is relaxed so that all the unpaired delays are brought to the front of the list, with the paired delays following. Within each of these two groups, delay order is maintained. This allows a less fussy presentation. The paired delays, starred to make them stand out, are specified completely independently of the unstarred ones, so that the paired delays may match other paired or unpaired delays. In light of (24), the B1m1 expansion can be rearranged into slice form, which, like (5), has the advantage of emphasizing the time dependence of the response components:

$$r(t) = \mathcal{B}_{(\emptyset)} + \sum_{d=0}^{T_{(0)}-1} \mathcal{B}_{(0)}(d) S_{(0)}(t-d) + \sum_{d_1=1}^{T-1} \left( \sum_{d=0}^{T_{(0,d_1)}-1} \mathcal{B}_{(0,d_1)}(d) S_{(0,d_1)}(t-d) \right) + \sum_{d_1=1}^{T-1} \sum_{d_2=d_1+1}^{T-1} \left( \sum_{d=0}^{T_{(0,d_1,d_2)}-1} \mathcal{B}_{(0,d_1,d_2)}(d) S_{(0,d_1,d_2)}(t-d) \right) + \ldots \quad (25)$$

where

$$\mathcal{B}_{(\emptyset)} = \mathcal{L}_{(\emptyset)} + \sum_{d=0}^{T_{(0)}-1} \mathcal{L}_{(0,0)}(d) + \sum_{d^*_1=0}^{T-1} \sum_{d=0}^{T_{(0,d^*_1)}-1} \mathcal{L}_{(0,0,d^*_1,d^*_1)}(d) + \ldots$$

$$\mathcal{B}_{(0)}(d) = \mathcal{L}_{(0)}(d) + \sum_{d^*_1=0}^{T-1} \mathcal{L}_{(0,d^*_1,d^*_1)}(d) + \sum_{d^*_1=0}^{T-1} \sum_{d^*_2=d^*_1}^{T-1} \mathcal{L}_{(0,d^*_1,d^*_1,d^*_2,d^*_2)}(d) + \ldots \quad (26)$$

$$\mathcal{B}_{(0,d_1)}(d) = \mathcal{L}_{(0,d_1)}(d) + \sum_{d^*_2=0}^{T-1} \mathcal{L}^*_{(0,d_1,d^*_2,d^*_2)}(d) + \sum_{d^*_2=0}^{T-1} \sum_{d^*_3=d^*_2}^{T-1} \mathcal{L}^*_{(0,d_1,d^*_2,d^*_2,d^*_3,d^*_3)}(d) + \ldots$$

etc.

Analogous $S^*$ in (24), the symbol $\mathcal{L}^*$ in (26) denotes the slice $\mathcal{L}$ with the ordered signature delay condition relaxed. Since the signatures in (25) are duplicate-free, the sum is finite. There are no terms of order greater than $T$. In fact, if we take each delay in a binary signature to specify a particular element in a set containing $T$ distinct elements, we see that there is a unique correspondence between binary signatures and subsets of the set of $T$ elements. It is well known that there are exactly $2^T$ of these, and so there are that many B1m1 kernels.

Note that if a binary system is obtained by restriction to B1m1 inputs of a Volterra system that has no duplicate-delay terms, then the right hand sides of Equations (26) reduce to their first terms. In that case, the B1m1 kernels are exactly equal to the Volterra kernels. *Only* in this case is Equation (25) valid for non-B1m1 inputs.

Note further that in the case of multiple inputs, B1m1 signatures are obtained by applying mod-2 association on each input signature *separately*; since input sequences are in general independent of each other, identical signature delays corresponding to *different* inputs must *not* be identified.

Finally note that although conversion to the *slice* representation is valid and makes sense, the *primitive* representation shown in (8) and (9) reflects an organization of kernels based on 2-for-1 equivalence of signatures, and so is at odds with the situation here where mod-2 equivalence is natural.

B1m1 kernels can be estimated quickly and robustly and have found wide application in analysis of the visual system (Anzai, Ohzawa & Freeman, 1997, Baseler, Sutter, Klein & Carney, 1994, Benardete & Kaplan, 1997, Chen et al., 1996, de Zwart, Silva, van Gelderen, Kellman, Fukunaga, Chu, Koretsky, Frank & Duyn, 2005, Fortune, Wang, Bui, Cull, Dong & Cioffi, 2003, Gardner, Anzai, Ohzawa & Freeman, 1999, Keating, Parks, Smith & Evans, 2002, Reid, Victor & Shapley, 1997, Slotnick, Klein, Carney, Sutter & Dastmalchi, 1999, Sutter, 2001, Sutter & Tran, 1992, Tabuchi, Yokoyama, Shimogawara, Shiraki, Nagasaka & Miki, 2002).

Before proceeding to detailed considerations, we want to stress that the B1m1 coding, although thoroughly reasonable, is not canonical. Other codings are possible, and different codings lead to different slice decompositions and reductions of the general Volterra model shown in Figure 4. The coding choice has implications for interpretation.

*8.2 M-sequences*

Much of the popularity of B1m1 analysis was due to the fact, recognized early on (Golomb, 1982, Zierler, 1959) that a particular kind of binary-valued pseudo-random sequence, the m-sequence, could be used to provide periodic experimental inputs that run through all binary states representing a given discrete memory before repeating. The presentation of these $2^T$ distinct states fully tests the binary system. M-sequences may be easily constructed numerically on the fly. The usual algorithm produces a non-repeating sequence of length $2^T - 1$ that omits the all 1s state. This can be extended to length $2^T$ by inserting a single 1 at the end of the run of $T-1$ contiguous 1's. The extended sequence clearly contains all the possible memory states, and it is this sequence that we mean when we use the expression m-sequence.

The discussion of Section 4 applies to m-sequences (see Appendix III for details). In particular, the columns of the S-matrix are orthogonal. In this connection, we note that since there are no duplicate delays in any binary signature, the argument we gave in the first paragraph of Section 3.1 shows that there are no binary slices with interaction order greater than $T$. Thus, the discrete memory equals the maximum interaction order. The exponent in the length of the m-sequence, here identified with the discrete memory, is commonly called the "m-sequence order". In view of the equality of $T$ and the maximum interaction order, confusion of the two uses of the word order is not too harmful.

The foregoing means that the computational speed advantage of the projection method is available for the extraction of the B1m1 kernels. However, one can do even better. Recall that the rows of an S-matrix are labeled by input states and the columns by kernel signatures. It is a basic result of matrix algebra that the solution of an equation of the form of (18) is unaffected by reordering the rows and columns of $\mathbf{S}$ if the column vector $\mathbf{R}$ is reordered the same way as the columns of $\mathbf{S}$, and the column vector $\mathbf{L}$ is reordered the same way as the rows of $\mathbf{S}$. It turns out that, in the present context, a particular reordering is helpful in that it converts S-matrix into one that may be solved by a "self-similar" method. Self-similar methods have a computational cost on the order of $p \cdot \log p$ FLOPS, where $p$ is the number of columns, here $2^T$ (e.g. Fast Fourier Tansforms; {Press, 1988 #2031}). For large $p$, this is effectively linear in $p$, which is an important saving over the $p^2$ of the projection method. (See Appendix V for additional details).

We end this section by noting that the self-similar methods deliver the entire set of kernels based on responses to a complete set of input states. To use these methods, one must go whole hog. If one *can* do that, computation time becomes negligible compared to experiment time. But often the experiment time gets out of hand. For example, with the system hypothesized in Section 3.4 requiring a step rate of 10 ms and having a memory of 1 sec, the number of binary states in a complete set is $2^{100}$. In practice, an experiment that presents something like $2^{16}$ states pushes the limit of feasibility. Computing the corresponding full set of $2^{16} \simeq 6 \cdot 10^4$ kernels by the ordinary projection method requires on the order of $4 \cdot 10^9$ FLOPS, which translates to just a few seconds. These numbers remove some of the motivation for applying the self-similar techniques.

In addition, there is no particular reason to expect that the maximum interaction order of a real system with binary inputs is any larger than that of a real system with arbitrary inputs, so having an analysis method that quickly computes a full set of high-order kernels may give us a capability we don't need or want. We might be better off directly targeting the low-order kernels that we expect to be present. M-sequence driven projection is optimal for that purpose. If we go that route, the cost calculations look rather like those in Section 3.4. If the discrete memory and interaction order are still too great to allow an experiment, we have no recourse but slow stimulation.

*8.3 AKAT: All the kernels, all the time*

The inputs in B1m1 analysis are taken to present only the values +1 and -1. As noted above, this means that evaluations of the individual signature-state products also always result in one of these values, and by extension, the functions $S_{\text{signature}}$ shown in (24) do also. Thus, as can be seen from the slice formulation of the B1m1 expansion (25), at any time $t$, every non-zero slice $\mathcal{B}_{\text{signature}}$ enters into the response $r(t)$ with either a positive or a negative sign. This is the phenomenon that we have called AKAT, "all the kernels (or slices) all the time". In the parallel subsystem decomposition (see Figure 4) implied by the B1m1 expansion (25), the output of the individual DSNLs are simply strings of plus and minus ones. Each subsystem passes the output of its DSNL through the filter specified by its B1m1 slice function. There are fewer subsystems in the B1m1 decomposition than in the corresponding Volterra decomposition because of the mod-2 reduction indicated in (26).

AKAT makes it hard to grasp the meaning of B1m1 slices – there is always just too much going on. For example, if the constant input comprising an infinite string of +1's is presented, the output of each DSNL itself has the constant value +1, which means that the filter output is constant with value equal to the sum of the elements of the filter function. The total response is then the sum over all the individual filter outputs, again a constant, as is required by the assumption of time-independence of the system that underlies both Volterra and B1m1 analysis. For more complicated, *packet-like*, inputs comprising, say, all +1's except for a number of grouped -1's, outputs of the DSNLs remain plus except when the group passes through the memory window, during which time interval both plus and minus signs appear. Though the sign variations caused by B1m1 packets are rather like varying absence or presence of input value caused by Volterra packets, the zero values there cause no new filter response, while in the B1m1 case, the filter is always being driven, sometime by +1 and sometimes by -1.

The interpretive complication implied by AKAT just indicated is made qualitatively worse by slow stimulation, as will be illustrated by analysis of simulation data presented below.

*8.4 Mod-2 reduction and boundary crossing*

When slow stimulation is combined with B1m1 analysis, the boundary crossing effects from the slow stimulation described previously interact with mod-2 signature reduction to further complicate the situation. Figure 10 shows a very simple illustration.

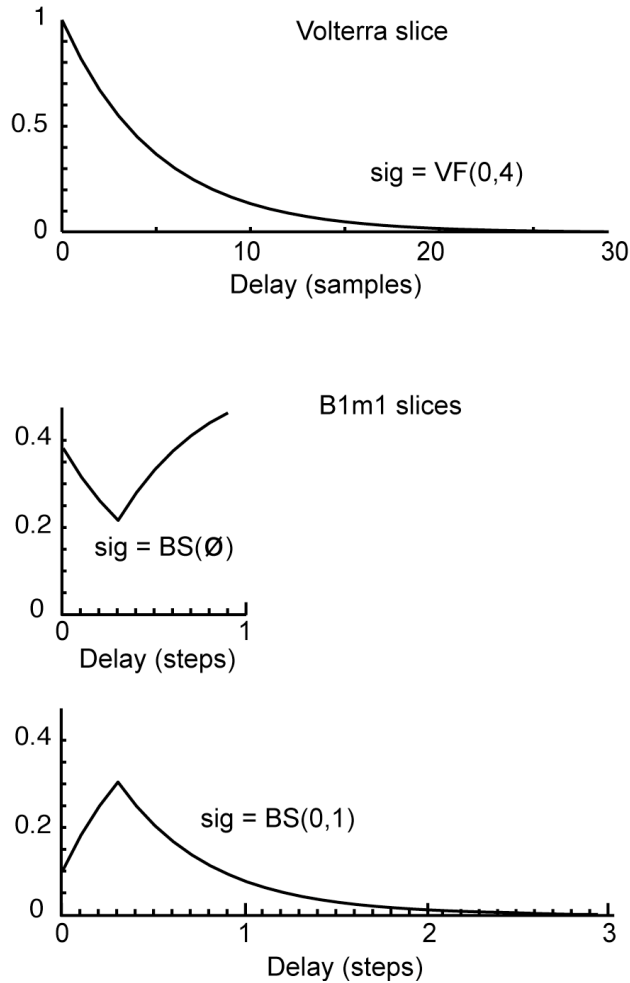

***Figure 10.*** *The non-zero slices of the binary system obtained by restricting the underlying Volterra system of Figure 8 to inputs having values plus and minus 1. The $VS(0,1)$-slices are similar in both figures, but the 2$^{nd}$ order $VS(0,0)$-slice of Figure 8 has been replaced by the 0$^{th}$ order bias slice, signature $BS(\varnothing)$, and the slice function has altered. These changes are the result of the mod-2 reduction that appears when Volterra signatures are converted to B1m1. The presence of significant features in the ubiquitous bias function complicates interpretation.*

The top panel repeats the top panel of Figure 8, and here too, it represents a single Volterra slice, signature $VF(0,4)$, that is taken to fully describe the underlying system. When this system is presented with B1m1 inputs and binary slices calculated, exactly two non-zero slices, having slow binary signatures $BS(\varnothing)$ and $BS(0,1)$, result. These are shown in the middle and bottom panels of the figure. The $BS(0,1)$ slice is precisely the same as the $VS(0,1)$ of Figure 8, but the $BS(\varnothing)$ slice shows new features. First, there is the signature itself. Binary analysis doesn't support repeated delays, and, as we saw in Section 8.1, Volterra slices having such delays contribute to B1m1 slices having mod-2 reduced signatures. The first of Equations (26) shows the reduction, $VS(0,0)$ to $BS(\varnothing)$, explicitly. Next, there are changes in the form of the slice function. One is quite radical, but forced by the definitions. The length of the $VS(0,0)$ slice, as shown in Figure 8, may be up to the full memory's worth of steps since it comprises the slow kernels having (non-primary) signatures $VS(0,0)$, $VS(1,1)$, $VS(2,2)$, etc. On the other hand, the slice corresponding to the empty signature is always just one step long. This is because the fast empty slice is just the single value $L_0$, and the interleaving process of slow analysis adds one value per relative sample. Thus, the long $VS(0,0)$ slice function must wrap around to produce the short $BS(\varnothing)$ slice function. This wrapping is apparent from the summation in the second term of the first equation in (26) together with the interleaving process of slow analysis. Note that the rise of the $BS(\varnothing)$ function at sample 4, which is characteristic of late onset typically produced by boundary crossing, is preceded by the period of decay that would be implied by the wrapping around of the tail of the long slice.

      Our example shows that, unlike the situation in Volterra analysis, boundary crossing in B1m1 may imply interaction order reduction. It also shows that in B1m1, the $BS(\varnothing)$ slice has complex *intra-step* structure, quite different from the $VS(\varnothing)$ slice, which is necessarily constant across its whole width. The non-constant nature of the $BS(\varnothing)$ slice is confusing if one forgets AKAT and tries to think, as is true in Volterra, of the empty-signature slice as the response to a constant input. Such inputs contribute to the $BS(\varnothing)$, but so do pulse inputs acting, as here, through $VS(0,0)$ and other diagonal slices. These higher-order slices know the fact of the input transitions occurring only at step boundaries, and so can be, and are, time locked to the steps, and thus capable of showing non-constant relative structure.

*8.5 Response reconstruction from B1m1 slices*

      Since B1m1 slices are complete, system responses can be reconstructed from the slices, but unlike the situation for Volterra slices, partial reconstructions from the various slices are not easily interpreted. As previously noted (Sutter, 2000), the

relationship between B1m1 slices and system response waveforms is *indirect*. Figure 11 depicts two response reconstructions of the underlying Volterra system we have been studying. The reconstructions use the slices shown in Figure 10. The left column of plots show slice contributions to the reconstruction of the response to constant 1's input, and the right column to that of a single -1 pulse. The 2$^{nd}$ row of plots shows the system responses, and the last row, the reconstruction from the full set of non-zero slices (of which there are only two in this example). Comparing these rows, we see that the reconstructions are correct. This is not surprising, since correct reconstruction is implied by completeness in an analysis scheme: Kernels are, by definition, things that when added up give back what you started with.

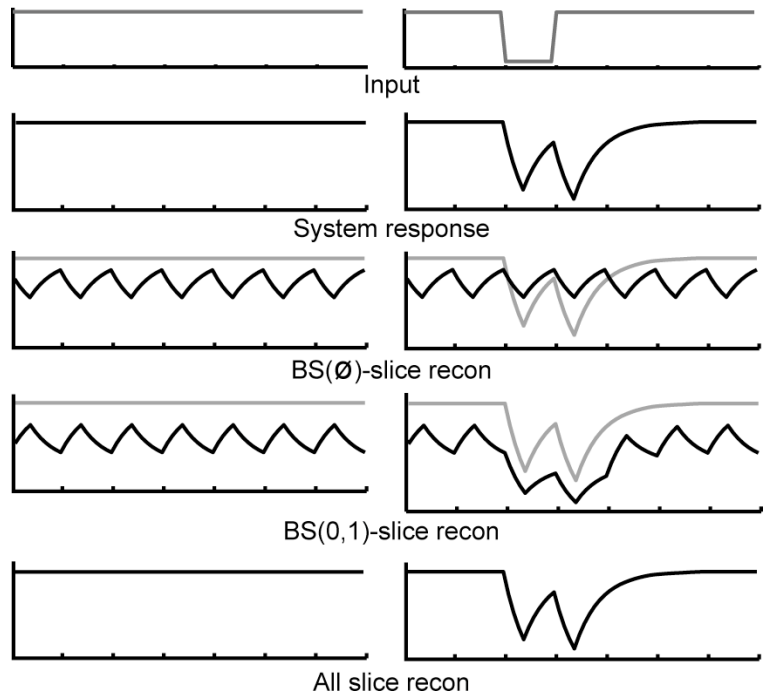

*Figure 11. Reconstructions based on the B1m1 slices of Figure 10. The left column pertains to the constant 1 stimulus, and the right column to a down-going pulse. The first row shows the inputs. A comparison of rows 2 and 5 shows that the full reconstructions recover the system response. Rows 3 and 4 illustrate how the complicated partial reconstructions cancel to produce the much simpler system responses. The gray traces in these rows are copies of the full system responses, shown to make comparison easier.*

Partial reconstructions, representing the contribution of the individual non-zero slices are shown in the 3$^{rd}$ and 4$^{th}$ rows. The gray traces in these plots are copies of the system responses, shown to make comparison with the partials easier. It is striking that the highly complicated partial reconstructions, clearly keyed to the stimulation step, cancel in just the right way to produce the much simpler system responses. The $BS(\emptyset)$-partial is just the periodic concatenation of copies of the $BS(\emptyset)$-slice shown in Figure 10. Since this partial reconstruction must be a part of every full reconstruction, other partial reconstructions, themselves complex, must be available to counterbalance the $BS(\emptyset)$-partial. In our example, this balance comes from the $BS(0,1)$-partial, representing the only other non-zero B1m1 slice.

This last observation applied to the left column of Figure 11 shows that B1m1 slices do not satisfy the subpacket-correction property that we found helpful in interpreting Volterra slices. But we knew that: AKAT and subpacket-correction are antithetical. Figure 11 shows only the tip of the AKAT iceberg. Partials formed from both non-zero slices must be added to get the full reconstruction for any input (Gerth, Sutter & Werner, 2003, Keating et al., 2002, Sutter, 2000).

*8.6 Multiple inputs*

We saw in the single input example of the previous two sections that mod-2 reduction led to incorporation of an order-2 Volterra slice into the B1m1 bias term, giving it structure that is difficult to interpret. B1m1 analysis of multiple input systems can yield results that are even more problematical. For example, suppose we have a 2-input Volterra system with a single slice with signature $VU\{(0),(0,0)\}$ (see Section 6 for notation). Then the assumption of B1m1 inputs and the resulting mod-2 reduction puts everything into the binary slice with signature $BS\{0,\emptyset\}$, which one might take to represent *linear* system behavior associated with input 1 alone. All other binary slices are identically zero. Looking at the slices, one would have no inkling of the interaction between the inputs of the underlying Volterra system. We shall return to this later in Section 9 in our discussion of B01 analysis.

As in the single-input case, multiple-input binary systems can be probed using m-sequence derived stimulus sequences for the inputs. If the goal is that the system be completely tested, that is, each possible input state be shown, efficient presentation takes some care. The detailed procedure explained in Appendix IV.2 is guaranteed to work.

Finally, we repeat the warning of Section 8.2 that it is often not feasible to do a complete experiment. The experiment duration, already frequently marginal with a single input, is much worse here since the duration goes up exponentially with the number of inputs. Our first recourse, as before, is to focus on a limited set of kernels which we expect to include all the non-zero ones, choose inputs that make the S-matrix columns corresponding to these kernels orthogonal, and analyze by projection.

## 9. B01 – A more interpretable form of binary analysis

As we have shown above, the relationship between B1m1 slices and response waveforms tends to be obscure. This is primarily because of the complicated sign alternation associated with a slice's contribution to response reconstruction. At every timestep, the slice must be added in its entirety, either upright or sign-reversed. This is hard enough to comprehend for a single slice, but when there are multiple non-zero slices, each has its own pattern of sign alternations, and the combinations quickly become unfathomable. The difficulty is compounded by AKAT, which requires that *every* non-zero kernel be used in every reconstruction.

In contrast, there is another form of binary analysis, B01, mentioned in the introduction and reflecting a different, but closely related system model, that ties slices to response waveforms in the most intimate possible way; that is, via the strong subpacket-correction property that we examined in the discussion of Volterra systems without duplicate-delay terms (Section 3.1). This alternate analysis was initially suggested by (Yokota & Usui, 1999) but has not previously been applied. We describe it in detail and explore its features.

As in the B1m1 analysis of Section 8.1, we assume a system having two-level inputs, but in doing B01 analysis we encode the levels differently, using 0's and 1's, instead of 1's and -1's. As before, we feed the coded input values into a Volterra expansion. SSPs evaluated on B01 states produce B01 results.

Since the product of any number of 1's is 1 and the product of any number of 0's is 0, it is clear that for any B01 state, the value of the SSP for any given Volterra signature is equal to the value of the SSP for the state and the duplicate-free Volterra signature that is 2-for-1 equivalent to the given signature. (2-for-1 equivalence was defined in Section 2.1.) This allows Volterra slices having 2-for-1 equivalent signatures to be summed and renamed, converting the Volterra expansion into a B01 expansion. This is analogous to what we did in Section 8.1 to form the B1m1 expansion. Equations (27) show the slice forms of the products that have identical B01 values.

$$1 = S_{(\varnothing)}(t)$$
$$S_{(0)}(t) = S_{(0,0)}(t) = S_{(0,0,0)}(t) = \ldots$$
$$S_{(0,d_1)}(t) = S_{(0,0,d_1)}(t) = S_{(0,d_1,d_1)}(t) \tag{27}$$
$$= S_{(0,0,d_1,d_1)}(t) = S_{(0,0,0,d_1)}(t) = S_{(0,d_1,d_1,d_1)}(t) = \ldots$$
$$etc.$$

The B01 slice expansion and its Volterra summands are shown in (28) and (29) below.

$$r(t) = \mathcal{H}_{(\varnothing)} + \sum_{d=0}^{T_{(0)}-1} \mathcal{H}_{(0)}(d) S_{(0)}(t-d) + \sum_{d_1=1}^{T-1} \left( \sum_{d=0}^{T_{(0,d_1)}-1} \mathcal{H}_{(0,d_1)}(d) S_{(0,d_1)}(t-d) \right) + \\ \sum_{d_1=1}^{T-1} \sum_{d_2=d_1+1}^{T-1} \left( \sum_{d=0}^{T_{(0,d_1,d_2)}-1} \mathcal{H}_{(0,d_1,d_2)}(d) S_{(0,d_1,d_2)}(t-d) \right) + \ldots \tag{28}$$

where

$$\mathcal{H}_{(\varnothing)} = \mathcal{L}_{(\varnothing)}$$
$$\mathcal{H}_{(0)}(d) = \mathcal{L}_{(0)}(d) + \mathcal{L}_{(0,0)}(d) + \mathcal{L}_{(0,0,0)}(d) + \ldots$$
$$\mathcal{H}_{(0,d_1)}(d) = \mathcal{L}_{(0,d_1)}(d) + \mathcal{L}_{(0,0,d_1)}(d) + \mathcal{L}_{(0,d_1,d_1)}(d) + \tag{29}$$
$$\mathcal{L}_{(0,0,d_1,d_1)}(d) + \mathcal{L}_{(0,0,0,d_1)}(d) + \mathcal{L}_{(0,d_1,d_1,d_1)}(d) + \ldots$$
$$etc.$$

Note that (28) is just (25) with a change of notation; $\mathcal{B}$ goes to $\mathcal{H}$. The input functions $S$ in (27) are the same as those in (24), but the groupings by equal value are different. This reflects the fact that the values, determined by the input factors $s$ defined in (7), have B01 values here, but B1m1 values in (24). Conversion between codes leads to a conversion between values of the products $S$, which can be parlayed into formulas expressing the B01 slices $\mathcal{H}$ in terms of the B1m1 slices $\mathcal{B}$, and vice versa. This is carried out in detail in (Yokota & Usui, 1999). For consistency when converting between B01 and B1m1, we shall always use the convention of Yokota and Usui that 1 goes to 0 and -1 goes to 1.

In the example of Section 3.1 we anticipated our present interest in B01 analysis. We showed that B01 slices satisfy the strict subpacket-correction property and that an S-matrix for computing B01 slices may be built on an impulse-packet based experiment.

An m-sequence experiment can also be used to build an invertible B01 S-matrix. This is because the complete set of B1m1 states delivered by an m-sequence becomes, on application of a consistent conversion convention, a complete set of B01 states. But the impulse experiment states are also complete. Thus, the m-sequence generated B01

S-matrix can be obtained from the impulse-packet S-matrix by row permutation, an operation that leaves column independence unchanged. Thus, for binary systems whose inputs are not inherently numerical, we can, without prejudice, run an m-sequence experiment and subsequently process it with either B01 or B1m1 coding.

It turns out that the ordering of states and signatures described Section 8.2 that converted the B1m1 S-matrix to a form amenable to solution by self-similar methods, works to do the same for the B01 S-matrix. Thus, full-kernel $p \cdot \log p$ FLOPS computations are available in B01 too.

*9.1 Reconstruction in B01 and B1m1*

The strong subpacket-correction property enjoyed by B01 makes reconstruction of impulse-packet responses from slices straightforward and easy to understand. In contrast, reconstruction from B1m1 slices must satisfy AKAT, which makes the procedure complex. In the B1m1 case, there is typically much partial cancellation among slices, making assignment of meaning to the individual slices difficult.

Figure 12 shows a comparison of non-slow-stimulus B01 and B1m1 reconstructions of the response of a simulated Volterra system to a two-pulse input packet having signature (0,2). In all the traces, the future is to the right, and data is localized at the timesteps; the connecting lines are linear interpolations. The left-hand column shows the reconstruction procedure for B01. The top plot displays the input pulse pair and the full reconstruction. Because of the completeness of the slices, the full reconstruction equals the Volterra response. We begin the reconstruction by adding the bias slice, which is one step long (thus constant) at every timestep. This is shown in the second plot. Next, at each of the two timesteps at which there is an input pulse, we add a copy of the linear slice, with more delayed, later arriving effects depicted to the right. The length of the linear slice is always no more than the system memory. These two additional contributions are shown in the $3^{rd}$ and $4^{th}$ plots. Finally, starting at the later-arriving pulse, we add the order-2 slice whose signature delay separation matches that of the input pair. This is shown in the $5^{th}$ plot. We have plotted one appropriately delayed slice copy for each sub-packet of the input packet, where both the empty packet at each time step, and the full packet are considered subpackets. The sum of these slice copies is the full reconstruction. Note that except for the bias slice, the B01 reconstruction contributions are identically zero before the pulse packet arrives and after it departs memory.

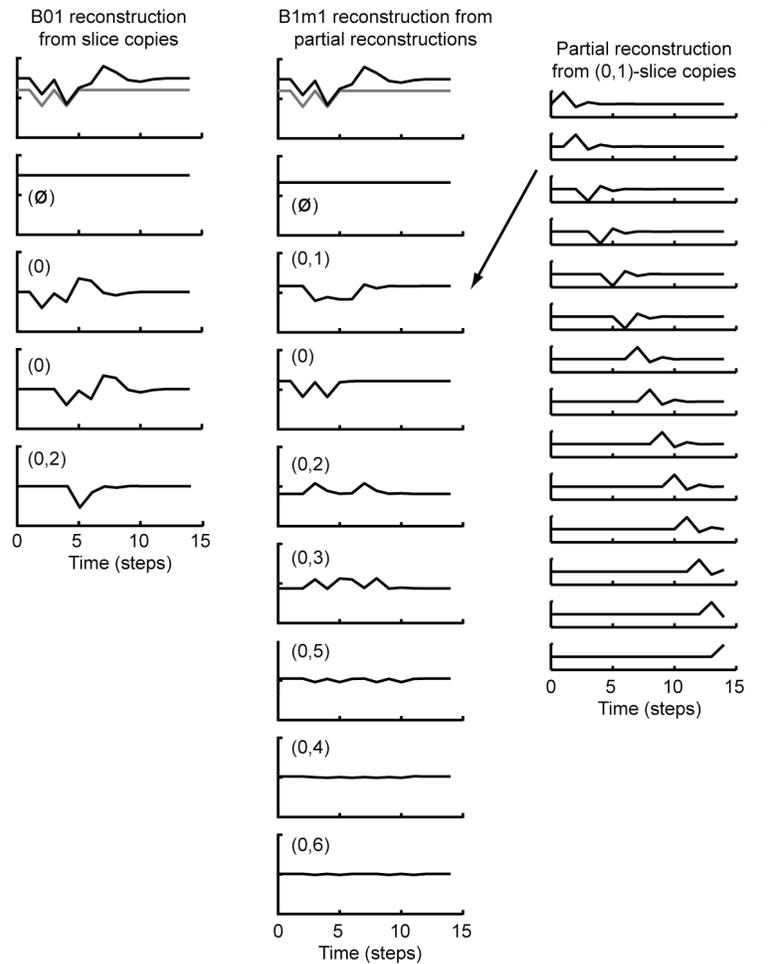

*Figure 12. Comparison of B01 and B1m1 contributions to reconstruction. The response of a binary system to an impulse packet can be written as a sum of suitably delayed slice contributions. In B01 (left-hand column), only slices whose signatures are "contained" in the impulse pattern are needed, and they are all summed with a plus sign to make the total response (subpacket-correction). The delayed slices are shown in the lower traces. They add up to the total response (black curve, top panel). This behavior is "strict subpacket-correction". B1m1 analysis (middle and right-hand columns) does not support such a simple scheme. Although the response to an impulse packet can be uniquely written in terms of B1m1 kernel slices, there is a thorough mixing in which every non-zero kernel contributes to the reconstruction (AKAT). As in the left-hand column, the total response in the middle column is a sum over lower traces, but these are not themselves B1m1 slices. Rather, each is a sum, with offsets and specific, complicated, sign changes, of a single B1m1 slice as illustrated in the right column, which shows the summing procedure for the trace in the middle panel pointed to by the diagonal arrow. The vertical scales are the same in all three columns.*

In the case of a more general input packet, we add an order-2 slice copy for each pulse pair contained in the input packet, an order-3 slice copy for each pulse triple, and so forth. In matching higher order signatures to subpackets, early arriving pulses have lower delays. Each slice is added with its start coinciding with the last arriving pulse in its subpacket.

The middle and right-hand columns of Figure 12 depict the components of the B1m1 reconstruction. The middle column shows partial reconstructions, each obtained by summing copies of a single slice, the copies being added with varying signs and starting points. The right-hand column shows the signed slice copies that combine to form the partial reconstruction shown in the 3$^{rd}$ plot of the middle column. The sign used for each slice copy is the SSP value corresponding to the slice signature and the input state whose most recent entry corresponds to the copy location. Note the general property that states before the pulse packet arrives and states after the packet goes out of discrete memory all contribute the same sign to the partial reconstruction. This makes the partial reconstructions constant (but typically not zero) in these time regions.

In case of slow stimulation, the previous discussion remains valid, except that each slice is added with its start at the *earliest* sample (that is, left-most) in the stimulus interval corresponding to the completion of its signature.

*9.2 Slice grouping in multiple input B01 and B1m1 analyses*

As we discussed in Section 7, we may divide two-input slices into four signature groups --- bias, input 1, input 2 and mutual. Here we show that for B01 slices, the group-based partial reconstructions of the system response are well-behaved and interpretable, but similar reconstructions from B1m1 slices are not. This difference in interpretability of the two kernel approaches strongly affects their utility. We illustrate group-based partial reconstructions for a simple two-input system comprised of a linear filter on each input followed by a parameterized sigmoidal non-linearity of the form $(f_1 + f_2)^2 / ((f_1 + f_2)^2 + k^2)$, where $f_j$ is the output of the filter on input $j$. A block diagram of the system is shown in Figure 13 and example group reconstructions are shown in Figures 14 and 15. Slow-stimulation was used to drive the system. Input values were restricted to 0 and 1. Input 1 was driven by a pulse two steps long and input 2 was driven by a one step long pulse commencing with the beginning of the second step of the input 1 pulse (see upper left panels of Figures 14 and 15). The solid trace represents input 1, and the dashed trace input 2. The gray trace, repeated in each of the panels of the figures is the simulated system response.

In the first example (shown in Figure 14), the sigmoid parameter k was chosen equal to 10.0, which put the input values in the range where the non-linearity is expansive. The relevant section of the non-linearity is shown at the top-right of Figure 14. The system response starts increasing with the appearance of the step on input 1 and then suddenly increases further with the advent of the second step, when both inputs become active. After the offset of both pulses, the response decays.

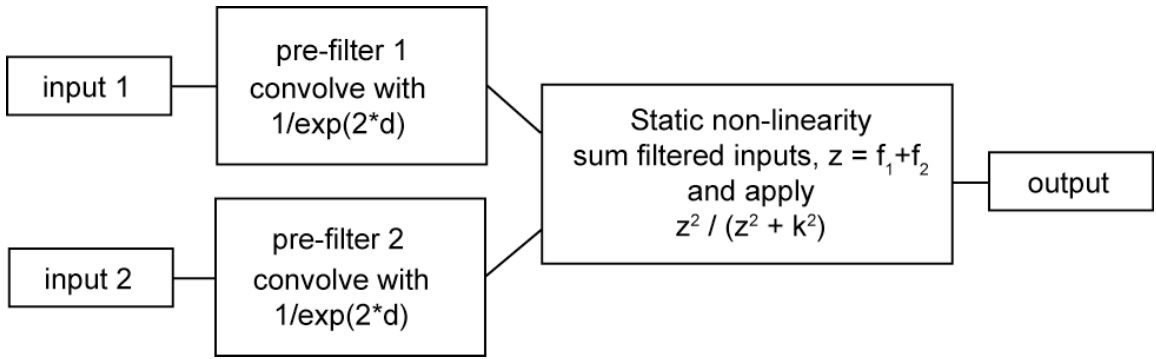

*Figure 13. Block diagram of LN system with two inputs. The general form of a 2-input linear-nonlinear (LN) system, with specific example values of the components is shown. The static non-linearity provides the interaction between channels, depicted here summation of the filtered inputs and application of a sigmoidal non-linearity. The result of the interaction is the system output. The parameter k in the non-linearity can be adjusted to emulate an expansive or compressive system for a given set of input levels.*

In the B01 case (left column of Figure 14), the bias contribution is identically zero. The input 1 group contribution rises smoothly during the 2-step long input 1 pulse presentation and thereafter decays. The input 2 group contribution is similar, rising during the single input 2 pulse, and thereafter decaying. Being driven for less time, it rises less than the input 1 contribution. The mutual group contribution begins at the onset of the second input pulse, and continues to increase as long as both inputs are active. The full system response is simply the sum of the group contributions. The decomposition makes sense for an expansive non-linearity: responses to two inputs considered separately are too small to explain the response when both are present. Something more must be added, and this is reflected in the mutual group, which adds with the same sign as the two single input contributions.

In the B1m1 case (right column of Figure 14), the bias contribution is not constant. Rather, it shows intra-step structure. Necessarily, from the definitions, the structure is periodic with period equal to the step. Since the system being simulated clearly has no response to the constant zero stimulus, the intra-step structure must be the result of mod-2 reduction of higher interaction order effects, ones that can be time-locked to the step onsets. The input 1 and input 2 group contributions are similar to those for B01, except that they display step-periodic structure both before and after the pulse presentations. That structure effectively cancels the periodic structure in the bias contribution, allowing the sum of all the contributions, which must equal to the full system response, to be zero away from the stimulus event. The mutual group contribution is more problematic. It goes negative with the beginning of the pulse on input 1 even though input 2 is not yet active. This upsets causality and defies simple interpretation. Beyond this, it is counterintuitive that the leading edge of the mutual contribution should be negative going rather than positive going. Finally, the mutual group contribution is not zero away from the input pulses.

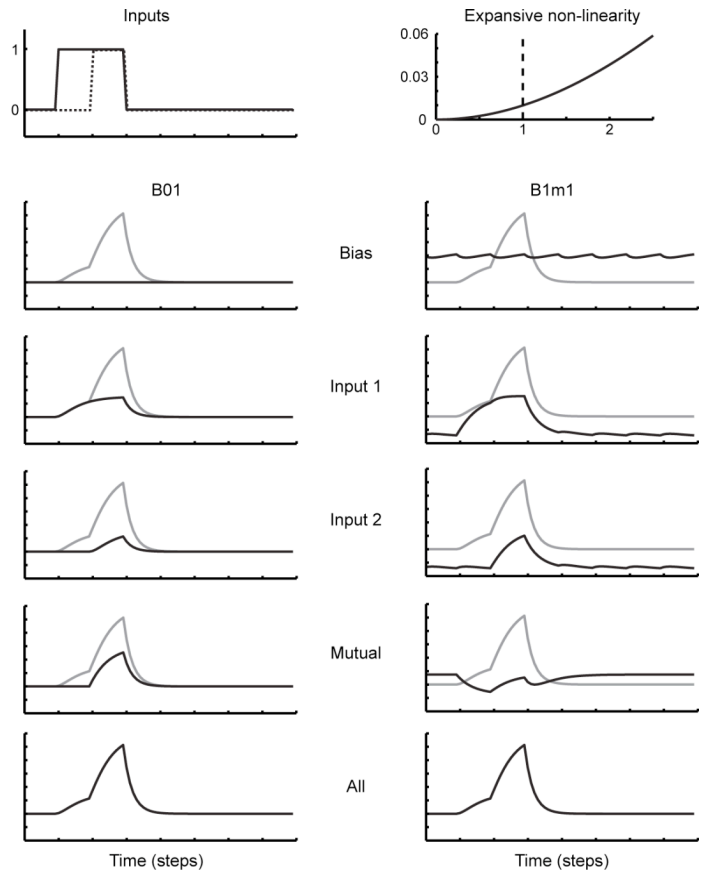

*Figure 14.* Comparison of group reconstructions for the system of Figure 13 with k set to emulate an expansive non-linearity over the range of inputs. The operative part of the sigmoid is shown in the top-right panel. The input, shown in the top-left panel, consists of a two-step long pulse on input 1 (solid black) and a one-step long pulse on input 2 (dotted black) that coincides with the second step of the input 1 pulse. The reconstructions are shown as black traces and the full, simulated system response is shown in gray, repeated in each panel to make comparison with the group reconstructions easier. The B01 group reconstructions are shown in the left column. They behave as one would expect: the input 1 and input 2 reconstructions are similar, starting to rise with their respective pulses, but the reconstruction for input 1 is longer and higher, reflecting the longer stimulation; the mutual reconstruction begins rising with the advent of the pulse on input 2 when both inputs become active. The sign of the mutual reconstruction agrees with those of the two separate input reconstructions, reflecting the expansive nature of the non-linearity. In contrast, the B1m1 group reconstructions shown in the right column are much harder to interpret. Most difficult is the fact that the mutual reconstruction begins to go active at the start of the input 1 pulse, one whole step before input 2 does anything.

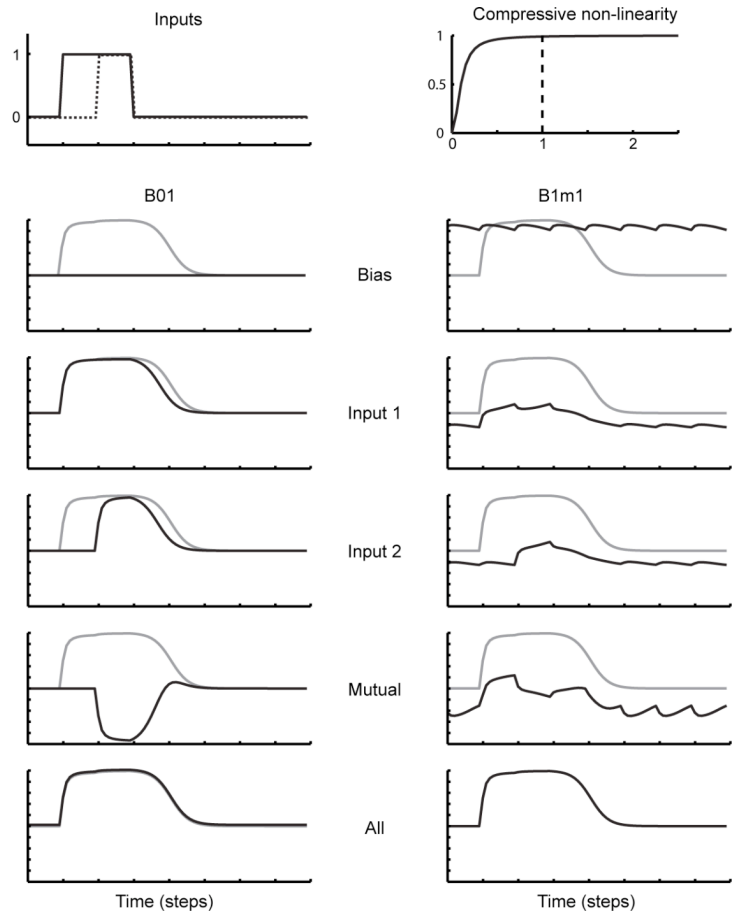

*Figure 15.* *The same system as in Figure 14, but with k set to emulate a compressive non-linearity. Again the B01 group reconstructions conform to our expectations. In particular, the mutual reconstruction is negative reflecting the saturation of the non-linearity in the region of activity.*

In the second example (shown in Figure 15), the sigmoid parameter k was chosen equal to 0.1, which put the top input value in the range where the non-linearity is compressive. The relevant section of the non-linearity is shown at the top-right of Figure 15. Now the system response is quite different: It starts increasing abruptly with the appearance of the step on input 1, then increases very gradually, even after the advent of the second step, when both inputs become active. There is a barely detectable change in the system response at the second step, in contrast to what was observed when the system was driven over its expansive range. After the offset of both pulses, the response rounds off and then decays.

In the B01 case, the bias is zero, the input 1 contribution is similar to the total response, and the input 2 contribution looks like a delayed and shortened version of the input 1 contribution. Input 1 and 2 contributions are each of similar magnitude to that of the total response. Therefore, it is not surprising that the mutual contribution is negative going and roughly of the same size, because the compression puts a limit on the magnitude of the system response for large inputs.

In the B1m1 case, the bias term is of large magnitude and shows intra-step structure. The input 1 and 2 contributions are small and highly structured. The mutual contribution again begins before the onset of the second step. It shows strong structure away from the inputs. It is not possible to discern the inherent saturating behavior of this compressive system from the contributions.

While this illustration shows the general fact that B01 analysis is more interpretable than B1m1 analysis, it also clearly illustrates the limitation of binary testing as a complete description of the system over its full operating range. The slice groups derived with one set of input values differ greatly from those derived with a different set of values. The effective change of input values was accomplished here by varying k, but in a real experiment, the system would remain fixed while the levels of stimulus intensity would vary.

## 10. Analysis of a two-input Visual Evoked Potential experiment

In this section, using data from a human Visual Evoked Potential (VEP) experiment, we illustrate the main features of binary kernel analysis that we have thus far demonstrated with simulations: reconstruction of system responses from slices and partial reconstruction from self and mutual slice groups. In the process, we see clear examples of the effects of slow stimulation and mod-2 reduction on the B1m1 slices. Details of the experimental set-up are provided in Appendix VI.

The specific illustrations come from a texture-segmentation experiment in which the observer viewed a two-part display consisting of a background texture, upon which appeared a set of "figures", 9 disks that segmented from the background either on the basis of a difference in orientation or in relative alignment of their textures. Both the figures and background were switched between horizontal and vertical texture orientations according to a two-input sampling of an m-sequence of order 8. The four states of the stimulus are shown in Figure 16. The figure region is considered to be input 1 and the background region input 2. For both inputs, the horizontal orientation is coded as the 0 input state for B01 analysis, and the 1 state for the B1m1 analysis. The vertical orientation is coded as the 1 state in B01 and the -1 state in B1m1. The details of the experimental set-up are presented in Appendix VI.

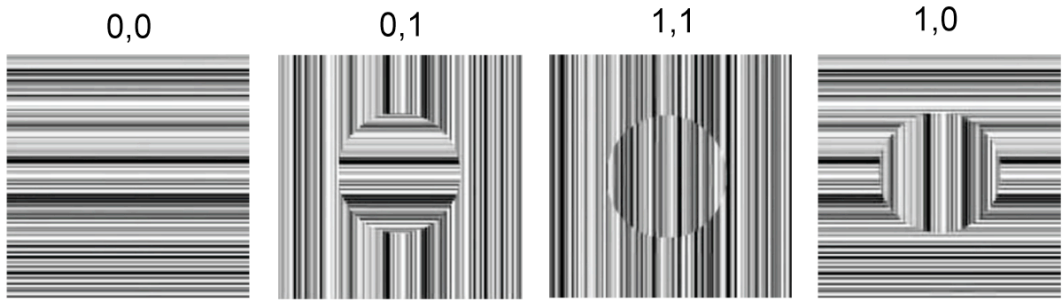

*Figure 16.* Stimulus states for two-input VEP experiment. One of the nine repeating sections of the display is illustrated. The figure region is considered to be input 1 and the background region input 2. For both inputs, the horizontal orientation is coded as the 0 input state for B01 analysis, and the 1 state for the B1m1 analysis. The vertical orientation is correspondingly coded as the 1 input state for B01 analysis and −1 for B1m1 analysis. The figure and background regions rotated in opposite directions when switching states.

In this experiment, we expected to observe responses arising from the background region, from the figure regions, and from non-linear interactions between them. Based on previous work with frequency-tagged stimuli (Norcia, Wesemann & Manny, 1999, Zemon & Ratliff, 1982, Zemon & Ratliff, 1984), we expected strong non-linear interaction when the figure and background regions were abutting and that the interaction could be sharply reduced by the introduction of a gap between regions. A similar proof-of-concept experiment using m-sequences has been presented by (Zhang, 2003).

Separate recordings were made with and without the gap present. For each recording, both B1m1 and B01 kernels were derived. These were then used to reconstruct the response to 1 Hz periodic alternations between two different pairings of the four states that had been presented during the m-sequence stimulation. The response to these periodic alternations was also measured directly in separate VEP experiments.

We first look at the reconstruction of the response when only the figure region is active (switched between horizontal and vertical) while the background region is fixed at horizontal. Figure 17 plots the group reconstructions (Bias, Inp 1, Inp 2 and Mutual; see section 7). The B01 reconstruction of this response (Figure 17, red curve) consists of the summation of the slices called for by the strong subpacket-correction property, here the $\{(\emptyset),(\emptyset)\}$-slice, the $\{(0),(\emptyset)\}$-slice, and the $\{(0,1),(\emptyset)\}$-slice (see Section 6 for notation). The input-2 parts of the signatures are empty since that input was held constant at value 0; therefore all these slices belong to the input-1 group. Thus the input-1 group reconstruction is the only one that is non-zero and nothing is needed from the other groups. The reconstruction of the response to the alternation of the two figure states (Figure 17, red curve, bottom row) is quite similar to the measured periodic response (Figure 17, blue, repeated in all rows). In contrast, the B1m1 group reconstructions *all* have marked temporal structure (Figure 17, right panels) and all of them are needed to reconstruct the response (bottom right panel; All). This was expected from the AKAT property of B1m1. Note that the bias term is large and has periodic structure. We know that the visual system should not have a strongly modulated response to a constant input. This becomes comprehensible if we recall that in Section 8.4 we saw that, because of mod-2 reduction, the bias term can be strongly structured. The structure reflects diagonal Volterra activity that is projected onto the B1m1 bias term. Similarly, there is no temporal variation on input 2, but again there is periodic structure in the input-2 group reconstruction. In this case the structure is due to a projection from mutual group Volterra elements.

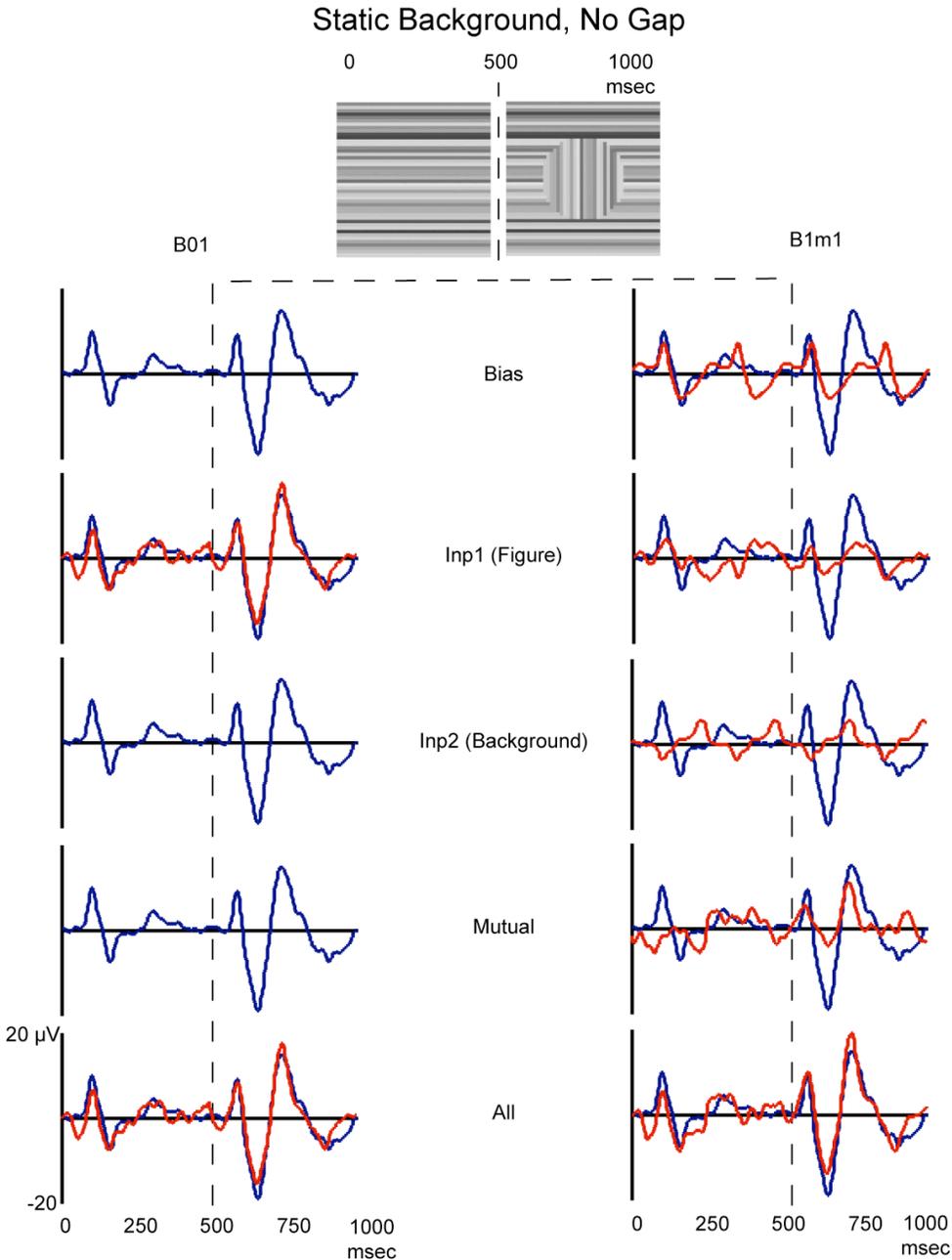

*Figure 17. Reconstruction of single input VEP responses. The icons at the top indicate the states of the periodic stimulus whose response (blue traces) was reconstructed (red traces) from either B01 (left panels) or B1m1 (right panels) slice groups. The B01 reconstruction of this response consists of the summation of the slices called for by the strong subpacket-correction property. The input-1 group reconstruction is the only one that is non-zero and nothing is needed from the other groups. The reconstruction of the response to the alternation of the two figure states (red curve, left panels) is quite similar to the measured periodic response (blue curve; left panels). In contrast, the B1m1 group reconstructions all have marked temporal structure (red curves, right panels) and all of them are needed to reconstruct the response (blue curve, bottom right panel; All). This was expected from the AKAT property of B1m1. Note also that the bias term is large and has periodic structure. Nonetheless, the full B1m1 reconstruction is accurate.*

     In the next example, we reconstructed periodic responses when both the figure and background regions were actively changing their orientation (see Figure 18). The periodic stimulus in this case has different properties: the segmentation of the figure and background occurs as a result of a difference in texture *alignment* of the two regions; in the previous case the segmentation resulted from a difference in *orientation*.

     The B01 input-1 group reconstructions of Figures 17 and 18 provide a useful example of the working of subpacket-correction. In general, the slices comprising the input-1 group are driven only by input packets that cause variation of input 1 while leaving input 2 fixed at horizontal. Thus, only these packets contribute to the input-1 group reconstruction. A consequence of this is that the input-1 group reconstruction reflects only the time variation of input 1. It doesn't matter what input 2 does. Since the input 1 variation is identical in the inputs of Figures 17 and 18, both input-1 group reconstructions should be equal, and they are. Of course, the same reasoning applies to the input-2 group reconstructions, but the input 2 time variations are different in the figures, so we would not expect input-2 group reconstructions to be the same.

     Comparing the input-1 and input-2 group reconstructions of Figure 18, we see that they are quite similar to one another. This was not at all necessary even though the input 1 and input 2 time variations are identical, since the group reconstructions depend

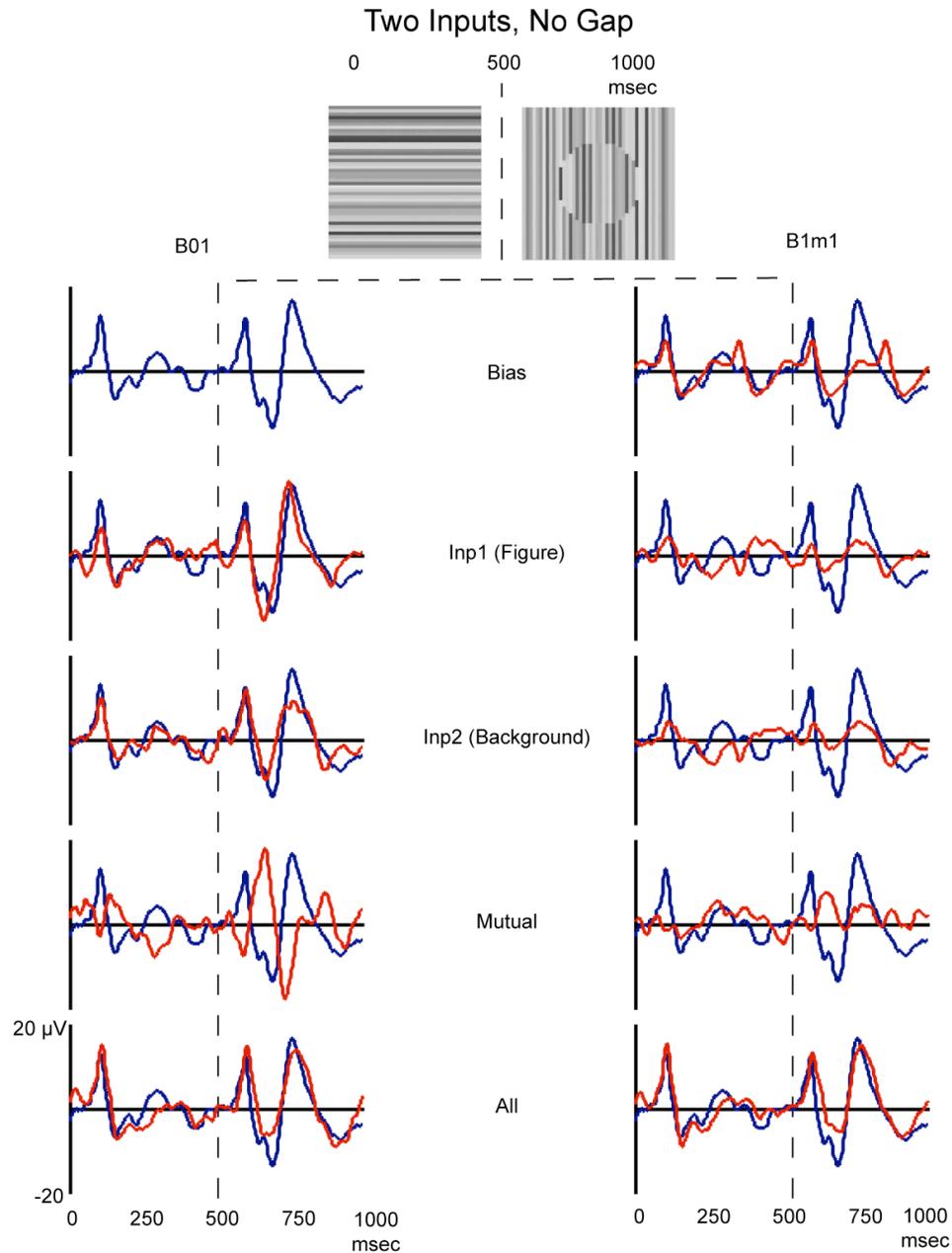

*Figure 18.* Reconstruction of texture-segmentation evoked responses, both inputs active, No Gap. Plotting convention as in Figure 17. The input 1 partial reconstructions are the same as those from Figure 17, but the input 2 and mutual reconstructions differ. The full reconstructions (red curves, All) are the same for B01 and B1m1 methods and are similar to the measured periodic response (blue curve). Note, however the very different structures of the group reconstructions of the two methods. In particular, the mutual group reconstruction is much larger for the B01 method.

on the slices as well as on the inputs. The similarity of the reconstructions might indicate similar brain dynamics have led to similar input 1 and input 2 slices.

Interestingly, the two self-group reconstructions also look somewhat similar to the measured response. A notable feature of the self-group reconstructions is the peak at 270 msec after the transition to the segmented state that is absent at the same relative time after the transition to the uniform state at time zero. This asymmetry is also present in the measured response (blue). However, the sum of the two self-group reconstructions is much too large to predict the measured response. Here is where the mutual group comes in. It is largely in anti-phase with the activity in both halves of the individual input-group reconstructions. Adding all of the group reconstructions together yields a good *quantitative* estimate of the measured response.

Turning to the B1m1 analysis of the no-gap data, we see a number of differences in the group reconstructions. First, there is a substantial bias reconstruction that has strong, time-varying structure. Second, the amplitudes of each of the figure, background and mutual group reconstructions are smaller than in the B01 case and are not obviously different between the first and second halves of the stimulus. Considered individually, none of the B1m1 partial reconstructions are interpretable in terms of the stimulus events. The individual figure and background input reconstructions have multiple small peaks in them that are distributed over the entire epoch, rather than being large and concentrated after the major transitions in the stimulus at 0 and 500 msec. Note also that the small peaks in the individual reconstructions cancel corresponding peaks in the bias term at approximately 350 msec after each image update. The spurious peaks are likely due to mod-2 reduction effects. In the B1m1 analysis, there is no activity in the input 2 partial reconstructions, despite the fact that input 2 is specifically driven. These features each make it more difficult to interpret the B1m1 partial reconstructions.

We expected that mutual slices, and thus the mutual-group reconstructions, would be important when there was no gap and greatly reduced when the gap was present. This is the case for the B01 analysis where the reconstructions from self and mutual groups are all of large amplitude when there is no gap (Figure 18, left panels). When the gap is inserted (Figure 19, left panels), the reconstruction based on the mutual slices drops in amplitude by a factor of approximately 7, while the sum of the two self-group reconstructions yields the measured response, more or less within the measurement noise. The adequacy of linear summing the figure and background group reconstructions is expected given the absence of a nonlinear interaction as indexed by the mutual group reconstruction. In addition, the measured response becomes symmetric between the first and second halves of the stimulus cycle. This is also consistent with the gap eliminating the interaction between figure and background. In the absence of a strong interaction, the response is dominated by the independent changes of orientation that occur for both the figure and background regions. We expect this orientation-change response to be symmetric, given that horizontal and vertical stimulus orientations are both strongly represented and not distinguishable in the mass response of the EEG.

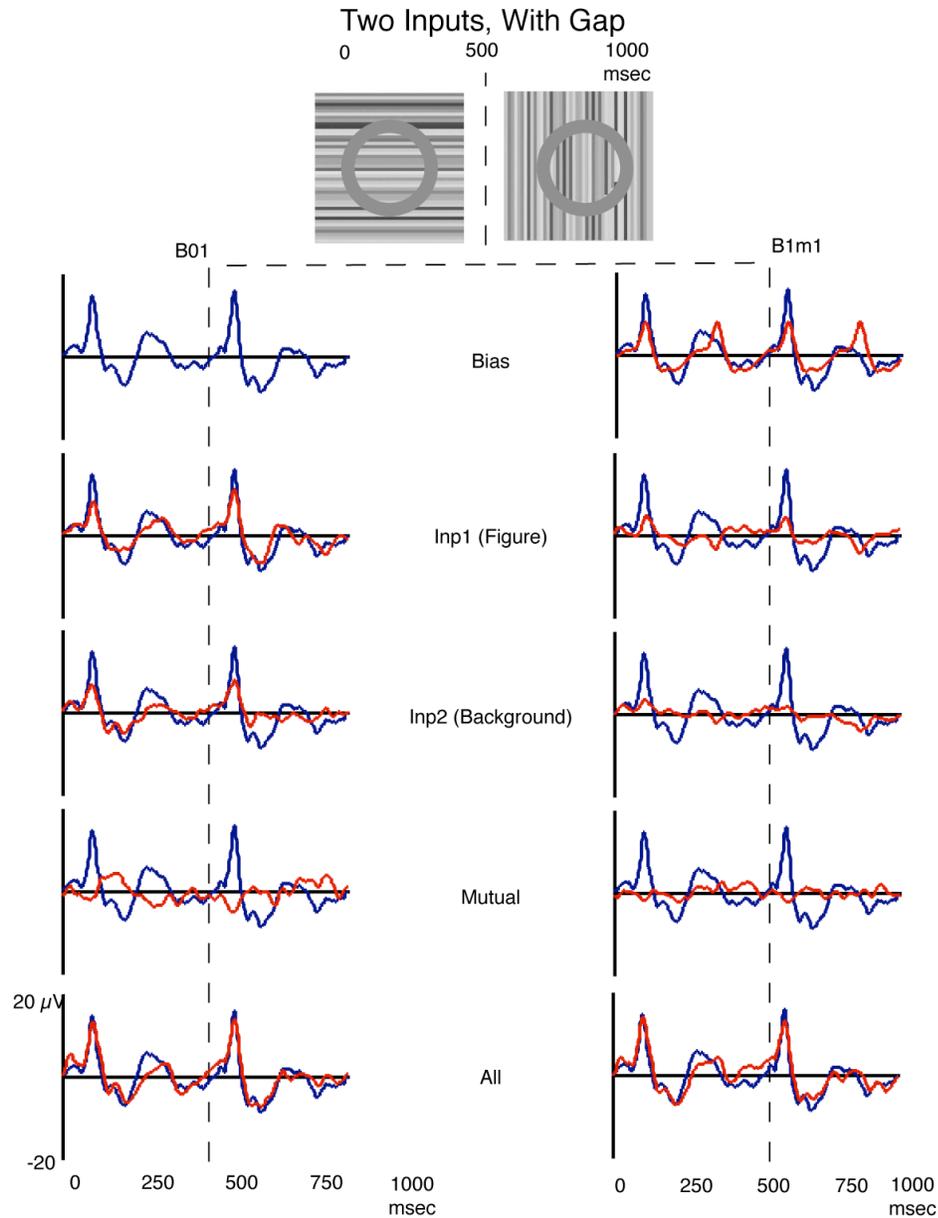

*Figure 19.* Reconstruction of texture-segmentation evoked responses, both inputs active, Gap stimulus. Plotting convention as in Figure 18. The presence of a gap between the figure and background regions causes the periodic response to be symmetric, rather than asymmetric, as is the case in Figure 18. The full reconstructions (red curves, All) for both B01 and B1m1 methods capture the effects of this stimulus manipulation. Note again that the partial reconstructions are very different between the two methods.

In the no gap condition, the B1m1 bias term is similar to that measured in the gap case (compare right panels of Figures 18 and 19). The figure and background group reconstructions are smaller than in the no gap condition and are poorly structured. Finally, the mutual reconstruction differs in size between the no-gap and gap conditions, but only by a factor of about 2. Nonetheless, the full reconstruction is quite accurate, so the diffuse structure of the group reconstructions is not simply noise. The difficulty here is relating the structure of the group reconstructions to events in the stimulus and ultimately to the measured response. As we have seen both theoretically and in the simulations, activity that in Volterra analysis would be represented in quadratic (or of higher even degree) slices in the mutual group can shift under mod-2 reduction into the binary bias group. Shifting can also occur from the mutual group to the self group. It is this shifting of activity that adds to the difficulty of interpreting B1m1 partial reconstructions by group.

The similarity of the B01 self-group reconstructions and the measured response (which is approximately the full reconstruction) might be telling us about segmentation. The display figure and background combinations associated with input packets that drive the two self group slices, and the combinations associated with the actual input in Figure 18, all involve a periodic change from segmented to unsegmented states. In the case of the self-group combinations, the segmented state is due to a difference in the *orientation* of the figure and background textures, while in the measured response, it is due to a difference in *alignment* of vertically oriented figure and background textures. This suggests that either orientation or alignment differences are similarly powerful cues for segmentation and that yield similar patterns of figure/background interaction.

**11. Noise and binary slices**

Systemic noise works its way into the slices, whence it contaminates reconstructions. Because of the linear relationship between the kernels and the responses from which they are derived, most clearly seen in Equation (18), additive noise contamination of the slices depends only on the noise and the S-matrix. In particular, it is independent of the system itself as represented by the uncontaminated slices. As usual, redundant response collection allows noise reduction. We touched on this Section 3.2 and in the discussion of overdetermined S-matrices in Section 3.3. However, even without overdetermination, B1m1 slices calculated from a complete m-sequence test look less noisy than the responses from which they are computed. This noise reduction is quite marked, and is uniform.  It may be understood as a consequence of the orthonormality of the S-matrix and the projection procedure for kernel calculation.  The noise at the each of the samples gets added with pseudo-random signs into each of the slices, so the consequent reduction goes with the square root of the m-sequence length.  Although it is undeniably better to have less noisy slices, the benefit from noise reduction via this path is somewhat limited, since upon reconstruction, because every kernel is need to reconstruct the response at each state (AKAT), the apparent gain in signal-to-noise is lost.  This phenomenon is mathematically necessary, since full reconstruction of the m-sequence input exactly recreates the noisy responses that were measured and used to construct the slices.  Of course, if one knows *a priori* that certain slices are identically zero, those slices, and their noise contamination need not be used in reconstruction, which leads to a real noise reduction.  But, in this case, the effective S-matrix is overdetermined, and the noise advantage might as well be ascribed to that fact.

In B01, the situation is quite different.  It was first investigated by (Yokota & Usui, 1999).  Here noise reduction in the slices is not uniform.  Instead, as derived from a m-sequence experiment, the noise in a kernel goes, remarkably, with the square root of the number of sub-signatures in the kernel signature.  This can be argued analytically from the particular lower triangular form of the S-matrix (19).  This argument is outlined in Appendix VII.  In an alternative approach, Yokota & Usui show how to compute the B01 slices from the B1m1 slices, and use that methodology to develop formulas for the signature-dependent B01 noise, most notably in the overdetermined case.  Figure 20 shows the results of a numerical experiment that confirms these theoretical conclusions.

In passing, Yokota & Usui observed the very useful fact that for a general binary system, all (noise-free) B1m1 slices of order greater than k are identically zero if and only if the same holds for all (noise-free) B01 slices.

Figures 21 and 22 show slices computed for an example 1-input, PLNL system.  The system was tested by a slow-stimulus, m-sequence experiment.  The order of the m-sequence was 4, and the number of samples per stimulus interval was 15.  Responses were simulated, and either left clean or contaminated by gaussian noise whose standard deviation was roughly 10% of the measured standard deviation of the noise-free responses.  The black traces in Figure 21 represent the noise-free B1m1

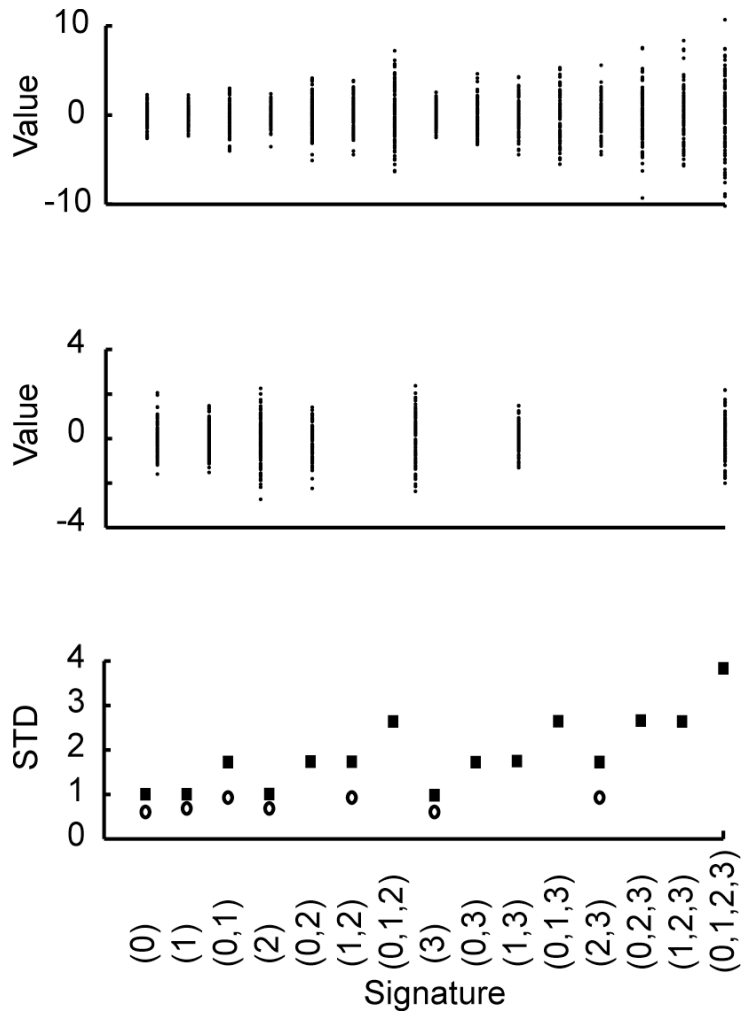

*Figure 20.* Noise affects on B01 slices are not uniform. Because of the linear connection between the slices and the responses, the effects of additive noise on the slices can be obtained by computing the slices from a response sequence drawn from the noise distribution. The figure illustrates the situation for a B01 system with memory length 4. The top panel shows the distribution of kernel values as a function of signature obtained from a numerical experiment of 100 trials, in each of which a sequence of $2^4-1$ normally distributed, standard-deviation-1 responses was converted to kernels by full S-matrix inversion. The square markers in the bottom panel show the standard deviation of the kernel distribution (actually from another experiment with more trials, to make the outcomes more consistent). In the theoretical limit, each standard deviation value is the square root of the number of sub-signatures in the corresponding kernel signature. The middle panel shows the results of the same experiment under the assumption that only the $(0)$-slice and $(0,1)$-slice contain non-zero kernels. These overdetermined slices were calculated by SVD inversion of the restricted S-Matrix. These slices are less noisy, as shown by the circles in the bottom panel.

slices. Note that only slices of order <= 2 are non-zero. The overlying gray traces in the figure represent the slices computed from the contaminated responses. As expected from the previous discussion, the differences between the gray and black traces look

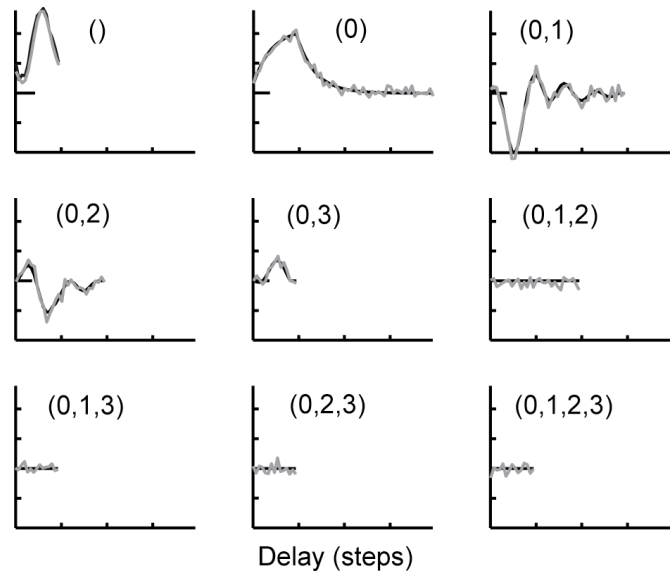

Delay (steps)

*Figure 21. The B1m1 slices for an example 1-input PLNL system showing the effects of additive noise. The black traces are the slices computed from noise-free responses, and the overlying gray traces are the slices computed from a single-pass m-sequence experiment whose responses were contaminated by normally-distributed noise of standard deviation 0.1 units added at each timestep. The standard deviation of the noise-free responses was on the order of 1 unit. The long tick on the vertical axis marks zero slice value. The plots are scaled identically. Note that for this system the noise-free slices of order > 2 are identically zero. The noise is distributed uniformly over slices, as can be clearly seen in the high-order slices and in the tails of the $(0)$- and $(0,1)$-slices.*

uniform and small. Figure 22 shows the analogous B01 kernels computed from precisely the same (noise-free and noisy) data. Again expectations based on the discussion above are realized. First, the only noise-free slices of order <= 2 are non-zero. Second, the differences between the gray and black traces are rather uniform on each slice but grow dramatically with slice-signature complication.

    Finally, Figure 23 shows the reconstruction of the response to an example input calculated from the slices. The black trace is the reconstruction from the noise-free slices. Both B1m1 and B01 reconstructions yield exactly the same trace, and that trace exactly equals the directly simulated PLNL system response to the same input. The gray trace is the reconstruction from the noisy slices. Again, both the B1m1 and B01 reconstructions are identical, and reproduce, input state by state, the noisy responses measured during the experimental test. This should not be a surprise, since, up to computational error, which is minute at the low order of the experiment, the full set of responses and the full set of kernels encapsulate the same information.

    The standard deviation of the samplewise difference between the gray and black curves in Figure 23 was calculated and found to be very nearly 0.1, which was to be

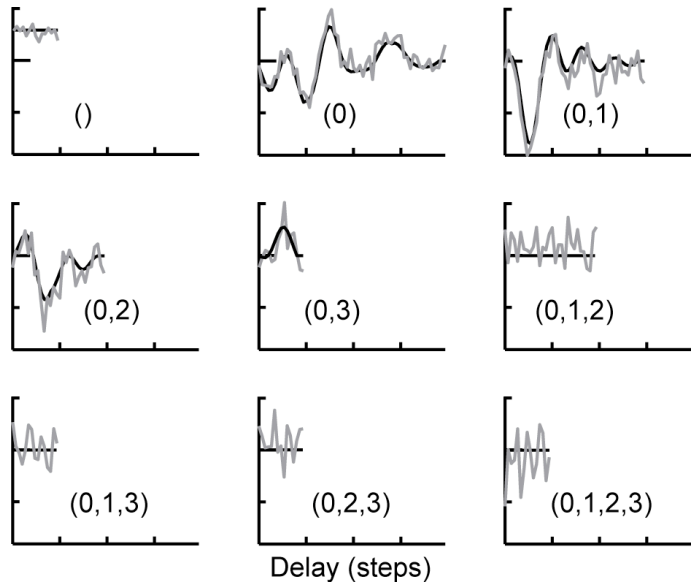

*Figure 22.* The B01 slices for the same system and experiment as Figure 21. The scales of all the plots here are the same, but the vertical scale is different from that in the previous figure. For interpretive purposes, only the relative magnitudes of the slices and noise are important. Note that again the noise-free slices of order > 2 are identically zero. This is not an accident. Yokota & Usui derive the generally useful fact that for an arbitrary system, all (noise-free) B1m1 slices of order > k are identically zero if and only if the same holds for all (noise-free) B01 slices. Unlike the B1m1 situation, the effect of the noise on the slices is not uniform, but instead, increases with signature complication as shown in Figure 20.

expected from the large total number of samples (16 steps, 15 samples per step) reflected in the figure.

Because the PLNL system produced non-zero slices only of order <= 2, we are free to recompute the B01 slices using an S-matrix from which the columns corresponding to signatures having order > 2 have been eliminated. Using that matrix and the full experimental data set, the non-zero slices become overdetermined, and as discussed previously, solution by SVD inversion leads to quieter slices. These quieter slices produce quieter reconstructions. When this procedure was carried out, the standard deviation of the difference between the noisy and quiet reconstructions of the input in Figure 23 was approximately 0.082.

In B1m1 analysis, one may simply compute all the slices, and then just use the non-zero ones in the reconstruction. This leads to about the same improvement as in the restricted B01 calculation.

We ran another experiment, based on an m-sequence of order 8 and the same rate of slow-stimulation. The increased m-sequence length together with restriction to the non-zero signatures makes the slice calculations considerably more overdetermined. The resulting B01 noisy slices were quiet enough so that all the noise-free slice features were clearly discernible. Using these slices, the difference between the noisy reconstruction of the input of Figure 23 and the noise-free one had a standard deviation of approximately 0.032.

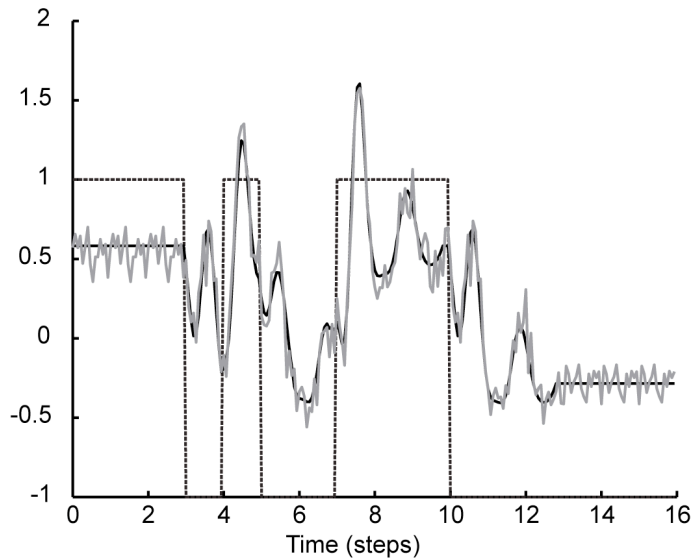

*Figure 23. An example reconstruction from the slices. This figure shows reconstructions from the slices derived from the noise-free (black) and noisy (gray) responses of the system and experiment of Figs. 21 and 22 to an example input (dashed black). Both B1m1 and B01 slices yield exactly the same reconstructions, and these are identical with the system response as measured during the experiment. This result is mathematically necessary. The standard deviation of the difference between the gray and black curves is very nearly 0.1, which is to be expected from the large total number of samples (16 steps, 15 samples per step) reflected in the figure.*

## 12. Conclusions

We have reviewed here the Volterra model and discussed the kinds of systems to which it can be effectively applied. The presentation highlighted the importance of input design choices. We emphasized the notion of subpacket-correction that is implicit in the Volterra representation of a non-linear system. We discussed the utility of a reduction to binary inputs in suitable cases. The subpacket-correction property leads naturally to the B01 formulation of the binary version of the Volterra kernels. We reviewed the relationship between the B01 formulation and the widely used B1m1 formulation and show the benefits of the B01 approach. We have presented a matrix-based framework for kernel calculation that makes the connections between the different kernel methods clear. This framework emphasizes kernel slices, which bind together kernel elements into more interpretable pieces. Within this framework, we showed how sampling considerations can effect the interpretability of kernels, pointing out a number of specific pitfalls that may be encountered in practical experiments. We emphasized the importance for B1m1 analysis of full reconstruction of responses to particular input patterns rather than examination of slices for interpretation of system responses. In contrast we showed that B01 analysis yields useful group reconstructions either at the group or even slice level that are more directly interpretable than B1m1 slices. We illustrated key points through simulation of model systems and a human VEP experiment. Finally, we discussed the quite different effects of noise in B1m1 and B01 analysis.

**Appendix I – Multilevel inputs and column independence**

The theorems in this section show that in the restricted Volterra world of this paper, set out in Section 1.2, S-matrices constructed from inputs limited to finite sets of amplitudes are adequate for kernel extraction, and that, in particular, the number of amplitudes needed depends on the maximum interaction order $n$ of the signatures under consideration, but not on the memory length $T$.

We begin with a well known historical result that is a special case of our main theorem and that provides a nice introduction to the general argument. Consider the matrix

$$\mathbf{S} = \begin{pmatrix} 1 & a_0 & a_0^2 & \cdots & a_0^n \\ 1 & a_1 & a_1^2 & \cdots & a_1^n \\ 1 & a_2 & a_2^2 & \cdots & a_2^n \\ \cdots & \cdots & \cdots & \cdots & \cdots \\ 1 & a_n & a_n^2 & \cdots & a_n^n \end{pmatrix} \quad (30)$$

Matrices of this form go by the name of Vandermonde, and have long been known to have non-zero determinant if and only if the $a_i$ are all distinct. Consider further the restricted Volterra system of memory $T = 1$ whose signatures comprise all interaction orders up to and including $n$. Since $T = 1$, input states for this system each comprise single amplitudes. If the amplitudes are drawn from a set $a_0, a_1, a_2, \ldots, a_n$, then $\mathbf{S}$ defined in Equation (30) is clearly the S-matrix for the system and the states. If the amplitudes are all distinct, the matix is invertible, and so is adequate to convert a set of observed responses for the system into a full set of kernels.

One can see the truth of the Vandermonde assertion by a very simple induction argument. If we temporarily think of the value $a_n$ in (30) as a variable $t_0$, we obtain the form

$$\mathbf{S}(t_0) = \begin{pmatrix} 1 & a_0 & a_0^2 & \cdots & a_0^n \\ 1 & a_1 & a_1^2 & \cdots & a_1^n \\ 1 & a_2 & a_2^2 & \cdots & a_2^n \\ \cdots & \cdots & \cdots & \cdots & \cdots \\ \hline 1 & t_0 & t_0^2 & \cdots & t_0^n \end{pmatrix} \qquad (31)$$

Computing the determinant of (31) by expanding in minors of the last row shows that the determinant $|\mathbf{S}(t_0)|$ is a polynomial of degree $n$ in $t_0$. Furthermore, by induction on $n$, the minor corresponding to the highest degree monomial, $t_0^n$, which is the coefficient $t_0^n$ in $|\mathbf{S}(t_0)|$, is non-zero. This means, in particular, that $|\mathbf{S}(t_0)|$ is not the identically-zero polynomial. Separately substituting each of the values $a_i$ for $t_0$ in (31), we obtain a matrix whose $i^{\text{th}}$ and $n^{\text{th}}$ rows are identical. Such a matrix has zero determinant. Thus we may conclude that each of the values $a_0, a_1, a_2, \ldots, a_{n-1}$ is a root of the polynomial $|\mathbf{S}(t_0)|$. If these values are distinct, we can further conclude that they represent *all* the roots of $|\mathbf{S}(t_0)|$, since a non-zero polynomial of degree $n$ can have no more than $n$ distinct roots. Thus, any other value, for example $a_n$, is *not* a root. That is, the determinant of (31) is not zero.

The main theorem of this section is that for a degree $n$ system, only $n+1$ input levels are needed to ensure column independence of the S-matrix. In particular: For any value of discrete memory $T$ and any interaction order $n$, if $\Sigma$ is a collection of signatures each comprising delays $d_j < T$ and having total number of delays (including multiplicity) $\leq n$, then for any $n+1$ distinct input levels $a_0, a_1, \ldots, a_n$, it is possible to find a set of of input states constructed entirely from these levels such that the columns of the S-matrix $\mathbf{S}$ formed from the states and signatures are all independent.

If there were no restrictions on input values, any real number could be appear in memory at each delay. Thus, the possible states of a memory $T$ system comprises a $T$-dimensional Euclidean space $\mathbf{R}^T$. Among the set of all states, the states satisfying the restriction that all input values are found among the distinct levels $a_0, a_1, \ldots, a_n$ comprise a grid $\mathbf{G}$ in $\mathbf{R}^T$ of $(n+1)^T$ discrete points. Most of the work in proving the main theorem is encapsulated in the following lemma: If $P(t_0,\ldots,t_{T-1})$ is a polynomal in $T$ variables of degree no greater than $n$, then if the restriction of $P$ to $\mathbf{G}$ is zero, that is, if $P\left(a_{i_0},a_{i_1},\ldots,a_{i_{T-1}}\right)=0$ for each $\left(a_{i_0},a_{i_1},\ldots,a_{i_{T-1}}\right)\in\mathbf{G}$, then $P$ is the identically zero polynomial. We prove the lemma by induction on $T$. Suppose we know it is true for $T-1$. Let $l$ be any straight line in $\mathbf{R}^T$. Then for some $j\in(0,\ldots,T-1)$, the projection of $l$ onto the $j^{\text{th}}$-axis of $\mathbf{R}^T$ is a 1-1 mapping. Let $\mathbf{R}_j^{T-1}$ be the $T-1$-dimensional subspace of $\mathbf{R}^T$ perpendicular to the $j^{\text{th}}$-axis. For each $i\in(0,\ldots,T-1)$, let $\mathbf{R}_{j,a_i}^{T-1}$ be the $T-1$-dimensional hyperplane of $\mathbf{R}^T$ parallel to $\mathbf{R}_j^{T-1}$ that contains $a_i$. By our projection assumption, the line $l$ intersects each of the hyperplanes $\mathbf{R}_j^{T-1}$ in a single point, and since the $a_i$ are distinct, so are all these intersection points. Now, note that $\mathbf{R}_{j,a_i}^{T-1}$ contains $\mathbf{G}_i$ the discrete $T-1$-dimensional subgrid of $\mathbf{G}$ comprising the points whose $i^{\text{th}}$ element is $a_i$. Thus, the restriction of $P$ to $\mathbf{G}_i$ is zero. Generally, restriction of a polynomial to a subplane in its domain is another polynomial whose degree cannot be higher than that of the original. Consequently, by induction, we have that the original polynomial is identically zero on each hyperplane $\mathbf{R}_{j,a_i}^{T-1}$. But this means that the restriction $P_l$ of $P$ to the line $l$ is zero at each of the $n+1$ intersection points. However, again, since restriction does not increase degree, the degree of $P_l$ is no greater than $n$, and since the intersection points are distinct, we conclude that $P_l$ is identically zero. Finally, since the choice of the line $l$ was entirely arbitrary, we have that $P$ itself is identically zero. This completes the proof of the lemma.

The proof of the general theorem is as follows. We argue by induction on the number $k$ of signatures in $\Sigma$. The matrix of interest is square and of the form

$$\mathbf{S}=\begin{pmatrix} m_1\left(a_{i_{1,0}},a_{i_{1,1}},\ldots,a_{i_{1,T-1}}\right) & m_2\left(a_{i_{1,0}},a_{i_{1,1}},\ldots,a_{i_{1,T-1}}\right) & \cdots & m_k\left(a_{i_{1,0}},a_{i_{1,1}},\ldots,a_{i_{1,T-1}}\right) \\ m_1\left(a_{i_{2,0}},a_{i_{2,1}},\ldots,a_{i_{2,T-1}}\right) & m_2\left(a_{i_{2,0}},a_{i_{2,1}},\ldots,a_{i_{2,T-1}}\right) & \cdots & m_k\left(a_{i_{2,0}},a_{i_{2,1}},\ldots,a_{i_{2,T-1}}\right) \\ \cdots & \cdots & \cdots & \cdots \\ m_1(t_0,t_1,\ldots,t_{T-1}) & m_2(t_0,t_1,\ldots,t_{T-1}) & \cdots & m_k(t_0,t_1,\ldots,t_{T-1}) \end{pmatrix} \quad (32)$$

where $m_j(t_0, t_1, ..., t_{T-1})$ is the monomial in the memory slot variables $t_0, t_1, ..., t_{T-1}$ corresponding to the $j^{\text{th}}$ signature in $\Sigma$. The $T-1$-tuples $(a_{i_{1,0}}, a_{i_{1,1}}, ..., a_{i_{1,T-1}})$, $(a_{i_{2,0}}, a_{i_{2,1}}, ..., a_{i_{2,T-1}})$, etc. are $k$ distinct states comprising amplitudes drawn from the set of distinct amplitudes $a_0, a_1, ..., a_n$. As in the Vandermonde case, expansion of the matrix by minors of the last row shows that the determinant of $\mathbf{S}$, $|\mathbf{S}(t_0, t_1, ..., t_{T-1})|$ is a polynomial in the several variables $t_0, t_1, ..., t_{T-1}$. Here, as previously, by induction, the choice of states may be made so that the minor corresponding to $m_k(t_0, t_1, ..., t_{T-1})$ is non-zero. Since the monomials in the last line of (32) are distinct, they are independent in the space of polynomials, so the fact that the coefficient of $m_k(t_0, t_1, ..., t_{T-1})$ in $|\mathbf{S}(t_0, t_1, ..., t_{T-1})|$ is not zero insures that $|\mathbf{S}(t_0, t_1, ..., t_{T-1})|$ is not the identically-zero polynomial. Again, as in the Vandermonde case, when each of the states $a_{i_{j,0}}, a_{i_{j,1}}, ..., a_{i_{j,T-1}}$ is substituted for the variables $t_0, t_1, ..., t_{T-1}$ in the last row of (32), the resulting matrix has two identical rows and consequently zero determinant, showing that each of the states is a root of $|\mathbf{S}(t_0, t_1, ..., t_{T-1})|$. Our task is to show that there is a choice of restricted state $(a_{i_{k,0}}, a_{i_{k,1}}, ..., a_{i_{k,T-1}})$ which when substituted for $(t_0, t_1, ..., t_{T-1})$ in the last row of (32) makes $|\mathbf{S}(a_{i_{k,0}}, a_{i_{k,1}}, ..., a_{i_{k,T-1}})| \neq 0$. But if for *any* choice $(a_{i_{k,0}}, a_{i_{k,1}}, ..., a_{i_{k,T-1}})$ the determinant were zero, we could apply our lemma to conclude that $|\mathbf{S}(t_0, t_1, ..., t_{T-1})|$ was the identically zero polynomial, which is a contradiction. This completes the proof.

**Appendix II – Repeated inputs**

Divide the S-matrix into blocks corresponding to each separate repetition. These are all identical, so the individual columns across blocks are identical, so any vector in the subspace $\mathcal{S}$, in particular $\text{proj}(\mathbf{R})$, has the same property. One minimizes the distance from $\mathbf{R}$ to $\text{proj}(\mathbf{R})$ by minimizing it for each state. Since for each fixed state $\text{proj}(\mathbf{R})$ has the same value for each repetition, and this number is the unique one that minimizes the sum of the squared distances to the responses to that state, but that means it must be the average of those responses.

**Appendix III – Random inputs**

Generally in this appendix, we restrict our attention to finite distributions. This is adequate for our purposes and always contains the gist of the the argument in the continuous distribution case as well.

*III.1 Overdetermination and likely column independence*

If the number of possible input states, which under our assumption is $C^T$, where $C$ is the number of distinct input values and $T$ is the discrete memory, is less than the number kernels (for example, in a mod-2 equivalent set), then the S-matrix constructed from all the states is under-determined and so the columns cannot be independent. Adding more rows cannot improve matters, since the new rows must duplicate ones already present. On the other hand, as shown in Appendix I, if the number of distinct non-zero amplitudes presented is no less than the interaction order of the columns, then for a generic (see Appendix I for definition) choice of amplitudes, a collection of states can be created from the amplitudes for which the columns of the associated S-matrix are independent. Clearly, as the sequence of inputs extracted from our distribution gets longer, the likelihood of finding all the required states among the inputs tends to certainty.

*III.2 S-matrix columns as delayed sequence products*

Let $s$ be a time-ordered, bi-infinite sequence (one that goes to infinity in both time directions) of real numbers $s_k$ which we take to be inputs to a reduced Volterra system. Let $(d_0, d_1, \ldots, d_{n-1})$ be a set of delays comprising a non-empty signature. Then we define the "signature-state product of the sequence" $s$ to be the sequence whose elements are the SSPs at each time $k$, $S(d_0, d_1, \ldots, d_{n-1}; k) = s_{k-d_0} \cdot s_{k-d_1} \cdots s_{k-d_{n-1}}$ (see Section 3). By definition, this sequence is the column of the bi-infinite S-matrix corresponding to the signature. It is useful to note that the SSP of the sequence can be re-expressed as the element-wise product of delayed copies of the sequence $s$, with delays determined by the signature. From this observation, in particular, it is immediate that the columns corresponding to kernels comprising a single slice are each just delayed copies of the column representing the primary signature of that slice.

*III.3 Statistical preliminaries*

For the manipulations we have in mind to work, we require an input sequence $s$ having the property that, for each non-empty signature $(d_0, d_1, d_2, \ldots, d_{n-1})$, the average value of the SSP of the sequence exists (that is, is defined and finite), and is non-zero if and only if all signature delays appear in duplicate pairs. Applying the definition of the average, and the characterization of the SSP as the product of delayed copies of the sequence, the requirement is given by the expression

$$\lim_{p \to \infty} (2p+1)^{-1} \cdot \sum_{k=-p}^{p} s_{k-d_0} s_{k-d_1} \cdots s_{k-d_{n-1}} \begin{cases} \neq 0, \text{ finite, all delays paired} \\ = 0, \text{ otherwise} \end{cases}. \quad (33)$$

Thus, when $n = 1$, the requirement translates to the statement that the mean of the sequence is zero, and when $n = 2$, that the autocorrelation function of the sequence is a constant multiple of the (discrete) delta function. This last is the usual definition of a sequence being "white".

In the white-noise technique, the elements of $s$ are drawn as a random sample from a probability distribution $\mathbf{D}$. The word "noise" in the name of the technique refers to this randomness. In the original formulation, $\mathbf{D}$ was the Gaussian distribution. Subsequently, other distributions have been used.

We note here that the relatively modest condition that $\mathbf{D}$ is non-zero and has exponentially decaying tails and zero mean is enough to guarantee that our requirement as expressed in (33) is satisfied for all non-empty signatures. This may be seen as follows: One form of the so called Law of Large Numbers (Feller, 1971) states that if $X_1, X_2, X_3, \ldots$ is a sequence drawn at random from a distribution $\mathbf{D}$ having finite mean $\mu$ and variance $\sigma^2$, then the sequence of sample means $\bar{X}_w = (X_1 + \ldots + X_w)/w$ limits "almost surely" to the mean of $\mathbf{D}$. That is, $\text{Prob}(\lim_{w \to \infty} \bar{X}_w = \mu) = 1$. We can apply this theorem directly to validate the requirement expressed in (33) in the case $n = 1$. In our situation, we have assumed that the tails of $\mathbf{D}$ decay exponentially which implies that the hypothesis that the first two moments are finite is satisfied. We also assumed that the mean of $\mathbf{D}$ is 0. We can, of course, take the sample groups to be indexed by $-p \leq k \leq p$. Applying the theorem, we conclude that the equality in (33) holds with probability 1. This is good enough for all practical purposes.

For $n > 1$, the samples indexed by $-p \leq k \leq p$ may be understood to come from distributions $\mathbf{P}$ that are products of powers of the distribution $\mathbf{D}$. The first and second moments of those distributions will exist if higher power moments of $\mathbf{D}$ exist, and that will be guaranteed if the tails of $\mathbf{D}$ decay exponentially.

Once one knows that the limits in (33) exist, the particular limit values required are gotten as follows: If all the delays $(d_0, d_1, d_2, \ldots, d_{n-1})$ are in duplicate pairs, then $\mathbf{P}$ is non-zero and non-negative, so has positive mean. If at least one of the delays is unpaired, then $\mathbf{P}$ may itself be taken as the product of a distribution with zero mean and another with finite mean, which makes the mean of the product zero.

*III.4 Rate of convergence*

When applying the white-noise method to stimuli that have only finitely many steps, knowledge of the *rate* of convergence to the limit is important. The same hypotheses required by the Law of Large Numbers quoted above, that the mean and variance of $\mathbf{D}$ exist, also imply (Bendat & Piersol, 1986) that $E\left[(\bar{X}_w - \mu)^2\right] = \sigma^2 / w$, which we interpret as saying that the expected variance of the sample means goes with $1/w$. In the context of (33), the variance $\sigma^2$ substituted into the formula depends on the number of delays in the signature, and the denominator is $2p+1$ which is of order $p$.

*III.5 Intercolumn angles*

Consider the space of bi-infinite column vectors $\mathbf{S}_i$ having entries $\mathbf{S}_{ki}$ for which

$$\lim_{p \to \infty} (2p+1)^{-1} \cdot \sum_{k=-p}^{p} \mathbf{S}_{ki}^{2} \text{ exists.} \tag{34}$$

The formula

$$\langle \mathbf{S}_i, \mathbf{S}_j \rangle = \lim_{p \to \infty} (2p+1)^{-1} \cdot \sum_{k=-p}^{p} \mathbf{S}_{ki} \cdot \mathbf{S}_{kj}. \tag{35}$$

is well defined (Cauchy-Schwarz) on this space and defines an "inner product" that makes the space into a Hilbert space which we denote $\mathbf{R}^\infty$. Following the usual conventions we use the symbol $\|\mathbf{S}_i\|$ as shorthand for $\sqrt{\langle \mathbf{S}_i, \mathbf{S}_i \rangle}$, the length (or "norm") of the vector $\mathbf{S}_i$.

Suppose we have a bi-infinite white-noise input sequence $s$ constructed according to the method of the previous section, and let $\mathbf{S}$ be the corresponding S-matrix. Then since (33) holds, the columns of the S-matrix $\mathbf{S}$ described above satisfy (34), as can be seen by substituting $s_{k-d_0} s_{k-d_1} \cdots s_{k-d_{n-1}}$ for $\mathbf{S}_{ki}$. Thus, the columns of S represent vectors in $\mathbf{R}^\infty$, and inner products of pairs of columns determine angles according to the usual formula

$$\cos(\alpha_{ij}) = \frac{\langle \mathbf{S}_i, \mathbf{S}_j \rangle}{\|\mathbf{S}_i\| \cdot \|\mathbf{S}_j\|}. \tag{36}$$

Substituting the delayed products for the two columns in (35) yields

$$\langle \mathbf{S}_i, \mathbf{S}_j \rangle = \lim_{p \to \infty} (2p+1)^{-1} \cdot \sum_{k=-p}^{p} s_{k-d_0} s_{k-d_1} \cdots s_{k-d_{n-1}} \cdot s_{k-d'_0} s_{k-d'_1} \cdots s_{k-d'_{n'-1}} \tag{37}$$

where the unprimed delays refer to $\mathbf{S}_i$ and the primed delays to $\mathbf{S}_j$. The next equation

$$\langle \mathbf{S}_i, \mathbf{S}_j \rangle = \lim_{p \to \infty} (2p+1)^{-1} \cdot \sum_{k=-p}^{p} s_{k-d''_0}^{m''_0} s_{k-d''_1}^{m''_1} \cdots s_{k-d''_{n''}}^{m''_{n''}} \tag{38}$$

is (37) where the unprimed and primed delays have been mingled and grouped. It is assumed there are no duplications among the $d''_j$. Each factor $s_{k-d''_j}^{m''_j}$, $k$ varying, represents an element-wise power of a delayed copy of the original sequence, and thus a random sample drawn from the $m''_j$ power distribution of the original distribution $\mathbf{D}$. Since the samples drawn from $\mathbf{D}$ at distinct delays are independent, the product $s_{k-d''_0}^{m''_0} s_{k-d''_1}^{m''_1} \cdots s_{k-d''_{n''}}^{m''_{n''}}$ is a random sample drawn from the joint-product distribution $\mathbf{P}$ of the individual power distributions. The right side of Equation (38) represents the mean of the distribution $\mathbf{P}$. If each of the $m''_j$ is even, and if $\mathbf{D}$ is not identically zero, then there are no negative values in $\mathbf{P}$, and at least one positive value, which makes the inner product $\langle \mathbf{S}_i, \mathbf{S}_j \rangle$ positive, and hence $\mathbf{S}_i$ and $\mathbf{S}_j$ are not orthogonal. Since the columns are not identical, the angle between them is not zero.

*III.6 Kernel dependence on distribution*

The result that columns corresponding to mod-2 inequivalent signatures are orthogonal can be extended to show that if $\mathbf{D}'$ is another distribution satisfying the same conditions as $\mathbf{D}$, then a column formed from $\mathbf{D}'$ and one formed from $\mathbf{D}$ are orthogonal if they have mod-2 inequivalent signatures. Combined with the last observation in Section 3.3, this shows that in the case of noise-free measurements, the dependence of the kernels on the distribution is intra mod-2 class.

If the distributions $\mathbf{D}$ and $\mathbf{D}'$ comprise the same finite set of amplitudes, then the sets of states presented to the system is the same in experiments based on the two distributions. Only the relative number of repeated states is different. Since the repeats may be averaged before kernel extraction, we can conclude that the kernels produced by the two experiments are identical

**Appendix IV – Pseudo-random inputs and m-sequences**

The pseudo-random sequences of interest here have relatively short periods, ideally somewhat longer than the number of kernels to be determined, but, crucially, having the property that Equation (33) holds without the limit, that is, that

$$\sum_{k=0}^{p} s_{k-d_0} s_{k-d_1} \cdots s_{k-d_{n-1}} \begin{cases} \neq 0, \text{ all delays paired} \\ = 0, \text{ otherwise} \end{cases} \quad (39)$$

for delays representing all signatures required for the reduced analysis. In interpreting Equation (39), elements $s_{k-d_i}$ for small $k$ are defined by the assumed periodicity of the sequence, or equivalently, taken from the run-up inputs. It is remarkable that such sequences exist. In fact, binary m-sequences, having values 1 and -1, always satisfy (39) without any restriction on the magnitude of the delays (Sutter, 1992). As we show in the next subsection, pseudo-random sequences having more than 2 levels may be easily constructed from m-sequences. However, there is a caution: The method described below produces pseudo-random inputs for which (39) holds for delays having limited magnitudes.

M-sequences are intimately related to the mathematical theory of finite (Galois) fields (Golomb & Gong, 2005). Study of these fields yields elegant proofs of the specific properties of m-sequences, and in addition shows how pseudo-random p-level sequences, p a prime number, having properties analogous to m-sequences (p=2), may be directly generated.

Pseudo-random inputs are not a panacea. The chronic drawback of long experiment times remains. If $T$ is large and the reduced analysis order even modest, a pseudo-random sequence exactly satisfying (39) can be impracticably long.

*IV.1 Multi-level*

It is possible to construct multiple-level pseudo-random inputs satisfying (39) from a suitable (extended) m-sequence. The method illustrated below produces inputs having $2^k$ levels, integer $k > 1$. Suppose the multi-level sequence is to have discrete memory $T$, and, as an example, is to have 4 distinct levels. Let $s_k$ be the elements of an m-sequence of order $2T$. The individual elements in the sequence have values 1 or -1. Consider the sequence

$$c_k = as_k + bs_{k-T} \qquad (40)$$

where $a$ and $b$ are positive real numbers. Clearly, for almost all choices of the real numbers, the sequence has the desired 4 distinct levels. Substituting our sequence into (39), and specializing to the case where there are just two factors, the summation becomes

$$\sum_{k=0}^{p} (as_{k-d_0} + bs_{k-d_0-T})(as_{k-d_1} + bs_{k-d_1-T}). \qquad (41)$$

Multiplying out and distributing the summation, we have

$$a^2 \sum_{k=0}^{p} s_{k-d_0} s_{k-d_1} + ba \sum_{k=0}^{p} s_{k-d_0-T} s_{k-d_1} + ab \sum_{k=0}^{p} s_{k-d_0} s_{k-d_1-T} + b^2 \sum_{k=0}^{p} s_{k-d_0-T} s_{k-d_1-T}. \qquad (42)$$

Here, $d_0$ and $d_1$ are in the range $(0,\ldots,T-1)$. If $d_0 = d_1$, by the fact that (39) holds for the m-sequence, the first and last terms are positive, and middle two terms, representing delays differing by $T$, are zero, making the entire expression positive. If $d_0 \neq d_1$, then all four summands are zero, completing the argument that (39) holds for $c_k$ when there are just two delays. The general case where there are several delays is confirmed similarly.

*IV.1 Multi-input binary*

For a system having discrete memory $m$ and having $k$ inputs, use of a single m-sequence of length $2^{m \cdot k}$ is adequate. The simplest procedure is to divide the length $m \cdot k$ state at each time $t$ into $k$ contiguous substates, each of length $m$, and use the most recent element of the substates to specify the $k$ input values at time $t$. If this is done, and if the system responses are processed against the order $m \cdot k$ m-sequence as though the system had only one input, then each single-input kernel actually equals a multi-input kernel. The association of multiple-input signatures (duplicate-free delays in the range $0,\ldots,m-1$ for each input) with single-input signatures (duplicate-free delays in the range $0,\ldots,m \cdot k-1$) obtained by multiplying, for each $j$, the $j^{th}$ input signature delays by $j$ and concatenating the results, identifies equal multiple- and single-input kernels.

It is important to note that only certain schemes for multiply sampling a single m-sequence (even if it is adequately long) to provide multiple inputs results in a complete multi-input test that can be used to determine a full set of multi-input kernels. One must check that the scheme used actually works. This can always be done numerically.

**Appendix V – B1m1 self-similarity**

The reordering leading to a self-similar matrix for B1m1 analysis is as follows: Since the delays in binary signatures are duplicate-free, we can use the set of delays in a signature to pick out bits (delay 0 is low-bit) to set in the binary representation of a non-negative integer. This procedure uniquely associates each binary signature to a number. Reorder the columns so that the numbers increase monotonically. Similarly, the delays in a state specification corresponding to B1m1 input value -1 (present in memory is delay 0) leads to a conversion of states to non-negative integers. Reorder the states so that these numbers increase monotonically. When this has been done, the S-matrix will have the desired form. This is illustrated in (43) in the case where $T = 3$.

$$\begin{pmatrix} +1 & +1 & +1 & +1 & +1 & +1 & +1 & +1 \\ +1 & -1 & +1 & -1 & +1 & -1 & +1 & -1 \\ +1 & +1 & -1 & -1 & +1 & +1 & -1 & -1 \\ +1 & -1 & -1 & +1 & +1 & -1 & -1 & +1 \\ \hline +1 & +1 & +1 & +1 & -1 & -1 & -1 & -1 \\ +1 & -1 & +1 & -1 & -1 & +1 & -1 & +1 \\ +1 & +1 & -1 & -1 & -1 & -1 & +1 & +1 \\ +1 & -1 & -1 & +1 & -1 & +1 & +1 & -1 \end{pmatrix} \quad (43)$$

Notice, as an indication of the self-similar structure, that three of the delineated 4x4 submatrices are identical, and the fourth is the negative of the others. The inductive possibilities of the form may seen by observing that the upper-left submatrix may itself be divided into submatrices having that have the same repetition pattern as the full matrix. Self-similar B1m1 kernel extraction was done, in the same spirit, but a little differently, by (Sutter, 1992) using the "Fast Walsh Transform" (FWT) which is the binary analog of the Fast Fourier Transform (FFT).

**Appendix VI – Experimental Methods**

The displays were generated with conventional bit-mapped graphics (800 X 600 pixels, 72 Hz refresh) on a Sony GDM-400 color monitor. The figure/ground presentation was based on one-dimensional random luminance bar textures. The smallest width of a bar was 6 arc min. Texture-defined "objects" were defined by the relative orientation or alignment of two regions of the display, the figure and background regions. In the main condition (Figure 17, No Gap) the figure region consisted of 9 disk-shaped patches of texture (4 deg diameter, 6 deg center-to-center on a square grid) that changed orientation between horizontal and vertical according to a pseudo-random binary m-sequence. The background consisted of a 20 by 20 deg region of uniform texture that filled the space between and surrounding the figure regions. The background orientation was modulated between horizontal and vertical with an independent m-sequence. When both regions were in the horizontal state (labeled 0 in Figure 17), the entire field was covered with a uniform texture. When the background was vertical (in its "1" state), a set of 9 orientation-defined figures was visible (Figure 16, second panel). When both figure and background regions were vertical (both in the 1 state, Figure 16, third panel), the figures were defined by a difference in relative alignment. In practice, the image texture in the figure regions was rotated by 90 deg in the opposite direction to the direction of the background region. When the figure was vertical and the background was horizontal, the figure was also defined by a relative orientation difference between the regions (the 1, 0 state, Figure 1d). The Figures illustrates only a portion of the display.

In a control experiment, the texture defining the figure region separated from the background texture by a mean luminance gray gap of 0.25 deg. This condition was run to test the extent to which responses depended on continuity in the image. The stimuli were otherwise identical to those of the main experiment. This condition and the main condition were run in one block of trials, where each trial lasted approximately 35 sec. The m-sequence for each input was updated every 500 msec.

Time averaged responses were also recorded from periodic image sequences that were based on an alternation between the first and second and first and third images of Figure 16 (see insets in Figures 17 and 18) or the the first and third image of Figure 16, but with a gap (see inset of Figure 19). Data from these conditions was compared to reconstructed waveforms that were generated from the kernels derived from the corresponding m-sequence runs. In the periodic runs, the images were updated every 500 msec (1 Hz full cycle) in trials that lasted 10 sec. A total of 20 periodic trials were collected per condition (20 sec total) and 24 trials lasting 35 sec each were collected in the two m-sequence conditions (840 sec per condition in total).

*EEG recording*

The electroencephalogram (EEG) was recorded with 128-channel HydroCell Sensor Nets (Electrical Geodesics, Eugene OR) that utilize silver-silver chloride electrodes embedded in electrolyte soaked sponges. The EEG was amplified at a gain of 1,000 and recorded with a vertex physical reference. Signals were 0.1 Hz high-pass and 200 Hz (elliptical) low-pass filtered and digitized at 432 Hz with a precision of 4-bits per microvolt at the input. Artifact rejection was done off-line. Raw data was evaluated according to a sample-by-sample thresholding procedure to remove noisy sensors which were replaced by the average of the six nearest spatial neighbors. Once noisy sensors were substituted, the EEG was re-referenced to the common average of all the sensors. Additionally, EEG epochs that contained a large percentage of data samples exceeding threshold (~25-50 microvolts) were excluded on a sensor-by-sensor basis.

**Appendix VII – B01 noise**

The argument that the level of additive noise contaminating a B01 slice depends on the number of sub-signatures in the kernel signature goes generally as follows: The number of sub-signatures equals the number of non-zero S-matrix entries in the row whose state delay pattern equals the signature at hand, but the actual argument requires additional subtlety. Straightforward, but involved, algebraic manipulation shows that each kernel is the sum of the sign-adjusted response at each state whose delay pattern is contained in the kernel delay pattern, where the sign is plus if the number of delays in the signature minus the number of delays in the sub-signature is even, and minus otherwise. This prescription gives a constructive method of computing the B01 kernels. Applied to a complete set of pure noise responses, the terms in the sign-adjusted sum have completely random signs, from which the assertion follows.

# References

Anzai, A., Ohzawa, I., & Freeman, R.D. (1997). Neural mechanisms underlying binocular fusion and stereopsis: position vs. phase. *Proc Natl Acad Sci U S A, 94* (10), 5438-5443.

Baseler, H.A., Sutter, E.E., Klein, S.A., & Carney, T. (1994). The topography of visual evoked response properties across the visual field. *Electroencephalogr Clin Neurophysiol, 90* (1), 65-81.

Benardete, E.A., & Kaplan, E. (1997). The receptive field of the primate P retinal ganglion cell, II: Nonlinear dynamics. *Vis Neurosci, 14* (1), 187-205.

Benardete, E.A., & Victor, J.D. (1994). An extension of the m-sequence technique for the analysis of multi-input nonlinear systems. In: V.Z. Marmarelis (Ed.) *Advanced Methods of Physiological System Identification,* 3 (pp. 87-110). New York: Plenum Press.

Bendat, J., & Piersol, A. (1986). Random data: analysis and measurement procedures. (New York: John Wiley and Sons.

Breitmeyer, B.G., & Ogmen, H. (2000). Recent models and findings in visual backward masking: a comparison, review, and update. *Percept Psychophys, 62* (8), 1572-1595.

Candy, T.R., Skoczenski, A.M., & Norcia, A.M. (2001). Normalization models applied to orientation masking in the human infant. *J Neurosci, 21* (12), 4530-4541.

Carandini, M., Heeger, D.J., & Movshon, J.A. (1997). Linearity and normalization in simple cells of the macaque primary visual cortex. *J Neurosci, 17* (21), 8621-8644.

Cavanaugh, J.R., Bair, W., & Movshon, J.A. (2002). Nature and interaction of signals from the receptive field center and surround in macaque V1 neurons. *J Neurophysiol, 88* (5), 2530-2546.

Chen, H.W., Aine, C.J., Best, E., Ranken, D., Harrison, R.R., Flynn, E.R., & Wood, C.C. (1996). Nonlinear analysis of biological systems using short M-sequences and sparse-stimulation techniques. *Ann Biomed Eng, 24* (4), 513-536.

de Zwart, J.A., Silva, A.C., van Gelderen, P., Kellman, P., Fukunaga, M., Chu, R., Koretsky, A.P., Frank, J.A., & Duyn, J.H. (2005). Temporal dynamics of the BOLD fMRI impulse response. *Neuroimage, 24* (3), 667-677.

Doyle, F.J., III, Pearson, R.K., & Ogunnaike, B. (2002). Identification and Control Using Volterra Models. (London: Springer-Verlag.

Feller, W. (1971). An introduction to probability theory and its applications. II (New York: John Wiley and Sons.

Fortune, B., Wang, L., Bui, B.V., Cull, G., Dong, J., & Cioffi, G.A. (2003). Local ganglion cell contributions to the macaque electroretinogram revealed by experimental nerve fiber layer bundle defect. *Invest Ophthalmol Vis Sci, 44* (10), 4567-4579.

Gardner, J.L., Anzai, A., Ohzawa, I., & Freeman, R.D. (1999). Linear and nonlinear contributions to orientation tuning of simple cells in the cat's striate cortex. *Vis Neurosci, 16* (6), 1115-1121.

Gerth, C., Sutter, E.E., & Werner, J.S. (2003). mfERG response dynamics of the aging retina. *Invest Ophthalmol Vis Sci, 44* (10), 4443-4450.

Golomb, S. (1982). Shift Register Sequences. (Laguna Hills, CA: Aegean Park Press.

Golomb, S., & Gong, G. (2005). Signal Design for Good Correlation. (Cambridge, UK: Cambridge University Press.

Hou, C., Pettet, M.W., Sampath, V., Candy, T.R., & Norcia, A.M. (2003). Development of the spatial organization and dynamics of lateral interactions in the human visual system. *J Neurosci, 23* (25), 8630-8640.

Keating, D., Parks, S., Malloch, C., & Evans, A. (2001). A comparison of CRT and digital stimulus delivery methods in the multifocal ERG. *Doc Ophthalmol, 102* (2), 95-114.

Keating, D., Parks, S., Smith, D., & Evans, A. (2002). The multifocal ERG: unmasked by selective cross-correlation. *Vision Res, 42* (27), 2959-2968.

Klein, S.A. (1992). Optimizing the estimation of nonlinear kernels. In: R.B. Pinter, & B. Nabet (Eds.), *Nonlinear Vision: Determination of Neural Receptive Fields, Function, and Networks.* (pp. 109-170). Boca Raton, FL: CRC Press.

Macknik, S.L., & Livingstone, M.S. (1998). Neuronal correlates of visibility and invisibility in the primate visual system. *Nat Neurosci, 1* (2), 144-149.

Marmarelis, V. (2004). Non-linear modeling of physiological systems. (Piscataway, NY: IEEE Press.

Nelles, O. (2001). Nonlinear system identification. (Berlin: Springer-Verlag.

Norcia, A.M., Wesemann, W., & Manny, R.E. (1999). Electrophysiological correlates of vernier and relative motion mechanisms in human visual cortex. *Vis Neurosci, 16* (6), 1123-1131.

Press, W.H., Flannery, B.P., Teukolsky, S.A., & Vetterling, W.T. (1988). Numerical Recipes in C. (New York, NY: Cambridge University Press.

Rashbass, C. (1970). The visibility of transient changes of luminance. *J Physiol, 210* (2), 165-186.

Reid, R.C., Victor, J.D., & Shapley, R.M. (1997). The use of m-sequences in the analysis of visual neurons: linear receptive field properties. *Vis Neurosci, 14* (6), 1015-1027.

Schetzen, M. (1989). The Volterra and Wiener theories of nonlinear systems. (Malabar, FL: Krieger Publishing.

Schiller, P.H. (1968). Single unit analysis of backward visual masking and metacontrast in the cat lateral geniculate nucleus. *Vision Res, 8* (7), 855-866.

Simoncelli, E.P., & Heeger, D.J. (1998). A model of neuronal responses in visual area MT. *Vision Res, 38* (5), 743-761.

Slotnick, S.D., Klein, S.A., Carney, T., Sutter, E., & Dastmalchi, S. (1999). Using multi-stimulus VEP source localization to obtain a retinotopic map of human primary visual cortex. *Clin Neurophysiol, 110* (10), 1793-1800.

Sutter, E. (2000). The interpretation of multifocal binary kernels. *Doc Ophthalmol, 100* (2-3), 49-75.

Sutter, E.E. (1992). A deterministic approach to nonlinear systems analysis. In: R.B. Pinter, & B. Nabet (Eds.), *Non-linear Vision: Determination of Receptive Fields, Function, and Networks* (pp. 171-220). Boca Raton, FL: CRC Press.

Sutter, E.E. (2001). Imaging visual function with the multifocal m-sequence technique. *Vision Res, 41* (10-11), 1241-1255.

Sutter, E.E., & Tran, D. (1992). The field topography of ERG components in man--I. The photopic luminance response. *Vision Res, 32* (3), 433-446.

Tabuchi, H., Yokoyama, T., Shimogawara, M., Shiraki, K., Nagasaka, E., & Miki, T. (2002). Study of the visual evoked magnetic field with the m-sequence technique. *Invest Ophthalmol Vis Sci, 43* (6), 2045-2054.

Victor, J.D. (1992). Nonlinear systems analysis in vision: overview of kernel methods. In: R.B. Pinter, & B. Nabet (Eds.), *Nonlinear vision: determination of neural receptive fields, function and networks* (pp. 1-37). Boca Raton, FL: CRC Press.

Victor, J.D., & Conte, M.M. (2000). Two-frequency analysis of interactions elicited by Vernier stimuli. *Vis Neurosci, 17* (6), 959-973.

Westwick, D., & Kearny, R. (2003). Identification of nonlinear physiological systems. (Piscataway, NJ: IEEE Press.

Wilson, H.R. (1999). Non-fourier cortical processes in texture, form, and motion perception. In: U.e. al. (Ed.) *Cerebral Cortex,* 13 (pp. 445-477). New York: Kluwer Academic.

Wu, M.C., David, S.V., & Gallant, J.L. (2006). Complete functional characterization of sensory neurons by system identification. *Annu Rev Neurosci, 29*, 477-505.

Yokota, Y., & Usui, S. (1999). Precision analysis of estimated kernels of functional series model with binary input. *Electronics and Communications in Japan, Part 3, 82* (5), 11-18.

Zemon, V., & Ratliff, F. (1982). Visual evoked potentials: evidence for lateral interactions. *Proc Natl Acad Sci U S A, 79* (18), 5723-5726.

Zemon, V., & Ratliff, F. (1984). Intermodulation components of the visual evoked potential: responses to lateral and superimposed stimuli. *Biol Cybern, 50* (6), 401-408.

Zenger-Landolt, B., & Heeger, D.J. (2003). Response suppression in v1 agrees with psychophysics of surround masking. *J Neurosci, 23* (17), 6884-6893.

Zhang, X. (2003). Simultaneously recording local luminance responses, spatial and temporal interactions in the visual system with m-sequences. *Vision Res, 43* (15), 1689-1698.

Zierler, N. (1959). Linear recurring sequences. *JSoc Indust Appl Math, 7*, 31-49.